\documentclass[article]{svmono}

\usepackage[greek,english]{babel}
\usepackage{marvosym}
\usepackage{graphicx,caption}        

\usepackage{booktabs}
\usepackage{threeparttable}

\def\begmarg{\par \begingroup  \leftskip2.5em \rightskip2em }
\def\endmarg{\par \endgroup }

\def\gt{{\rm \;>}}
\def\lt{{\rm \;<}}
\def\gt{{\rm \;>}}
\def\lt{{\rm \;<}}

\def\3H{{3\over2}}
\def\simlt{\lower.5ex\hbox{$\; \buildrel < \over \sim \;$}}
\def\simgt{\lower.5ex\hbox{$\; \buildrel > \over \sim \;$}}

\def\degs{$^{\circ}$}

\def\secspt{$\buildrel{\prime\prime}\over .$}
\def\minspt{$\buildrel{\prime}\over .$}

\def\simlt{\lower.5ex\hbox{$\; \buildrel < \over \sim \;$}}
\def\simgt{\lower.5ex\hbox{$\; \buildrel > \over \sim \;$}}

\def\bcol{\begin{column}}
\def\ecol{\end{column}}
\def\bcols{\begin{columns}[]}
\def\ecols{\end{columns}}

\usepackage[a4paper]{geometry}

\begin{document}
{\large

\pagestyle{plain}
\ \ 
\vspace{1cm}

\begin{center}
{\huge {\bf }}
\bigskip
  
{\LARGE {\bf  CORNELIS EASTON:}} 
\vspace{0.7cm}

{\LARGE {\bf THE MILKY WAY AS A SPIRAL GALAXY }}
\vspace{1cm}

\noindent
{\Large Pieter C. van der Kruit},\\
{\large Kapteyn Astronomical Institute, University of Groningen,}\\
{\large P.O. Box 800, 9700AV Groningen, the Netherlands}.\\
{\large vdkruit@astro.rug.nl; www.astro.rug.nl/$\sim$vdkruit}
\vspace{2cm}

\end{center}

{\large
\noindent
\vspace{2cm}

\noindent
This manuscript has been accepted for publication
by the {\it Journal of Astronomical History and
Heritage}. This preprint version has been produced in \LaTeX.

}
\newpage

\counterwithout{figure}{chapter}
\counterwithout{section}{chapter}
\noindent
{\Large {\bf Abstract}}\\
\bigskip

Cornelis Easton (1864--1929) studied French and obtained qualifications to teach that language in secondary schools, but in the end he became a journalist and newspaper editor. As a teenager he became interested in astronomy together with two schoolmates, inspired by the writings of Camille Flammarion. During most of his career he was active as an amateur astronomer and contributed important papers in international, professional astronomical journals on aspects of the Milky Way. Much material in this paper is based on a hand-written {\it Notebook}, which in an autobiographical account of his astronomical work.  This concerned three areas.

The first of these  derived from his enthusiasm as a youth of mapping the Milky Way.  Mapping the Milky Way was induced by Friedrich Argelander’s book (translated into Dutch by the son of Frederik Kaiser, founder of the new Leiden Observatory), who  recommended it as an observational project for astronomy without a telescope. The resulting book, {\it La Voie Lactée dans l'hémisphère boréal}, which he published in 1893, made some impression. Since it had in addition to drawings of the surface brightness of the Milky Way, also extensive descriptions and discussions of features in the structure and a comprehensive, essentially complete, listing and discussion on everything that had ever been published on the Milky Way, this stood out as by far the most impressive book on this subject. Later he produce an isophotal chart and used photographs to improve the map. Comparison to Pannekoek’s similar work showed excellent agreement.

While observing the Milky Way for many hours, Easton had been struck by the idea that what he saw was actually a spiral nebula, of which he had seen illustrations in books,  but then seen edge-on. He proposed a form for the spiral in the Milky Way System that resembled these nebulae with the center in the direction of the constellation Cygnus. His publications on this `Theory of  the Milky Way’ in the {\it Astrophysical Journal} in 1900 and 1913 drew much attention, although many astronomers, including Jacobus Kapteyn, kept quite some reservations.

The third area concerned correlations between surface densities of stars and surface brightness of the Milky Way. Easton maintained that there was such a correlation for relatively bright stars, in particular for stars of  magnitude nine in the {\it Bonner Durchmusterung}. This would mean that the  Milky Way clouds would have to be rather close by. This contradicted Pannekoek’s work, who maintained that such a correlation only started around apparent magnitude 14 or so, and pointed at a much larger Galactic system.

There is a publication Easton failed to discuss in his {\it Notebook}, even to include in his list of his publications, and that is an 1894 short article in {\it Nature}. He noted that the elliptical companion  NGC205 of the Andromeda Nebula between pictures in 1874 and 1889 had rotated by 15\degs. This observation in fact was correct; the later photograph goes to fainter surface brightnesses and the now known isophote twists are responsible for the apparent rotation.

In 1903 Kapteyn and the University of Groningen bestowed an honorary doctorate upon Easton. This special honor was granted on the occasion of Kapteyn’s 25-th anniversary of professor in Groningen. For Easton it meant an important recognition of his work by the professional astronomers.

During the visit of Harlow Shapley and his wife  to the Netherlands in 1922 after the first General Assembly of the International Astronomical Union in Rome, and a discussion session in Leiden, Easton turned out the only one to oppose Shapley’s view of a large Galactic system by defending his correlation studies.

After the 1900 publication of the spiral nebula theory, Easton directed an increasing fraction of his attention to climate studies. Contributing, he felt, to observational astronomy was no longer possible without professional equipment. For a number of years he chaired the board of the amateur astronomers and meteorologists society and the editorial board of its magazine {\it Hemel \&\ Dampkring}, saving it from bankruptcy and giving it a healthy financial basis.

{In reaction to the request of one of the referees of this paper I provide in an end-note translations into English of titles that originally were in Dutch, French, German, etc. }
\bigskip


\section{Introduction}\label{sect:Introduction}

This paper concerns Cornelis Easton (1864--1929), Dutch amateur astronomer.
Easton, with a demanding job as newspaper editor, has been in addition a quite accomplished amateur astronomer. having published, over a time span of decades, important papers in major professional journals, such as {\it Astronomische Nachrichten},  {\it Monthly Notices of the Royal Astronomical Society} and {\it Astrophysical Journal}, as well as in {\it Nature}. Over the last decade I have published biographies and biographical articles on a number of Dutch astronomers that studied the Milky Way and the nature and structure of our Galaxy. This paper on Cornelis Easton is the final one. To finish off this series I have added at the end a short discussion on the in my view remarkable absence in early studies of observational stellar dynamics, until this was introduced by Kapteyn.

In 1893 Easton published a privately financed book containing drawings of the brightness distribution in and extensive description of the northern Milky Way, a {catalog} of objects, and  a comprehensive discussion on all that has ever been published on the Milky Way. On the basis of the detailed appearance on the sky, Easton had hypothesized already in the late 1880s that the Sidereal System actually was a spiral nebula, much like the Whirlpool Nebula in Canes Venatici (although spiral galaxies were then not yet seen as island universes). And he had worked extensively on possible correlations between counts of individual stars and surface brightness of the Milky Way. He has in 1903 been awarded an degree of {\it Philosophical Doctor honoris causa}) by the University of Groningen with Jacobus Cornelius Kapteyn (1851--1922) as honorary ‘promotor’, on the occasion of the latter’s jubilee of 25 years of university professor. Kapteyn was at that time already the most influential astronomer in the Netherlands.

In this paper I will examine who was this person, who as an amateur astronomer was judged by no lesser person than Jacobus Kapteyn to be worthy of an honorary doctor’s degree. But first I summarize some biographical details

Cornelis Easton was born in Dordrecht on 10 September 1864. After the secondary school  and interrupted studies in Delft, he went to Paris before re-turning to Dordrecht and obtained qualifications for teaching French at secondary schools. He became a correspondent of the newspaper {\it Nieuwe Rotterdamsche Courant} in 1888 and in 1891 editor of the {\it Dordtsche Courant}, the newspaper in his native Dordrecht. From 1895 to 1906 he was editor of the {\it Nieuwe Rotterdamsche Courant} in Rotterdam, and from 1906 to 1923 he was editor-in-chief of the {\it Nieuws van den Dag} in Amsterdam. In 1923 he left for Den Haag (The Hague), where he became editor of the {\it Haagsche Maandblad} in 1924. These were all quite respected daily newspapers and (in the last case) monthly current affairs magazine. His journalistic career was quite successful. He was set to retire in the fall of 1929, but he died after an extended illness on June 3, 1929. 

Easton’s view of the spiral structure of Galactic System, seen from ‘above’, published in the {\it Astrophysical Journal}, first schematically (Easton, 1900) and then in more detailed form (Easton, 1913), has been used extensively in the years after they appeared by others in textbooks and other publications on the Sidereal System. As everyone else at the time Easton had not taken interstellar extinction into account, so in the course of the 1920s it fell into disrepute.  The discovery of Galactic rotation by Jan Hendrik Oort (1900--1992) -- see my biographies of him (van der Kruit, 2019, 2021b) --, that indicated a center of the system in a direction about perpendicular to where Easton had put it, {making it even more difficult to accept the latter’s model}. Yet, it prompted Oort to add as late as 1935, an admiring footnote to his inaugural lecture as professor of astronomy in Leiden. This lecture has appeared in Dutch as Oort (1936), and has been translated into English in Appendix B3 of van der Kruit (2019). The footnote reads (p.655):
\begmarg 
A courageous attempt to do this has been undertaken by our fellow Dutchman Easton, who interpreted the structure in the Milky Way as the projection of a spiral-like, wound-up system. The better insight that has been gained for the absorbing matter in the Galaxy takes away much from this work. Furthermore, we now know that in the case that if our System indeed does have spiral structure, it must be completely different from that in Easton’s interesting attempt.
\endmarg

\section{Easton’s autobiographical Notebook}

In 2023 I was working on an article on the Galactic work of Anton Pannekoek (1873--1960), including his extensive mapping of the brightness distribution of the Milky Way. My article was eventually published in this journal as {\it Pannekoek’s Galaxy} (van der Kruit, 2024a). Pannekoek compared his visual isophotal maps, which turned out very good in terms of calibrated surface brightness, to similar ones produced by Cornelis Easton, and the comparison also was excellent.

I was quite familiar with Easton whom I had already met extensively while writing a biography of Kapteyn and an article on honorary doctorates awarded in Groningen by him (van der Kruit, 2014, 2021a, 2021c). While writing about Easton in the context of the Pannekoek paper, I remembered that I had found by accident, at least ten years earlier, on the Web a set of scans of a hand-written notebook by Easton. I had forgotten completely about these scans and never had  looked at them again, but when remembrance had returned -- and the Pannekoek paper had been finished -- , I searched through my electronic files (by that time it was May 2024). I found the scans again and it turned out that I had downloaded them in July 2012. The title of the  notebook scanned would translate into: {\it Notes on my Astronomical Work, compiled in 1903, 1913, 1928} (see Fig.~1). It was in the form of 190 jpegs, copies of individual pages (135 of text plus 4 with a bibliography, and listing of archive files and brief curriculum vitae), and in addition a number of figures and a number of clippings from newspapers, magazines, etc. The lay-out appeared to be that Easton wrote texts on the righthand pages and reserved the left hand pages for pasting in illustrations, with drawings or pieces of text taken from various sources. Or he simply left these open and then they were not part of the scans. It was an autobiographical account of his astronomical researches. 

\begin{figure}[t]
\begin{center}
\includegraphics[width=0.465\textwidth]{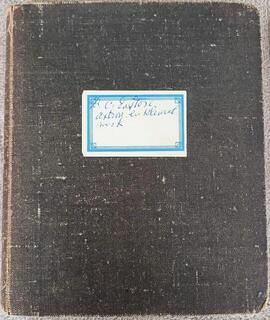}
\includegraphics[width=0.455\textwidth]{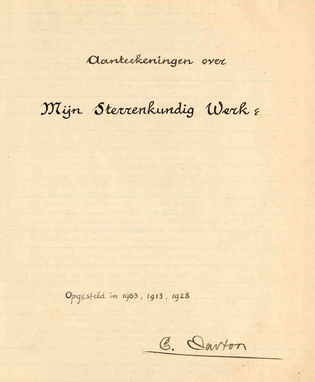}
\end{center}
\caption{\normalsize  Cover and title page of \textit{Easton’s Notebook}. The dimension is $22\times 19$ cm. }
\label{fig:}
\end{figure}

Was this an up to now unknown source? I went back to published accounts of Easton’s astronomical work. When I re-read the relatively long obituary on Easton in the Dutch amateur periodical {\it Hemel \&\ Dampkring} (Stein, 1929), it became clear that the author had used these same notes. This was Johannes Wilhelmus Jacobus Antonius Stein (1871--1951), astronomer and Jesuit priest, who was then working at Leiden Observatory, but in 1930 was to be become director of the Vatican Observatory. Stein most of the time merely paraphrased and quoted literally from these Easton notes, without adding much comment or discussion. So, although not entirely new in the literature (but never presented in English), there is likely more to be learned from these notes and I decided to look further into this.

But it was unclear where on the Web I had found these scans in 2012. Searches failed to identify a site that contained a description of this notebook and/or the links to download the scans. After much searching, I found an astro-blog about Easton (www.astroblogs.nl/2013/02/13/cornelis-easton-in-het-zonnetje-gezet/). To that one Rob Vlaardingerbroek had responded  and had provided a link to the scans.  The astroblog was from 2013, and the Vlaardingerbroek comment of early 2014, so I must have found that site earlier and not through this blog. Vlaardingerbroek mentioned in his comment that his wife was a great-granddaughter of Easton and had the booklet in her possession. However, in 2023 this link to the scans had (long ago?) been disabled and they were no longer available on the Web. Further searching using the name Rob Vlaardingerbroek yielded no useful information, so that is where the trail ended.
\bigskip

Looking into the genealogy of Cornelis Easton and his descendants, I found on the Web an interesting explanation of how he came to have an English surname. This was on a site concerning persons that died in the fighting during World War II, which had a page on a descendant of the older brother of Cornelis Easton (www.oorlogsslachtofferswestbetuwe.nl/wing-easton.html). The story is as follows. In 1796 the first Dutch Easton was born on the Frisian island of Terschelling, after the minister’s daughter there became pregnant by an English sea captain Nicolas Wing Easton. The child was `acknowledged’ by him, meaning that Easton had certified to be the father. The mother of the child  was Maria van Westingen, of which no more details could be uncovered. The Website states that she waited seven years for father Easton to return  (as a woman was supposed to do in such a situation) and then married a man from Terschelling. The child was named Nicolaas Wing Easton (1796--$\lt$1870), became a sailor as well, and in 1827 married in Amsterdam one Martha Anna van Ammers (1804--$\gt$1879). They had a son Johannes Jacobus Easton (1828--1872), who was a captain in the merchant navy and in 1858 married Margrieta Wilhelmina Ridderhof (1828--1881). The bride was born in the city of Dordrecht and the couple settled there. They had two sons, Nicolaas Wing Easton (1859--1937) and the subject of this paper Cornelis, who was born in 1864. From these years we see that Cornelis lost his father in 1872 at age about seven or eight and his mother early in 1881 when he was sixteen. 

 I now follow Cornelis Easton's descendants. He was married in Rotterdam on February 5, 1891 to Elisabeth Theresia Visser (1863--1945). They had a daughter, Titia Margaretha Elizabeth Easton (1892-1991), who was married in 1927 to Harm Henrick Kamerlingh Onnes (1893--1985). This Harm was a well-known painter and ceramist; he was -- as one would suspect -- indeed related to the physicist and Nobel Prize winner Heike, who was an uncle of his. Cornelis and Elisabeth also had a son, Jan Cornelis Easton, born in 1895, but he died as a teenager in 1910. So descendants of Cornelis Easton exist only through the daughter Titia.

 The Royal Library in the Netherlands has a site, called Delpher and located at delpher.nl, that -- quoting from its description -- `features millions of historical newspapers, books and periodicals from the collections of libraries, archives and academic institutions’. The very large collection of electronic copies of newspapers can be searched for names of persons, so one can find family notifications, usually deaths, sometimes births or weddings,  of individuals if published in a newspaper. The newspaper notifications of the deaths of both Harm and Titia were easily located this way. These were published in the newspaper {\it NRC Handelsblad} of May 22, 1985 and September 13, 1991. From these it followed that Harm and Titia had had two daughters, only identified by their initials. Further research showed that these were Elizabeth Theresia, born in 1928, and Mensina Cornelia born in 1931. The oldest daughter appears to have been unmarried and have no offspring, and the younger, Mesina was married in 1955 to Jan Carel de Knegt (divorced 1993), born in 1927, ophthalmologist in the city of Gouda. These had three daughters, the oldest named Titia, who was married to a Rob. So it follows that the {\it Notebook} had gone from daughter to daughter to daughter ending up with  Titia Vlaardingerbroek-de Knegt. Rob and Titia Vlaardingerbroek have, according to the newspaper clippings, a son, named Kristian. I have located him through LinkedIn. He gave me permission to use these scans as I saw fit.
\bigskip

\begin{figure}[t]
\begin{center}
\includegraphics[width=0.43\textwidth]{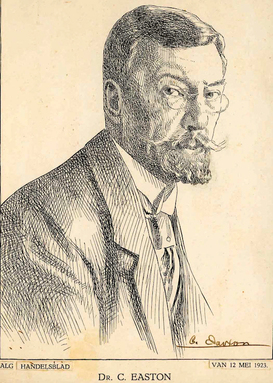}
\includegraphics[width=0.49\textwidth]{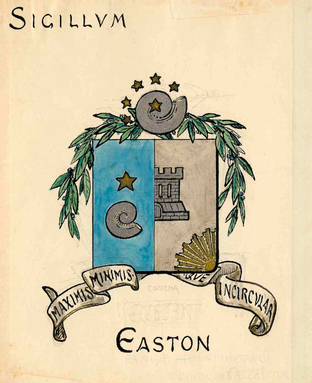}
\end{center}
\caption{\normalsize  First pages of the \textit{Notebook}, before the title page. Left: Drawing of Easton without source and credit. This has been taken from an article on Easton’s journalist career in the newspaper \textit{Algemeen Handelsblad} of 12 May 1923. A week earlier, the newspaper \textit{Nieuws van de Dag} had been terminated, of which Easton was editor-in-chief, merging with another newspaper, \textit{De Courant}. Right: Sigillum (seal) of Easton. According to experts, the text ‘maximis mininis que in circular’ is not an existing Latin motto (it is also not very good Latin) and must have been composed by Easton himself. It possibly means something like ‘in the large and in the small everything in the form of spirals’, on the left in the Universe, typified by the star, and in the small below that in the snail. Both symbols are also at the top of the seal in the form of a snail that continues as a spiral of stars. With thanks to classicists Manfred Horstmanshoﬀ and Jacqueline König.}
\label{fig:}
\end{figure}

I have transcribed this important source and subsequently translated it into English. I have provided all of this on the Web as an e-book in pdf-format, containing, in addition to an Introduction in Dutch and English, three parts:\\
{\bf I}. Chapter I is a complete transcript of the entire notebook  (in Dutch, retaining Easton’s spelling, which is an official, but old form of spelling\footnote{There has been a major revision of Dutch spelling initiated in 1934 by ministerial decree. This was followed by a formal adoption (with further adaptions) by law much later, in Belgium in 1946 and the Netherlands in 1947. It took until 1954 before it was formalized after further changes. The law formally prescribes obligatory use only in schools financed by the government and in documents issued by state and local authorities, but in practice its use is universal. There have been further significant, but less comprehensive changes in 1996 and 2005.}), including all figures, and in addition all material provided in Easton’s frontmatters and backmatters (for the first two pages see Fig.~2). The frontmatters concern some poems and illustrations, the backmatters include a list of Easton’s astronomical publications, followed by a list of his meteorological publications, another list of the titles of {\it Cahiers} (his archival rubrics), and a brief resum\'e  in the form of years with significant events. It concludes with a complete reprint of the special ‘Easton Issue’  of {\it Hemel \&\ Dampkring} of June 1928. Throughout I have copiously provided footnotes on persons, books, historical events, etc. mentioned by Easton, and I have translated English, French and German texts into Dutch.\\
{\bf II}. Chapter 2 provides a translation of the whole into English, for clarity again adding all the cuttings and drawings, but when possible image-processed for better definition or replaced by better resolution versions. Material supplied by Easton in Dutch, French or German, including the {\it Hemel \&\ Dampkring} ‘Easton Issue’  have been  translated into English as well. All  footnotes in the first part are also provided in English translation.\\
{\bf III}. The Appendix presents the full set of original scans.

This adds up to some 450 pages and 450MB. It is available on the Web with reduced resolution in the reproductions, which still means a roughly 62 MB pdf-file. However, the full resolution version is  available as three separate files. All these files are available though my  Kapteyn homepage, accompanying my Kapteyn biographies  (van der Kruit 2015, 2021a), at www.astro.rug.nl/JCKapteyn/Easton.html.
\bigskip

Most of Easton’s scientific archives are kept at the Rijksmuseum Boerhaave (specializing in the history of science and technology) in Leiden, including the `Cahiers’ (thin, soft-cover. notebooks), manuscripts of his major publications, reprints of astronomical and meteorological papers, letters, etc. Among the hand-written documents is a cover-less notebook {\it Aanteekeningen betreffende mijn wetenschappelijk en ander werk} (Notes concerning my scientific and other work), held together by sellotape (Scotch tape). It has been written in 1927, and covers 50 pages of text. Most of it duplicates the {\it Notebook} just described, but there is a little bit more on his personal life, professional and social activities. I have not transcribed that, but will use it below and then refer to it as the {\it Notes}.

\section{Early interest in astronomy}

Very little has been recorded on Easton’s youth. He was born in 1864 in the city of Dordrecht, some 20 kilometers south southeast of Rotterdam in a strategic position in the delta’s of the rivers Waal, Merwede, Meuse and Scheldt.  Among the smaller rivers connecting various branches is the Thuredrith. This name, which is the root of the name Dordrecht, means ‘pull through’ since ships could not navigate this narrow river by themselves. Dordrecht lies on an island since the {\it Saint Elisabeth's flood} of November 1421, which flooded most of southern Holland, the part of the Netherlands along the western coast of the North Sea. 'Holland' ends not far south of Dordrecht (the part containing Vlissingen, Flushing, is called Zeeland). Dordrecht is the oldest city of Holland, having had city rights since around the year 1200 and was a major trading port throughout most of its history, and in addition a center of shipbuilding. Its strategic location also meant that Dordrecht played an important role in the defense of Holland and Zeeland, so it was also an important site as a garrison town of the army. In Easton’s days Dordrecht was a center of trade and shipping.  In the eighteenth century Rotterdam took over in importance in trade, but Dordrecht remained important for the military and defense because of its strategic location. 

As mentioned Easton's father died in 1872, when the boy was about eight years old. His older brother Nicolaas Wing Easton, was about thirteen. It is likely that they grew up further in a one-parent household. There is no mention in the {\it Notebook} of what happened to Cornelis after the death of the mother (in 1881); he was only sixteen years of age. His older brother was about twenty-two, and was a student in the Polytechnical College (now Technical University) of Delft, where he studied to become an mining engineer, but also became an expert in geology and volcanism. He spent most of his career in the Dutch East-Indies, working on the island of Borneo. After his retirement he wrote various books on the subjects of mineralogy and geology; the Polytechnical College Delft bestowed on him an honorary doctorate in 1928. 

Cornelis Easton (most likely like his brother before him) went through the type of secondary school called HBS (Higher Citizens School), a new type of secondary education introduced in 1863 as an alternative to the gymnasium by Johan Rudolph Thorbecke (1798--1872), founder of the Liberal movement in the Netherlands. Thorbecke was responsible for the new constitution of 1848, which limited the power of the King Willem II. He was Prime Minister in three periods between 1849 and 1872; among his accomplishments was a revision of the school system in the Netherlands. This started with elementary schools and was followed by the establishing of the HBS in 1863.

At the HBS no Greek and Latin were taught -- contrary to the Gymnasium, where these languages formed a major part of the curriculum -- , the aim of the HBS being to educate young men (at that time not women) for leading positions in industry and commerce, rather than prepare them for academic professions. There was much emphasis on the natural sciences and the modern languages English, French and German. The level of education was very high; teachers in mathematics, physics and chemistry usually held a Ph.D. In fact the relatively large number of Nobel Prizes in physics and chemistry for the Netherlands in the first years it was awarded (from 1901 onward) is often quoted as a direct result of the introduction of the HBS.  For more, including the subsequent expansion of funding and personnel at the universities, see Willink (1991, 1998).

\begin{figure}[t]
\sidecaption[t]
\includegraphics[width=0.62\textwidth]{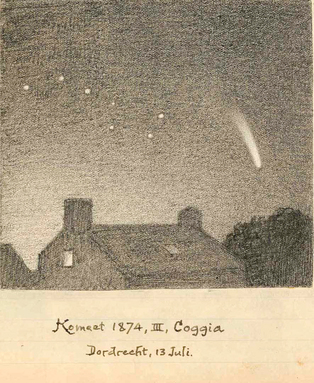}
\caption{\normalsize Comet Coggia in a drawing from the \textit{Notebook}. There is no mention by whom this was produced, but almost certainly it was Easton himself at age ten. }
\label{fig:coggia}
\end{figure}

Easton will have entered this at age twelve in 1876 and passed the final exam in 1881 -- as he mentioned in his {\it Notebook} a few months after his mother’s death. One may speculate why the Easton sons did not go to the gymnasium. From their paternal side males were sailors, although in leadership positions. Classics or academia were not part of a family tradition. The same held for his mother, Easton’s maternal grandfather was a typesetter at a printer/publisher. After leaving the HBS, Easton in fact took extra lessons to learn Greek and Latin. In his {\it Notebook} Easton’s eludes to his ineptitude for mathematics (p.130):
\begmarg
 When I began to think and learn seriously, geography and meteorology attracted me more than numerical series [...]

Later, unfortunately, my too little inclination {toward} mathematics stood very much in my way.
\endmarg
He might not have liked it, he unavoidably in the HBS had extensive lessons in mathematics and physics, but he never made very much use of this (in fact none whatsoever) in his astronomical writings.
\bigskip

According to his own recollections, Easton’s first encounter with astronomy took place on probably July 13, 1874, when he viewed Comet Coggia (see Fig.~3). Easton had turned ten a month before that. Coggia was a very bright comet, designated the {\it Great Comet of 1874}. Jérôme Eugène Coggia (1849–1919) was an astronomer at the Observatoire de Marseille. The comet had gone through perihelion on June 9, 1874; its very elongated orbit (eccentricity barely different from 1.0, (0.9988 according to Kronk, 2003) ensures it will not come back very soon — the orbital period is estimated at about 14,000 years.

The drawing in Fig.~3. has been taken from Easton's {\it Notebook}. It is not referred to in the text, nor is there any other remark on it. It almost certainly has been drawn by Easton himself as a boy aged ten. The position relative to the constellation Ursa Major is correct for the date indicated according to other published drawings of the comet {(see e.g. www.dspace.cam.ac.uk/handle/1810/224265)}.

The starry night sky kept catching his attention and he has described how he formed his own constellations. Furthermore, as a teenager in 1879, it were -- as happened to many youths -- the writings of Jules Verne (1828--1905) that awakened his desire to learn more about the universe. The decisive experience, however, that hooked him to astronomy definitely, came from a book for a general audience, namely Camille Flammarion’s {\it Astronomie populaire}. Nicolas Camille Flammarion (1842--1925) wrote a number of popular books on astronomy, but this {\it  Astronomie populaire} (Flammarion, 1880) was the best-selling one. It has been translated into many languages, including English and Dutch -- actually I read the Dutch version as a teenager. In 1882 he wrote a supplement, called {\it Les étoiles et les curiosités du Ciel} (Flammarion, 1882). The first translations into Dutch of these books appeared only in 1884  and 1885 (Flammarion, 1884, 1885); it is unlikely Easton read these translations, because he records having been captivated by Flammarion already a few years earlier; his French must have been very good already then.

In the summer of 1880, his schoolmate  Adriaan de Groot had come in possession of the {\it Astronomie populaire} book. How this happened is not described by Easton, but in France it had just appeared that year and  it must have been a French version. This book, which they read together, captivated the youths enormously, resulting in them founding in August 1880 a {\it Société Flammarion} with them two as the only members. In that same month, undoubtedly inspired by the Flammarion tome, on August 10, 1880 to be precise, the two started observing the sky themselves. They decided to view the `Laurentius shower of meteors' from the attic and the roof of Easton’s parental home.  These are the Perseids, sometimes referred to as the `Tears of Lawrence’ after the saint whose name day falls on 10 August, because it is said that on that date in the year 258 he had been tortured to death on a red-hot gridiron. Unfortunately for the boys, the sky became overcast when the Sun set, but the exercise  was repeated the next day with more success.

\begin{figure}[t]
\sidecaption[t]
\includegraphics[width=0.62\textwidth]{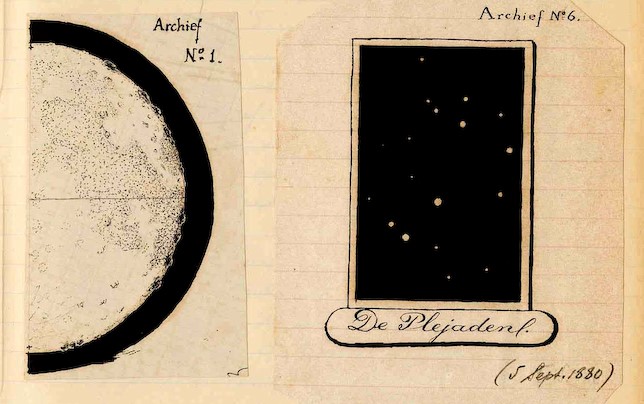}
\caption{\normalsize   Drawings by Easton in August and September 1880 from observations of the Moon (left) and the Pleiades from the Archives of the \textit{Société Flammarion}. These were obtained with a `telescope’ built by the three members from cardboard and a discarded shaving mirror. From the \textit{Notebook}.}
\label{fig:Pleiades}
\end{figure}

The {\it Société Flammarion} was extended with another schoolmate as third member, Laurens Vuyck, on October 8 of the same year 1880. Of this period in his youth, Easton wrote in his {\it Notebook} 
\begmarg
Rarely will boys have mixed so much sincere zeal and naive foolishness as when we wanted to make a large telescope from a discarded shaving mirror and a piece of cardboard – after all, Herschel had also made his own telescopes -- or drew the 14, indeed 14, Pleiades, which were visible in our telescope, for the greater honor of Science. All such beauties moved to the ‘archive’ and were (I was almost the only one who drew them) certainly not judged too favorably by my fellow members. But what did not end up in the archive and was not boyishly foolish: the delight of seeing a beautiful cluster of stars twinkling in the azure for the first time, or of seeing the fine ring of Saturn with one’s own, unspoiled eyes.
\endmarg

Some drawings of observations obtained with this telescope are in Fig.~4.  The {\it Société Flammarion} did not last long; in fact no more than a few months, in spite of formal bylaws having been drawn up. The {\it Archives} at the Museum Boerhaave  contain the Flammarion archives which has two versions of these bylaws with the longest having no less than 75(!) articles and signed by President de Groot and Secretary Vuyck. The members soon pursued their own interests. Adriaan de Groot (1863--??) studied in Delft at the Polytechnical College where he obtained the diploma of engineer in 1887. After some years in the Dutch East-Indies, China and Japan he returned to the Netherlands, where he introduced the burning of household waste, re-using the energy of the heat and the remaining slag as building material. Laurens Vuyck (1862--1931) studied biology in Leiden, specializing at first in botany of plants in the dunes, later in botany of the tropics (without ever visiting the East-Indies). He taught in secondary schools (HBS and Gymnasium) before joining the State Agricultural College in Wageningen  (later Agricultural University) and from there transferred to the Colonial Agricultural Secondary School in Deventer, of which he became director.
\bigskip

\begin{figure}[t]
\begin{center}
\includegraphics[width=0.46\textwidth]{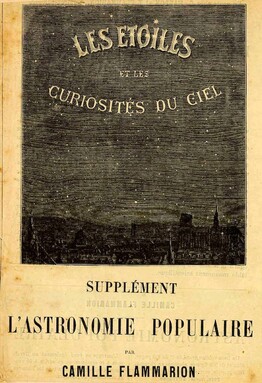}\includegraphics[width=0.46\textwidth]{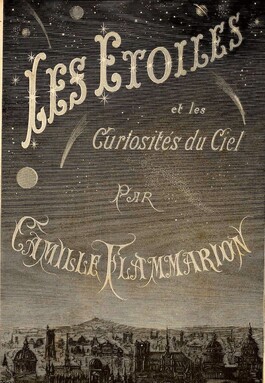}
\end{center}
\caption{\normalsize Covers of apparently two installments of Flammarion’s \textit{Les étoiles et les curiosités du Ciel}. From the \textit{Notebook}.}
\label{fig:}
\end{figure}

Easton’s last year at the HBS (1880-1881) was a difficult one, mainly because of the protracted illness of his mother, who died in February 1881. He noted that concentrating on astronomy gave him much distraction, comfort and solace. This involved reading about astronomy; the sequel to Flammarion’s {\it Astronomie populaire},  {\it Les étoiles et les curiosités du Ciel}, with altogether over 800 pages was a major investment of time. In his {\it Notebook} Easton wrote (p.140):
\begmarg
What I do remember is the pleasure that reading astronomical books gave me: a whole undiscovered world with new surprises every time. How I enjoyed, like a gourmand in a corner, every new installment of Flammarion’s {\it Les Étoiles},  the Supplement to his {\it Astronomie}.
\endmarg

Apparently the book appeared in instalments. In the {\it Notebook} there are two cover illustrations (I presume cut from a publisher’s {\it catalog} or so; see Fig.~5).

\begin{figure}[t]
\sidecaption[t]
\includegraphics[width=0.62\textwidth]{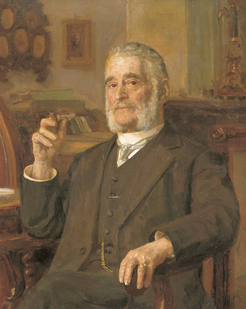}
\caption{\normalsize   Dr. A.S. van Oven, teacher of physics and chemistry of Easton at his HBS and council afterwards for many years. This painting has been produced by his son Coen van Oven in 1912. Courtesy De Stichting `Coen van Oven', reproduced with permission.}
\label{fig:vanOven}
\end{figure}

Throughout his years at the HBS, the period of the {\it Société Flammarion}, and in fact over many years after this, Easton profited from the wise council of his `fatherly friend’  Dr. A.S. van Oven (Fig.~6), teacher at, later director of, his HBS. Not only did he provide ‘ephemerides’  for the {\it Société Flammarion} (I presume what to see and where to look), he gave advice on important choices or decisions that lay before Easton up to many years later. Dr. Adolph Samuel van Oven (1837--1915) had studied mathematics and physics in Leiden and obtained a PhD on the galvanic gas battery, an early type of fuel cell, in 1862. He then became a teacher of mathematics and physics at the Gymnasium in Dordrecht, and -- after the HBS had been  introduced  -- he first became a teacher and later director at the one in Dordrecht. He was particularly well-known for his passionate teaching style. As a result, he was a source of inspiration, but also a source of information for Easton. See stichtingcoenvanoven.nl/dr-adolph-samuel-van-oven/ for more on him.

\begin{figure}[t]
\sidecaption[t]
\includegraphics[width=0.58\textwidth]{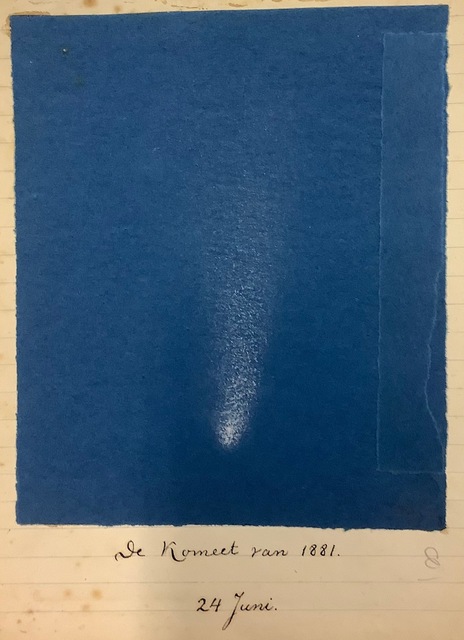}
\caption{\normalsize Drawing by Easton of the Great Comet of 1881, as it appeared to Easton on June 24, 1881. From Easton's \textit{Archives}.}
\label{fig:}
\end{figure}

Also Easton made naked eye observations of the sky, except when he at one time used a borrowed telescope. He made brightness estimates of bright stars and for example made a drawing of the ‘Great Comet of 1881’ (Fig.~7), the one he noted that prompted David Gill to photograph the southern sky and was the basis of the {\it Cape Photographic Durchmusterung}. Charting was a great hobby of Easton, almost compulsory he drew whatever was suitable for this ({\it Notebook}, p.145):
\begmarg 
My actual drive, however, was the same that had previously made me make a situation drawing of our garden, with all the apple and pear trees and flower beds, and if I had had nothing better to do, I might have made a map of the puddles that remain on the beach at low tide with the same zeal. [...]

At that time I only vaguely realized the importance of such Milky Way research. I thought I knew that there was no Milky Way map that met the simplest requirements, and I  had also read that Argelander attached value to such drawings, by a ‘serious enthusiast’, but I do not believe that I ever wondered what a man of the trade would say about my work. [...]

That same spring of 1881 I bought from a book stall the {\it Kosmos} by von Humboldt, a Dutch translation by Beima, and W.F. Kaiser’s translation of Argelander’s {\it Aufforderung an Freunde der Astronomie}. In this last booklet, among the important observations in the field of astronomy, drawings of the Milky Way are mentioned -- and that led me to my life’s work.
\endmarg

{\it Kosmos -- Entwurf einer physischen Weltbeschreibung} by Friedrich Wilhelm Heinrich Alexander von Humboldt (1769--1859) was a very influential series of books on science and nature (von Humboldt, 1845--1862). The fifth part had not yet been completed at the time of his death. Writer Elte Martens Beima (1801--1873), curator at the National Museum of Natural History in Leiden, who received a PhD in Leiden in 1842, produced an influential translation into Dutch.

The Argelander book is  a sort of manual for amateur astronomers by Friedrich Wilhelm August Argelander (1799--1875), the driving force behind the {\it Bonner Durchmusterung}.  This publication (Argelander, 1844) had been translated into Dutch and published  in 1851; the Dutch title, translated into English, is {\it Invitation to friends of astronomy to make observations on several important branches of celestial science that are as interesting and useful as they are easy to perform, translated from the high German by W.F. Kaiser, with a preface, notes and appendices by F. Kaiser}. This translator is Willem Frederik Kaiser (1836--1916), son of Frederik Kaiser (1808--1872), founder of Leiden Observatory. The observations referred to in the book concern the aurora, zodiacal light, shooting stars, twilight, Milky Way, magnitudes and colors of stars, and variable stars. For more interesting background, including a amusing quote from the Preface, see section 6 of my paper on Pannekoek’s Galactic research (van der Kruit,  2024a).

\section{Mapping the Milky Way}

Dordrecht, when Easton grew up, had of order 15,000 or a bit more inhabitants. Street lighting was still minimal and the Milky Way must have been visible easily from the center of towns like this. The middle of the nineteenth century saw the replacement of street lighting from burning various types of oil (in the Netherlands mostly `patent olie’ which is refined rapeseed oil) with lamps that were burning gas. The ensuing increase in the number and intensity of street lamps made observing the Milky Way more difficult, but this process must have been slow. Easton reports the production of maps of the Milky Way from the (one would think opened) windows of the house he lived in, or nearby bridges, up to about 1890.

The plan to draw up a map of the northern Milky Way dates back to shortly after the days of the {\it Société Flammarion}, when Easton was a boy of seventeen. He developed a special skill and adaption to the task ({\it Notebook}, p.143):
\begmarg
Once I had conceived the plan to ‘devote a few evenings to observing and drawing the Milky Way’ as is modestly stated in the introduction to my first description of the Milky Way [...]  two things were of the greatest use to me: 1. I had been doing map recognition for years as a hobby, 2. I had not thought for a moment about publishing my drawings. [...]
  
But it is certain that if I had been concerned with anything other than the drawings themselves, I would have had to sacrifice something of the demand for utmost perfection that I set for myself. And my special skill in drawing maps was now especially useful to me. Remarkably, I would have been less suited for this purpose if I had been able to draw more technically skilled or somewhat artistically. The monastic work of drawing meticulously precise, in which I had practiced myself so much on the maps, was exactly what I needed for such Milky Way drawings. As a result, my eyes had become naturally very sensitive, it seems, and particularly trained. My patience had been cultivated no less – almost to the point of becoming sick! – and since my enthusiasm and my seriousness were sufficient, I finally succeeded. [...] 

 Presumably I could have become a second-rate cartographer at best, and certainly never an artist. But with the kind of drawing that I could do, I found a promising field of work waiting.
\endmarg

The first drawings of the Milky Way were made in 1881 by Easton and Vuyck as a kind of after-effect of {\it Société Flammarion}. After passing their final exams for the HBS in that year, de Groot had left for Delft and was away most of the time. Vuyck and Easton had stayed in Dordrecht. This gave them quite a lot of free time. They had decided to get some private lessons in Greek and Latin.  Whereas de Groot had qualified to enroll into the Delft  Polytechnical College, both Easton and Vuyck did not have technical aspirations. Eventually Easton studied French and Vuyck biology and at that time the HBS diploma did not qualify for entering university for those studies. Additional lessons and an entrance exam in the classics were required. Arranging for these lessons did not come easy ({\it Notebook}, p.144):
\begmarg
 I may even say that the long stubborn delay of the curators of the Dordrecht Gymnasium to allow Dr. Warren (now rector of the Erasmian Gymnasium) to give private lessons in the ancient languages to Vuyck and myself, indirectly gave rise to my throwing myself into the study of the Milky Way.
\endmarg

Dr. S.J. Warren was rector of this well-known gymnasium (also called `Gymnasium Erasmianum’)  in Rotterdam, the third oldest in the Netherlands, until 1910, but before that he was teacher and rector of the gymnasium in Dordrecht. Vuyck and Easton did apparently pass this test, since eventually they studied the subjects of their choice at university level. From the {\it Notebook} it appears that eventually Vuyck gave up astronomical observations, but Easton did not. The way Easton worked is that he made numerous small drawings that the later merged into a single map of a larger part of the Milky Way. He was probably busy with these first drawings until the middle of 1882 `just about everywhere where the sky was clear and the place was suitable for observation’: ‘somewhere outside, or from a bedroom window, or with a lantern in a city garden’. He  did not get the entire northern Milky Way belt finished; starting from the Aquila and Cygnus areas  (with Vuyck)  he did not get up to  Cassiopeia or Perseus, so in particular did not cover the Orion region. For reference here and later in this article I present in Fig.~8  the full panorama of the Milky Way, produced by Serge Brunier and the European Southern Observatory (ESO, 2009), based on photographs with a digital camera from the Paranal and La Silla Observatories in Chili and from La Palma, Canary Islands, on the occasion of the International Year of Astronomy 2009, with the names of the constellations added.

\begin{figure}[t]
\begin{center}
\includegraphics[width=0.90\textwidth]{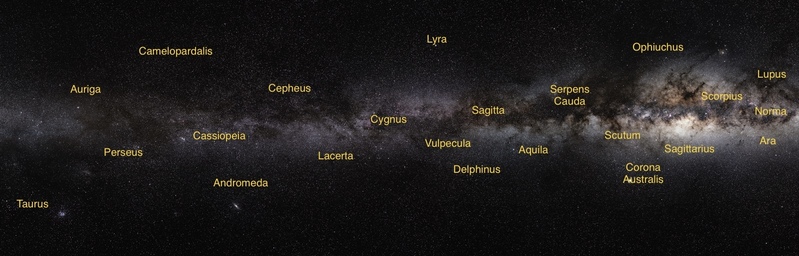}
\includegraphics[width=0.90\textwidth]{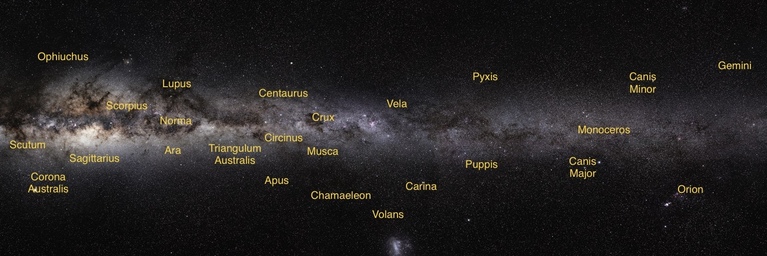}
\end{center}
\caption{\normalsize  Map of constellations in on near to the Milky Way on the the ESO Milky Way Panorama (ESO, 2009) in two parts with the area of the Galactic center repeated for clarity. Some noteworthy features are readily identified. The bright ‘spot’ to the right of `Taurus’ is the Pleiades star cluster. The clusters h and $\chi$ Persei are the two dots to the lower left of the C in `Cassiopeia’. The Andromeda Nebula is below the word ‘Andromeda’. The bright spot above `Centaurus’ is the Galaxy NGC5128 (radio source Centaurus A). Below `Volans’ the Large Magellanic Cloud (the Small Cloud is to the left below `Apus’ beyond the bottom of the picture). And on the right one can recognize the constellation Orion and the bright Orion Nebula. Credit: ESO/S. Brunier.}
\label{fig:}
\end{figure}

In the course of 1882, Easton moved to Delft, not to study engineering but at a school for training boys for administrative service in the Dutch East Indies. It had been instituted in 1864 and was called the ‘Indische Instelling’ (East Indies Institution), and has been closed in 1901. The curriculum at this school included in addition to modern languages as French, English, German and Italian, and state economy, etc.,  typical Delft subjects such as surveying, geodesy, architecture, mechanics, chemistry and physics. And of course typical subjects for service in the East Indies such as Javanese and Malay language and {Islamic} and native law. It is not made clear by Easton why he felt attracted to this educational program, and it not obvious from what little he did mention about himself. But for some reason or reasons, some  very likely associated with his distaste or ineptitude of technical or mathematical matters, he quit the studies in 1884.  He switched to studying French, but more about this is presented later in this paper.
\bigskip

Before continuing with the Delft period I note that Easton does mention in his {\it Notebook} what he described as `completely worthless’ observations for his pleasure of celestial objects other than the Milky Way. He  refers to the use of a ‘very small telescope of Gregory' to observe sunspots and solar eclipses (through a filter or projected on a white sheet of paper). It seems logically he also used the telescope for night observing. How he acquired this Gregorian telescope\footnote{The Gregorian telescope is a reflecting telescope designed by Scottish mathematician and astronomer James Gregory (1638--1675), and first actually built in 1673 by Robert Hooke (1635--1703). It employs a concave secondary mirror that throws the light through a central hole in the also concave primary mirror. The secondary mirror is placed \textit{beyond} the focal point of the primary mirror. It has been superseded by the Cassegrain design, which employs a convex secondary mirror placed \textit{before} the primary’s focal point, allowing large effective focal lengths and plate scales. This design is credited to French catholic priest Laurent Cassegrain (1629--1693).} or was allowed its use is not recorded.

\begin{figure}[t]
\begin{center}
\includegraphics[width=0.422\textwidth]{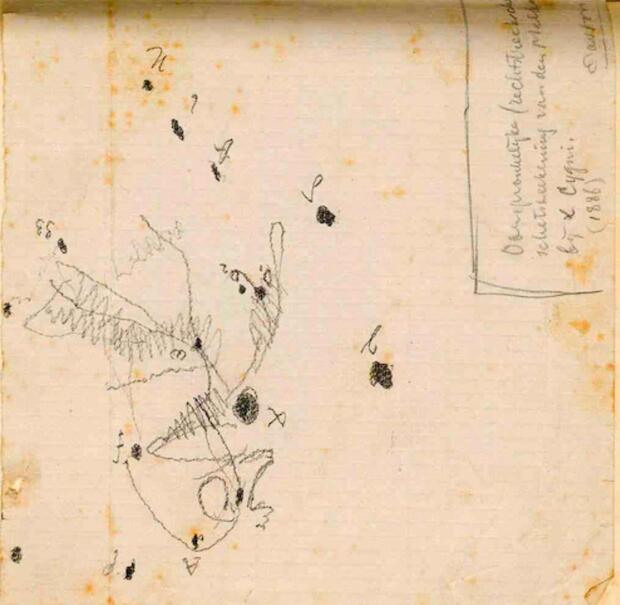}
\includegraphics[width=0.338\textwidth]{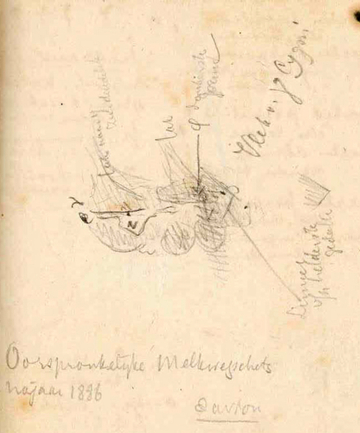}
\includegraphics[width=0.76\textwidth]{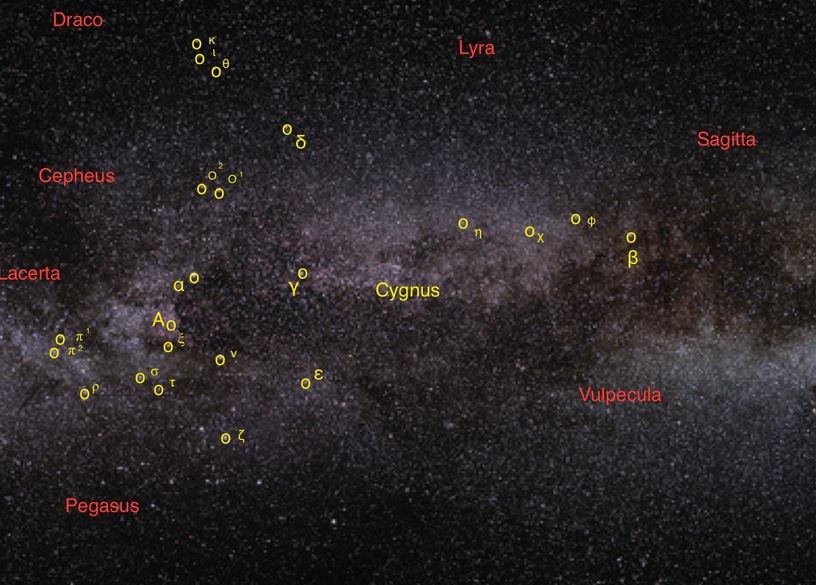}
\end{center}
\end{figure}

\begin{figure}[t]
\caption{\normalsize At the top two raw drawings by Easton of small sections of the Milky Way. Both are located in the constellation Cygnus, the Swan. In the lower panel the part of the Milky Way in Cygnus from the ESO Panorama (ESO, 2009). The stars of Cygnus have been indicated in yellow following the chart given by the International Astronomical Union (IAU, 2022). The background of the designation for the star A Cygni (referred to later in this paper) is explained in the text. The neighboring constellations are indicated in red. The top panels have been rotated as closely as possible to the orientation in the lower panel. As for orientation: the line between $\alpha$ Cygni (Deneb) and $\delta$ Cygni is close to east-west, so north is to the top-left and east to the bottom-left. From the \textit{Notebook}, and courtesy: ESO/S. Brunier.}
\label{fig:}
\end{figure}

In the {\it Notebook} there are a few examples of drawings produced while observing the Milky Way. These contain impressions indicated by shadings and sketches, but also notes. Later he used these to produced a ‘final’ map of the Milky Way. In Fig.~9. two examples are shown together with the view of the same area in the ESO panorama. The drawings in the upper part have been oriented as close as possible to the lower panel. On the left the area around $\alpha$ (Deneb) and $\gamma$ Cygni and above is shown. In the left drawing we have $\gamma$ Cygni on the left. The arrow with $\varphi$  does not point to a stellar object; the two dots on the left must be $\phi$ and $\beta$ Cygni. The letter ‘Z’ probably indicates ‘zwak’ (faint). The designation A Cygni (referred to later in this paper) has the following background. The star is sometimes named A(68) Cygni. It was cataloged A Cygni in the star atlas {\it  Uranometria Omnium Asterismorum} (Bayer, 1603) of Johann Bayer (1572--1625), the first when he ran out of Greek letters (so number 25 of the constellation). John Flamsteed (1646--1719) included the star as 68 Cygni in his {\it Historia Coelestis Britannica} (Flamsteed, 1725), published posthumously by his wife, the 68-th in order of Right Ascension in the constellation. 
\bigskip

Easton’s further activities in astronomy during his stay in Delft were very limited. I quote from the {\it Notebook} (p.146):
\begmarg
As long as I stayed in Delft, there was no question of any serious astronomical work. I kept my magazine: {\it l’Astronomie} – that was all. I did, however, establish an Astronomical Student Society with some other students: the {\it Society Copernicus} (See Delft Studenten Almanac of 1884). I remember that I once gave a lecture there, the title of which was: ‘Why is astronomy practiced by laymen?’ which sounds a bit like an apology. That Society was not very successful: except for the young Bosscha and myself, the astronomical enthusiasm was not very strong. In my memory, the Society Copernicus cannot compete with the Société Flammarion.
\endmarg

The `young Bosscha’ was in fact Karel Albert Rudolf (Ru) Bosscha, (1865--1928), the tea planter that later would be the founder of the Bosscha Observatory in the Dutch East Indies near Bandung on the island of Java.
\bigskip

The collecting of drawings to build up a map of the visible Milky Way continued throughout the 1880s. Easton described it as follows ({\it Notebook}, p.148):
\begmarg 
The year 1886 was a turning point in my life. The loss of my small fortune – worrisome in itself, and still lingering for many years in a vague fear of misery – had a beneficial effect on a character like mine in more than one respect. I now received an incentive to use my strengths, and by a coincidence that I cannot praise enough, I was transported into the midst of an environment that was the opposite of the barren and everyday one in which I had spent the last few years.
\endmarg
The remark about the loss of his fortune has been explained in the {\it Notes} as mismanagement of his financial affairs by his legal guardian, forcing Easton to work for his living, while still pursuing studies. {The custodian, the legal guardian designated to be responsible for and to make legal decisions concerning the child, was probably Hermanus Gunneweg, in whose house Easton lived for a number of years (see below).} Continuing the quotation:
\begmarg
So it happened that while I was working for my French M.O. certificate in the latter half of 1885 in order to successfully pass the exam in the fall, I still found the opportunity, amidst all sorts of household distractions and pleasantries, to supplement my Milky Way studies in a few months before and after the exam and to thoroughly revise those already made, so that on 1 January 1887 the whole thing was effectively completed. Everything I did on my Milky Way drawings was limited to revision and addition, the broad foundation laid between September ’86 and January ’87 remained intact. And all that happened without me getting the impression that I was working particularly hard. I still don’t understand how that was possible, in any case it proves again what a healthy and young person can accomplish if the desire to do so prevails.
\endmarg
After quitting his studies in Delft, he at least part of the time lived in the house of {artist} Hermanus Gunneweg {and his wife Caroline}, where he probably remained until in 1891. {These persons will be introduced with more details below. In that year 1891 Easton married Elizabeth Theresia Visser. When and how he met her and how their courtship developed has not been recorded. I requested more on her from Kristian Vlaardingerbroek and his mother no more information, ir a photograph on which she appears, but no further details were available.} From the {\it Notebook}, p.148:
\begmarg
My observation method had now improved and I was more practiced than before (also through drawings in other areas); I was familiar with the fainter stars of the galactic strip, with the beacons and markers that always come in handy with such a ‘topography of the sky’. Mostly I did my observations from our windows with the wonderfully wide view over the Bleekersdijk; sometimes -- the method is certainly recommendable -- Caroline Gunneweg fulfilled the post of secretary. I could now go into much more detail in my drawing and description than before, when I still had to absorb the main features; in the mean time I remained on my guard against seeking doubtful details and tracing shades on the edge of the light belt, and that was my good fortune, because in this way -- as it later turned out -- my maps were not duplicates of the very detailed, but in my opinion useless maps of Boeddicker.
\endmarg
The Bleekersdijk is just outside the old center of Dordrecht and must have presented a clear view of the sky {toward} the south from central Dordrecht.

Otto Boeddicker (1853--1937) was a German astronomer, who spent most of his career as the assistant of Lawrence Parsons (1840--1908) , the 4th Earl of Rosse at Birr Castle in Ireland, known for his large telescope, with which he discovered the spiral structure of the Whirlpool Nebula. Boeddicker (1892) published his drawings and mapping of the Milky Way in a separate publication: {\it The Milky Way from the North Pole to 10\degs\ south declination drawn at the Earl of Rosse’s observatory at Birr Castle} (Boeddicker, 1892). The remarks at the end in the citation above should be read to mean that Easton was of the opinion that much in Boeddicker’s drawings was incorrect as a result of too much attention to detail so that the large-scale structure had disappeared from view.

From the {\it Notebook}, p.157:
\begmarg
In the summer and late summer of 1887, partly in Hilversum, the large Milky Way map ‘Via Lactea’ was completed, which contained the whole of my observations. When I had completed it in October ’87 -- it now hangs, framed, in my room -- I had the satisfaction (in the midst of very poor living conditions) of knowing that for the first time a proper comparative representation had been given of the majestic phenomenon of the Milky Way. I was not mistaken. Boeddicker’s map, the only one that could be compared to mine, was not completed until 1892.
\endmarg

There is no indication why he happened to be in Hilversum (about 20 km southeast of Amsterdam). Easton often worked extensively and in a focussed way without daily work or other distractions on his astronomical programs during vacation periods, often staying in the house of friends; this might be a first occurrence of such a routine, although it is unknown what made him go to Hilversum. 
\bigskip

The completion of the map was not the end-result. Easton went as far as collecting his notes in a manuscript during the 1880s. Quoting again from the {\it Notebook} (p.157):
\begmarg
... tidying up of my Milky Way drawings, the editing of the description in French, (at that time, although still vague, a possible publication of my studies seems to have been in my mind), the collecting of everything in a single map, even a figure of nomenclature of the light and dark spots of the Milky Way -- among great and small celebrities a tribute to my future wife found a place, in the names Elisabeth and Theresia, Cygnus region – yes, there was even a table of contents and an introduction, the latter first dated 10 April 1889.
\endmarg
In fact, Easton had {\it two} manuscripts prepared, with titles {\it La Voie Lactée boréal} and {\it La Voie Lactée dans l'espace}. The latter concerned his hypothesis of spiral structure in the Milky Way System.

\section{Non-astronomical matters}

Easton is not very specific about his terminating his studies in Delft. In his very brief {\it curriculum vitae} in the {\it Notebook} there are only two brief items: ‘1882–1884 — Student in Delft’, and ‘1884–1895 — at Dordrecht’. That he obtained a diploma ‘French M.O.’ in 1885 is not mentioned as part of the cv, nor a period he spent in Paris (see below). I will comment first on his period in Dordrecht in this period and then on the meaning of this certificate.

From the {\it Notes} we can learn a few more details (p.1-3):
\begmarg
June 1881 final exam [HBS], one year of private lessons Latin, etc. Sept. '82 to Delft. Spring 1885 to Paris, Dec. '86 diploma M.O. French.

In the spring of 1886, due to mismanagement by my guardian (my father $\dagger$ 1872, my mother $\dagger$  1881), I was forced to provide for my own living. I did that mainly with private lessons French; I did not give classroom lessons — except for cases in which I 'substituted' for quite some time at the Schreuders Institute in Noordwijk [a private school for secondary education, on the coast near Leiden], at the Dordtsche HBS. and Gymnasium.

I gladly accepted the vacant correspondent position of the N.Rott.Ct. [Nieuwe  Rotterdamsche Courant] in Dordrecht. I tried to make something of the Dordrecht reporting, started writing reports on art, was often sent by the Director of the N.R.Ct  was to political meetings and events, as an assistant reporter, among others at the funeral of King Willem III in 1890. I acted as an advocate of the then much-contested `youth’ in painting in Dordrecht (and elsewhere), sometimes following the lead of Jan Veth; became a literary chronicler in the monthly magazine ‘Europa’, in which I (already before 1890) recommended the Nieuwe Gids literature: ...

... in 1890 I was offered the position of foreign affairs correspondent; Jan. 1891 followed my appointment as co-editor of the D.Ct. [Dordsche Courant]
\endmarg 
Jan Pieter Veth (1862--1925) became a very well-known painter of portraits, in particular in the current context of Kapteyn (see van der Kruit, 2024b).

It appears that at least part of this period Easton lived with Hermanus Petrus Antonius Gunneweg (1846--1904) and his wife Anna Maria Caroline Roterman (1850--1910) in Dordrecht. This evidence comes from the portrait of painter Gunneweg and his wife in Fig.~10. This drawing was produced by Gunneweg himself, after a photo taken by Easton. The latter  probably lived with the Gunnewegs in Dordrecht after leaving Delft in 1886, apart from a stay in Paris in 1885, until his marriage in 1891. Gunneweg is described as aquarellist, painter draughtsman. He initially worked as an art painter and draughtsman at a metal goods factory, but at age 50 he devoted himself entirely to art.

\begin{figure}[t]
\sidecaption[t]
\includegraphics[width=0.62\textwidth]{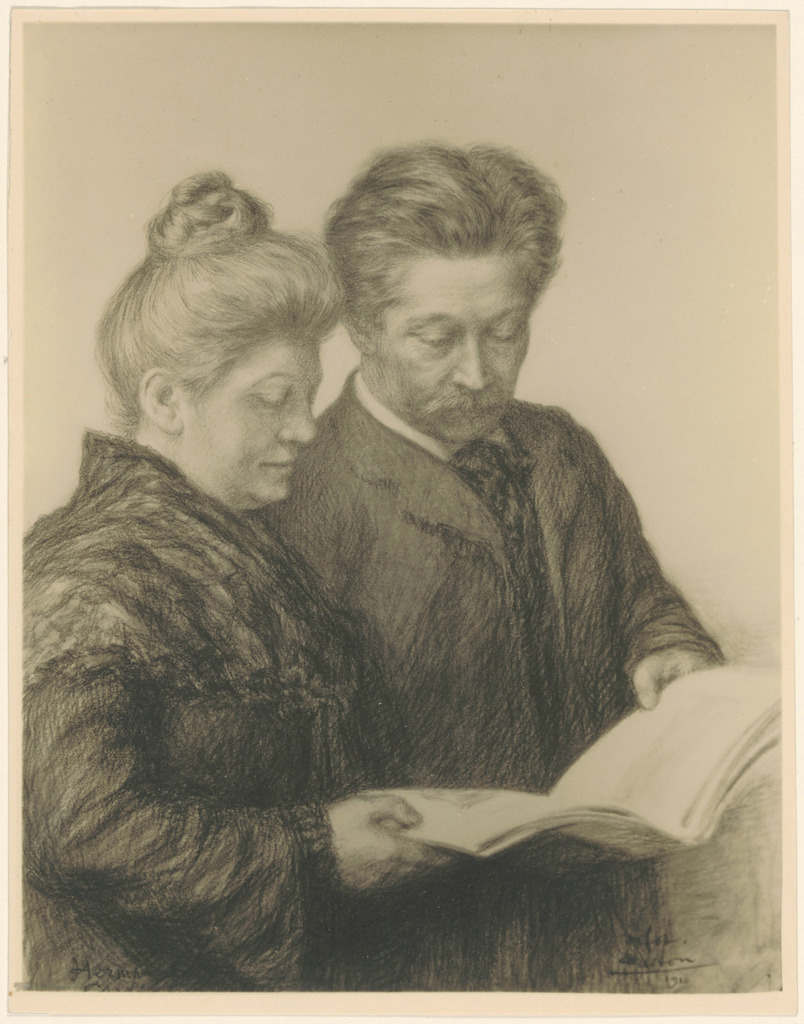}
\caption{\normalsize  Portrait of the painter Hermanus Petrus Antonius Gunneweg  and his wife Anna Maria Caroline Roterman. The Wikimedia description mentions (my translation from the Dutch): `Photograph of a drawing, by Gunneweg, after a photograph made by the teacher, journalist, astronomer, and housemate of the Gunnewegs Cornelis Easton'. After Wikimedia (2019).}
\label{fig:}
\end{figure}

The Gunnewegs were married in April 1873 in Rotterdam. The wife (Caroline Roterman) was born and raised in Dordrecht, but at the time of the marriage lived in Kralingen (now part of Rotterdam), without profession. They must have moved at some time to Dordrecht. I have not found when but it must have been before 1880, since the site ‘Dordtenazoeker’, specializing in making available all kinds of archival data concerning Dordrecht, has a page on the diaries of Hermanus Gunneweg (www.dordtenazoeker.nl/hulpbronnendordrecht/gunneweg\_hpa/2\#zijmenu2) with an entry of 1880 April 16 in the evening at 11 o'clock:
\begmarg 
A few words, and then I go to bed, this evening I have the house alone as my dear wife has with Jufv. Ridderhof gone to Rotterdam to attend the concert. They will return tomorrow evening.
\endmarg

‘Jufv. Ridderhof’ must have been Easton’s mother, Margrieta Wilhelmina Ridderhof. In 1880 she was a widow and therefore Gunneweg refers to her as `jufvrouw’ (or juffrouw), which was used for  unmarried women, followed by her maiden name. It is then also likely that the apparent friendship between the Gunnewegs and Mrs. Easton-Ridderhof resulted in the former taking care of young Cornelis of sixteen, when in 1882 his mother died. The Gunnewegs remained childless. 

It seems likely that the Gunnewegs had been friends of Cornelis’ mother and been taking care of him when his mother had died in 1881, up till he moved to Delft in 1882. After the Delft and Paris interval in 1884 he then had moved  back with the Gunnewegs and stayed there until his marriage in 1891.  Then Gunneweg may have been the {custodian} causing Easton’s fortune to evaporate.
\bigskip

`In the latter half of 1885’ Easton obtained an M.O. degree in French. M.O. stands for ‘Middelbaar Onderwijs’ or secondary education. There is no mention where he did these studies, but it would have been at a university, as I will explain shortly. In the  {\it Biographical Encyclopedia of Astronomers}, Adriaan Blaauw (2014) remarks that Easton studied French at the Sorbonne but it is not made clear where that information comes from. In the citation from the {\it Notes} Easton mentioned his departure for Paris after Delft and it is likely (even probable from the context) that he went there to study. I have seen no evidence independent of Blaauw that it was at the Sorbonne and for how long. Whatever the places where he studied and attended lectures, he could use all of this to prepare for the M.O. exam in the Netherlands. The {\it curriculum vitae} in the {\it Notebook} does not mention Paris.

I now turn to the teaching of modern languages at Dutch universities in order to understand Easton’s further studies to become a teacher of French at secondary schools. For this I have to go back to the founding of the Kingdom of the Netherlands in 1815 and the instituting of academic education. During  the Napoleonic times, starting in 1795, the Dutch universities were closed except those of Leiden (founded in 1576) and Groningen (1614). After this, in 1815, of the remaining ones only that in  Utrecht (1636) was again re-instituted and re-opened.  This was the time higher education had to be organized in a more uniform manner and this was done in the  so-called Organiek Besluit -- `organizational resolution’ -- of that year 1815. In this important document it was stipulated that in the universities there were to be five Faculties: one of which of philosophy and literature. In this Faculty  one could study Dutch and classical languages, but modern languages were not an academic discipline. Until then, mastery of French, German and English was considered a skill that an academic had to master himself. The {Organizational} Decree of 1815 offered the possibility of appointing lecturers to provide the education in modern languages, but these still  had no academic status.

The `Secondary Education Act’ of 1863 had introduced a program of the teaching of modern languages and literature at the HBS and Gymnasium. Especially at the HBS, with its aim to provide education preparing boys for important functions in commerce and industry, solid command of these modern languages was an important aim. The problem was to find qualified teachers. For mathematics,  physics and chemistry this meant that PhDs in these disciplines en mass started a career as teacher at an HBS, but since English, French and German were not an academic discipline this did not happen there.

It took until 1876 before a next step was taken with  the `Higher Education Act’ that stipulated that at least one university had to teach French, English and High German language and literature. Inge de Wilde (2007) has described how the curators of the University of Groningen and the municipal council, ahead of all others, did not hesitate for a moment to seize on this opportunity. The university was going through a difficult period since the number of students was dropping dramatically and the future of the university looked bleak. The Mayor and Aldermen of the city of Groningen therefore took the initiative to make funds available for the recruitment of private lecturers in modern languages to attract more students. Three privaat-docents were to be appointed, with a salary provided by the municipality of Groningen. It started with Barend Sijmons (1853--1935), who was appointed in 1878 for German (and until one was appointed also for English; see also van der Kruit, 2024b). This was a unique situation, since a privaat-docent was not supposed to be paid for his teaching, but {was} appointed {merely} to lecture (usually a subject that was not yet taught). Normally they had some other primary paid appointment at the university or elsewhere as well, such as teacher in secondary education. The decision to appoint full-time {privaat}-docents  gave an enormous boost to the teaching and research in modern languages and literature in Groningen. And this provided qualified teachers in secondary education,

The 1876 Higher Education Act also opened the possibility to appoint full professors in modern languages and again it was the University of Groningen that  in 1882 had Sijmons appointed as  Professor in German language and literature. The other modern languages followed, and so did other universities, although often in the lower rank of Lector (Lecturer) rather than Professor. It still was not possible to obtain a doctorate in French, German or English. This lasted until 1920.

So, how did teachers of modern languages obtain their qualification? When no Doctor’s thesis in modern languages could be submitted at a university, there had to be a possibility to do some other exam for a certificate of competence as teacher at secondary schools. For these there were `state examinations’, organized by the government. Easton only mentioned he obtained that certificate for French. It is likely that he did most of his studies after returning from Paris from home, but he might have attended lectures at the University of Leiden, which was the nearest for him. In any case he managed to prepare sufficiently well for the exam to pass. There were two certificates; the first, designated MO-A qualified for the lower classes and the other, MO-B, for all of secondary education, Easton does not specify this, but since he mentioned an MO certificate only once, it is likely he only obtained MO-A, and stopped there.

\section{La Voie Lactée dans l’espace}

 While observing, drawing and mapping the Milky Way, Easton’s mind sometimes drifted off reflecting on related questions ({\it Notebook}, p.151):
\begmarg
What I observed, drew and described so diligently, was the image of the Milky Way, the projection of a swarm of stars against the vault of heavens. But what was the actual shape of that star cloud?
\endmarg

Easton asked himself the question what the Sidereal System’s three-dimensional shape would be? Surely some flattened structure manifesting itself on the sky  as the band of light that is the Milky Way.

Frederick William Herschel (1738--1822) pioneered the study of the stellar distribution by obtaining star counts at different positions along a circle perpendicular\footnote{The great circle followed by William Herschel’s `star gauges’ indeed is almost perpendicular to the Milky Way plane. This is not obvious from his description that from geographic latitude 55\degs\ north this circle is traced out by the horizon when the star $\tau$ Ceti culminates. It means that it misses the Galactic poles by only about 5\degs\ and crosses the Galactic equator at  longitudes $\sim 45$\degs\ and $\sim 225$\degs\ (van der Kruit, 1986).} to the Milky Way (Hoskin, 2008, for a review; see also more recently Cunningham, 2018, and particular the chapter in this by Steinicke, 2018), and assuming that the distribution was extremely uniform and that he could see all stars out to the edge, he derived his famous cross-cut through the Sidereal System (Herschel, 1785). This flattened structure, however was abandoned and replaced with a flat stratum with unlimited extent later, when his deeper observations showed more stars and that space appeared filled with clouds and clusters of stars with voids in between. This led to a {catalog} of objects, completed after his death by his sister Caroline Lucretia Herschel (1750--1848). Herschel arranged nebulous objects into a sequence along which one could see the  development of contraction into stars. His son John Frederick William Herschel (1792--1871), extending the star counts to the southern hemisphere, and cataloged clusters and nebulae further. His work supported the view, that the stellar distribution was far from uniform, but showed a considerable amount of structure. In his view there are one or two outer rings.

\begin{figure}[t]
\begin{center}
\includegraphics[width=0.98\textwidth]{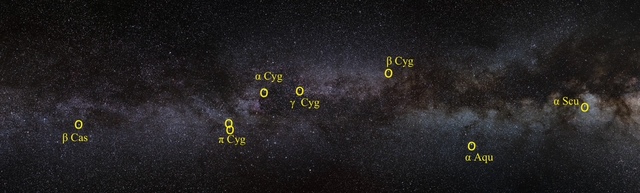}
\end{center}
\caption{\normalsize   Part of the ESO Milky Way Panorama to illustrate the remarks by Easton how the idea of a spiral structure of the Milky Way system occurred to him. Some stars mentioned have been indicated. From left to right these are $\beta$ Cassiopeiae, $\pi ^1$ and $\pi ^2$ Cygni, $\alpha$ Cygni (Deneb), $\gamma$ Cygni, $\beta$ Cygni, $\alpha$  Aquilae (Antares), and $\alpha$ Scuti. Credit: ESO/S. Brunier.}
`\label{fig:}
\end{figure}

But what if the Galaxy were seen ‘from above’? The realization came quite suddenly. This crucial event deserves to be quoted extensively; for images of the Milky Way and stars or constellations Easton referred to, see Figs.~8, 9 and in particular 11. From the {\it Notebook} (p.154):
\begmarg 
It is a pity that the origin of such, sometimes fruitful ideas can rarely be re-established in later times. This is also the case here. But I can still trace the two main motives.

The first factor was: the view of the Milky Way in the vicinity of Cygnus. As I said, I liked to survey the broad features of the Milky Way image, while on the other hand I knew all the visible stars by their {catalog} names, and even the light spots and dark rifts by the names which I myself had provisionally given them: I felt as much at home there as in the streets of my native town. Now, first of all, the brightness and variety of this part of the light belt, compared with the duller part on the Auriga side, had quickly struck me. But the more I studied that region in Aquila, Cygnus, and Cassiopeia, the more fascinated I became by their remarkable structure: that large spot between $\beta$ and $\gamma$ Cygni, so strangely situated to one side, yet connected with the smaller spot above $\alpha$ Cygni, then the subtler parts that rise in an arc as far as $\beta$ Cassiopeiae, and there unite with `the weevil’ (the large bright spot between $\pi$ Cygni and $\beta$ Cassiopeiae) -- so called because of its strange shape -- between which, just north of Deneb ($\alpha$ Cygni) , that strange dark `northern coalsack’, continuing southwards in a dark cleft, `the handle of the pan’, and further {toward} the horizon the slowly swelling branch of Aquila and Scutum. That great spot of light in Cygnus looked like a lens, seen from the side, and sending out arcs of light, which returned there, along the lens, downwards ....

Once, by a sudden inspiration, I must have thought of the pictures in my books, where the great nebula of the `Hunting Dogs’ (M51 in the constellation Canes Venatici) was depicted, in various forms, as viewers of all greater optical power have shown the spot successively: as a sphere with a circle around it, then -- as John Herschel saw it -- that circle split over half the circumference (so that this drawing was indeed cited as a type of the idea that one formed of our galaxy in its true form: an enormous sphere with irregularly distributed stars in it. With the half-split Milky Way belt closing around it ....). But that same nebula had resolved itself in the mighty telescope of Lord Rosse and developed as a beautiful spiral! --  And the Cygnus region, as I saw it in the sky, also had something of a spiral but seen from the side ....
\endmarg

\begin{figure}[t]
\sidecaption[t]
\includegraphics[width=0.62\textwidth]{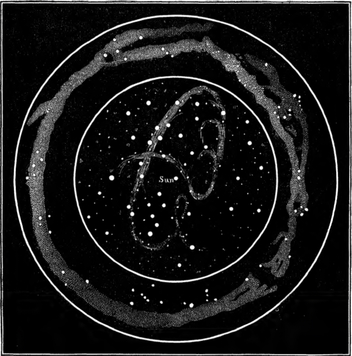}
\caption{\normalsize  The Milky Way as a stream within a larger Sidereal System is shown in the inner part of the figure. Proctor had determined this using the view of the Milky Way by John Herschel in the outer ring, where some prominent constellations can be identified. In his \textit{Notebook} Easton refers to this figure at one occasion as ‘Proctors Krakeling’ (Proctor’s Pretzel) to emphasize it was not an early version of a spiral form, for which Easton claimed the credit.  From Proctor (1869).}
\label{fig:}
\end{figure}

It may come as a surprise that the step to link the structure of the Milky Way system to spiral galaxies had not been taken before, at least not in writing. A notable step in the right direction was taken by Richard Anthony Proctor (1837--1888), British astronomer, who never worked at a university or observatory. He wrote books and articles and initially lived off inherited wealth, until financial misfortune left him without funds. Using mapping of the full-sky Milky Way from the north by William Herschel {and} the southern Milky Way by John Herschel during a five-year stay at Cape Town, Proctor proposed  `A new theory of the Milky Way' (Proctor, 1869). Fig.~12 has been taken from Proctor’s paper. The outer ring depicts a sketch of the Milky Way as outlined by the Herschel observations. Now, Proctor regards this as inconsistent with a flat ring, and instead proposed the central part of Fig.~12 as the view from ‘above’. The Sidereal System consists of `streams’ of stars. In later times Proctor’s work has sometimes been credited as a forerunner of the spiral theory of the Milky Way Galaxy. This must be linked to Proctor describing the small stream that passes close to the Sun `as the form of a spiral'. The structure shows a main stream with minor and major branches and has nothing to do with spiral structure as we use the term now.
\bigskip

\begin{figure}[t]
\begin{center}
\includegraphics[width=0.435\textwidth]{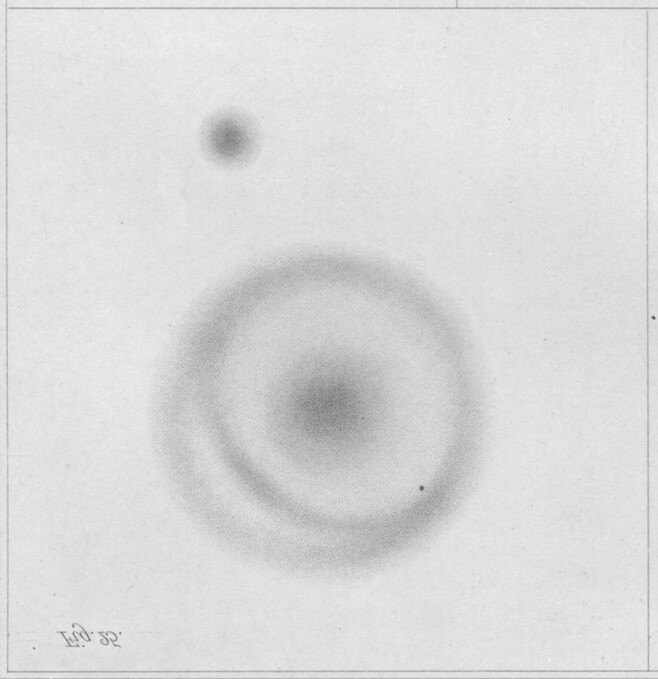}
\includegraphics[width=0.355\textwidth]{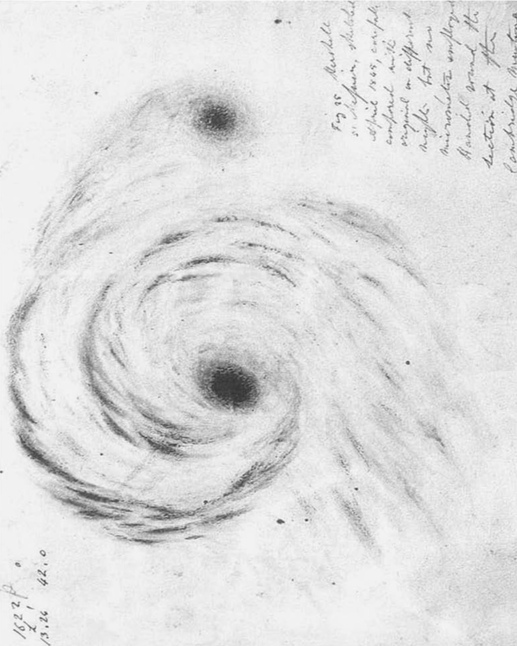}
\includegraphics[width=0.43\textwidth]{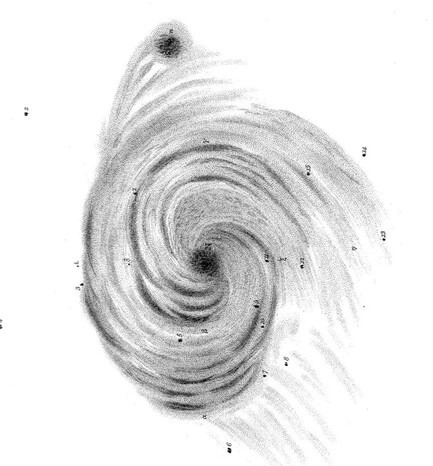}
\includegraphics[width=0.35\textwidth]{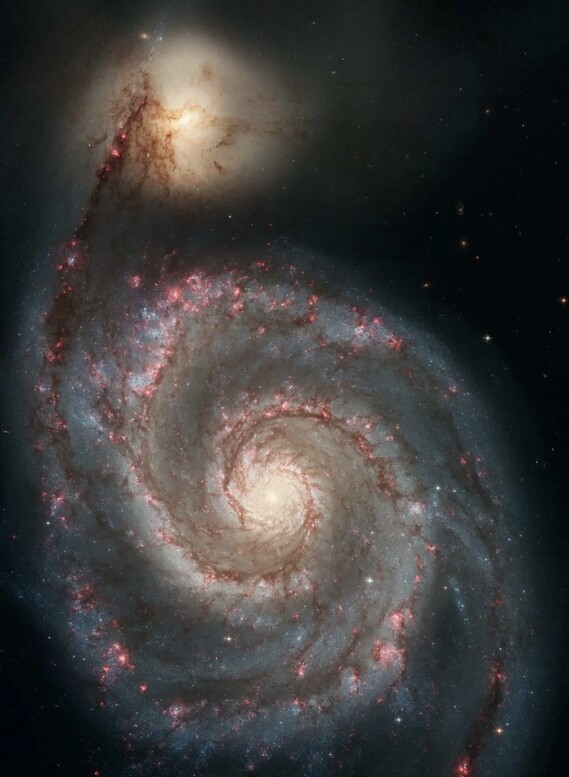}
\end{center}
\caption{\normalsize  Drawing of the Whirlpool galaxy M51 in 1833 by John Herschel (top-left) and by Lord Rosse in 1845 (top-right) and 1848 (bottom-left), and optical image by Hubble Space Telescope (bottom-right). Reproduced from Herschel (1833), Steinicke (2008) and Rosse (1850). Credit line HST image: NASA, ESA, S. Beckwith (STScI) and the Hubble Heritage Team (STScI/AURA)}
\label{fig:}
\end{figure}

Then Messier 51 in Canes Venatici and the description of its structure. Easton mentioned changing views of M51 with improved vision through better and better telescopes. The discovery of its spiral structure by Lord Rosse is well-known (Hoskin, 1982; for a recent, detailed discussion, including on work by others, see Tobin \&\ Holberg, 2008). John Herschel (1833) had viewed the system through his telescope and his drawing of it shows a bright center surrounded by a ring of light, partly branching in two parts (see Fig.~13, top-left panel). Interestingly, this is reminiscent of his view of the Milky Way. The discovery of spiral structure in that galaxy by the fourth Earl of Rosse (William Parsons, 1800--1867) followed in 1845 (see top-right panel of Fig.~13) and in 1858 (bottom-left). In the publication where this illustration comes from (Rosse, 1850), he reports 28 observations with the `6-feet instrument' and earlier ones with the `3-feet instrument'.
\begmarg
September 18, 1843. - Observed with the 3-feet instrument; […] a great number of stars clearly visible in it, still Herschel's rings not apparent, at least no such uniformity as he represents in his drawing. [...]

April 26, 1848. — 6-feet instrument. Saw the spirality of the principal nucleus very plainly; saw also spiral arrangement in the smaller nucleus.
\endmarg
In the paper Rosse gives a list of 38 remarkable nebulae, of which  14 in the category spiral or curvilinear. He noted: `Spiral structures in nebulae were no exception'. This was before great surveys of nebulae by photographic means. 

\begin{figure}[t]
\begin{center}
\includegraphics[width=0.98\textwidth]{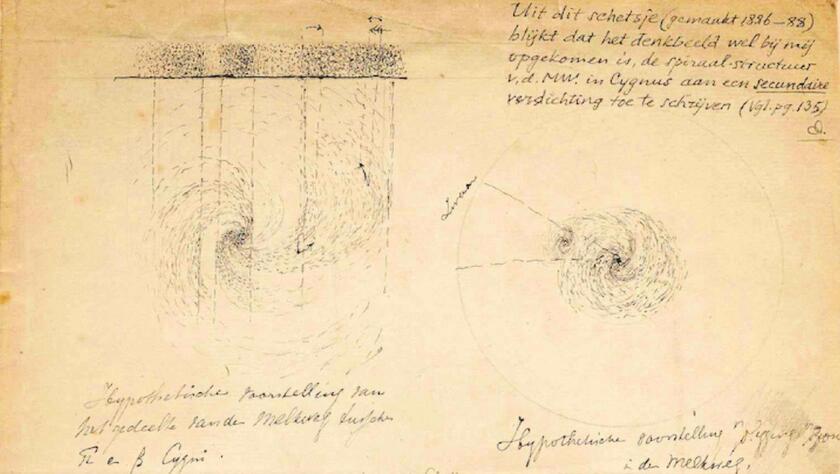}
\end{center}
\caption{\normalsize  Fold-out from the \textit{Notebook} showing how Easton arrived at the idea of spiral structure of the Milky Way system.  The text on the left says: `Hypothetical representation of the part of the Milky Way between $\pi$ and $\beta$ Cygni.' On the top-right: The hand-written note reads `This sketch (produced in 1886--88) shows that the idea had occurred to me to ascribe the spiral structure in Cygnus to a \underline{secondary} density increase.'.}
\label{fig:}
\end{figure}

By the time Easton was mapping the Milky Way the drawings by Rosse had reached textbooks and popular writings as a matter of routine, so it is no surprise Easton was well aware of them. The {\it Les étoiles et les curiosités du ciel} (Flammarion, 1882) for example, had an early (before 1830) drawing showing only the two nuclei (pages 122 and 123), and then the two drawings of Fig.~13, and that is what Easton would have seen. Fig.~14 (from the {\it Notebook}) shows how Easton illustrated his deduction of the spiral structure in the Milky Way system. As he wrote on the top-right, it illustrates how he proceeded to outline the spirals. The text at the bottom says that the drawing on the left represents the Milky Way between $\pi$ (double according to  IAU, 2022) and $\beta$ Cygni. Comparison with Fig.~9 and/or Fig.~11 shows that on the left the extent is actually a bit further, although not all the way out to $\beta$ Cassiopeiae. The procedure was then to interpret bright parts of the Milky Way as resulting from looking along a spiral arm and dark areas from looking through interarm regions.

The sketch on the right is an impression of what the situation in space would look like. The note on the top-right is clear on what this is. At that time Easton only attributed a spiral to the Milky Way light distribution around Cygnus as a {\it secondary}  density enhancement in that direction, the secondary small spiral (see Fig.~14 on the right; the secondary spiral is reminiscent of NGC5195, the companion to M51). The large circle would be the projected sky, where Easton has written ‘Zwaan’, which is Dutch for Swan or Cygnus. The Sun then is located in the center of a larger spiral ({\it Notebook}, p.155):  
\begmarg
And the Cygnus region, as I saw it in the sky, also had something of a spiral but seen from the side .... With a mental leap I must then have extended one or two of those spiral windings, through the southern hemisphere, and returning through Monoceros and Auriga, to where, in Perseus, an interruption in the light belt can almost be seen. It had been found -- the entire Milky Way itself was a spiral, and the Sun only one of the stars situated between its windings!
\endmarg

The small spiral that Easton had deduced from the structure in the Milky Way around Cygnus could in fact be a companion to a larger spiral encompassing the whole of the Milky Way,  even including the by Easton unseen southern part. The spiral Easton presumed our Galaxy looks like is quite different from the Whirlpool in spite of that galaxy being the inspirational source. He draws a three-armed spiral with much opener arms, quite unlike M51. I will return to this later.

Easton produced a manuscript, {\it La Voie Lact\'ee dans l’espace, vue de son pôle boréal}, on this, although — like the  first manuscript {\it La Voie Lactée dans l’hémisphère boréal} — he had no plans for actually publishing it. Both manuscripts have been turned in a final form during 1888, being completed early 1889 at the latest. And subsequently safely stored in a drawer. 
\bigskip

From the {\it Notebook} (p.157):
\begmarg
The following years were again a slack period for my astronomical work. The large map hung on the wall, and my manuscripts ‘Voie Lactée’ and ‘La Voie Lactée dans l’espace’ were lying, ready for printing, in the cupboard. They would remain there for a long time, however. I realized that reproduction and publication would be quite a task, and I was not sure if the scientific world would find them important enough for that. And when I thought about it, I consoled myself with the reflection that it was still a testimony of my work, something that would continue to exist; that when I died, it would be found, and people would then say: look, that’s nice, what Easton did.
\endmarg 

\section{La Voie Lactée dans l’hémisphère boréal}

Easton’s two manuscripts remained undisturbed for quite some time. In the mean time, in 1888 he became the Dordrecht correspondent to {\it De Nieuwe Rotterdamsche Courant}. This newspaper dates back to 1843 and is still one of the major Dutch ones. In 1891 {he} became editor of  {\it De Dordrechtsche Courant}. This is the oldest newspaper of Dordrecht: the first issue appeared in 1789.

On February 5, 1891, Easton married Elizabeth Theresia Visser. She had been born on  September 15, 1863 in Rotterdam. Her birth certificate mentions the mother, Margaretha Rosalia de Hoog, at that time twenty-one years of age, but there is no father. However, at the mother’s marriage on June 1, 1864, to Florinus Visser, sail-maker, the child is recognized. Bride and groom both by then were twenty-two. Possibly the child had another biological father, while Florinus Visser agreed to legally assume this role, but it is also possible that Florinus in fact was the biological father, but that he and Margaretha de Hoog  did at first not get permission to marry from either or both sets of their parents. In the New Civil Code of 1838  the minimum age for marriage was set  at 18 for men, and 16 for women, but both bride and groom needed parental consent to marry until age 30. This was only changed in 1970. The newly wed couple  settled in Dordrecht.

It is not specified when this occurred, but at some time Easton decided to show the manuscripts to his mentor A. van Oven, who at first let them rest in {\it his} drawer, before having a look and deciding to show them to the director of Leiden Observatory, Hendricus Gerardus van de Sande Bakhuyzen (1838--1923). The ‘Sterrewacht’ had been opened in 1861, having been built after the example of Pulkovo Observatory near St. Petersburg by Frederik Kaiser (1808--1872). Van de Sande Bakhuyzen was the second director. In Easton’s words ({\it Notebook}, p.157): 
\begmarg
At some time I gave my manuscripts to my friend and protector v. Oven to read. For a long time they remained in quiet rest in his writing desk, and I do not know how they came to exchange that berth with a period of equally undisturbed rest in the writing desk of Prof. Bakhuyzen. But finally Prof. Bakhuyzen looked through them, [...]
\endmarg
This was in June 1892, some four years after the manuscripts were completed. Van de Sande Bakhuyzen was very much impressed, writing back to van Oven among others ({\it Notebook}, p.159):
\begmarg
At last I have begun to examine this meritorious work more closely and I want to give you my opinion about it. It can only but be favorable, especially about the first part, containing the observations and images of the Milky Way. It is a difficult task to observe and represent the faint light shades well and as far as I can judge Mr. E. has succeeded well.
\endmarg

This was the trigger Easton needed to seriously consider publishing his manus\-cript on the Milky Way. The manuscript on the spiral structure was to be published separately according to the advice of van de Sande Bakhuyzen. The mappings appeared as a book, {\it La Voie Lactée dans l’hémisphère boréal: cinq planches lithographiées, des\-cription detaillée, catalogue et notice historique}. It was printed by Blussé \&\ cie. in Dordrecht and published by  Gauthier-Villars et fils in Paris (Easton, 1893). How this came about will be documented next. First I discuss the important aspect of reproducing the drawings in print, which was not a trivial exercise. In those days including pictures in printed materials was still very much in its infancy.  The original way of including pictures in newspapers had been using woodcuts or engravings in wood. The idea of halftone printing, pictures consisting of dots of varying sizes, was patented in 1881 by American inventor Frederic Eugene Ives (1856--1937), although the precursor of using photographic screens or veils that divided up the picture in small pieces, was patented in 1852 by William Henry Fox Talbot (1800--1877), an English inventor and early experimenter with the use of photography. Halftone printing was used in newspapers, but was absolutely unsuitable for Easton.

Easton first went to Henricus Jacobus Tollens (1864--1936), one of the first Dutch professional photographers; his studio included a main branch in Dordrecht. But the results were not satisfactory. He next came in contact with Walterus ‘Walter’ Lambertus Antonius Ruttenberg (1859--1921), who although he was a manufacturer (not clear of what), was also a painter, draftsman and lithographer, in which capacity he produced illustrations for books. He lived and worked in Dordrecht between 1883 and 1903. Easton  came in contact with him at the Gunnewegs, who apparently saw each other regularly.  Ruttenberg suggested the use of lithography. The principle of lithography, dating back to 1796 by Johann Alois Senefelder (1771--1834), is to represent the image on stone with hydrophilic or hydrophobic parts, which was then transferred to paper with water-ink (or oil-based ink on wet stone). The problem was of course how to transfer Easton’s pictures onto stone. Ruttenberg proposed  Easton would copy the figures on {especially} prepared paper and he would then transfer that onto stone for lithographic reproduction. Although this process did work, Easton still had to do quite a bit of retouching of the final figures before they could be added to the printed books.  

The printing was done by  Blussé \&\ Cie. in Dordrecht, who used a new typeface for this.  This printing firm was founded by Abraham Blussé (1726--1808), a bookseller, publisher, printer and poet in Dordrecht. His printing company  and publishing business merged, in 1808, through a marriage, with a bookseller/publisher firm in Dordrecht founded by  Johannes van Braam (1677--1751), and continued as Blussé \&\ van Braam. As remarked above, the Dordrechtsche Courant's first issue appeared in 1789. Since 1795, the newspaper was printed and  published by Blussé, later Brussé \&\ van Braam. It is therefore not surprising that Easton chose this printer. The printed book refers to Blussé \&\ Cie. (‘compagnie’), so the addition van Braam was not always used. As a matter of worthy interest I add that painter Vincent van Gogh (1853--1890) worked in the bookstore of Blussé \&\ van Braam for a short period (January to May 1877) as a clerk.

The publisher was Gauthier-Villars, a French publishing house that  dates back to 1790, when it was founded in Paris by printer and bookseller Jean Louis Marie Courcier (1790--1811) in a house described by him as a ‘bookstore for mathematics, physics, chemistry, mechanical arts, and the sciences that depend on it’. Jean-Albert Gauthier-Villars (1828--1898) bought it from him in 1864 and changed the name to his own. It established itself quickly as a publisher of scientific journals and books. By the time Easton had his book published there it was the prime scientific publisher of France and very much respected also outside France.
\bigskip

The book as it eventually appeared consisted of an Introduction with the details on how the maps were produced. This is followed by a chapter reviewing every publication that described astronomical aspects of the Milky Way back to Aristotle, with three appendices. The first addressed the discussion of the Milky Way in the {\it Almagest} of  Ptolemy (Claudius Ptolemaeus, ca.90--ca.168). The second appendix was a French translation from the German  of part of   a text  by H.J. Klein. Hermann Joseph Klein (1844--1914) was a German astronomer and meteorologist, known for popular writings and the publication of a star atlas. He observed from Cologne (Köln) and Easton noted that `we owe the first serious attempt at a detailed description of the Milky Way to him’. He only published a short article, `Sur l'aspect de la Voie Lactée’, on the brightness distribution of Milky Way (Klein, 1867; not listed in the NASA/SAO Astrophysics Data System ADS, but available in Google Books). Easton heard from him he had terminated the project to describe the Milky Way from Köln with its `lack of pure air’, and Easton decided to translate a part of his description into French. And finally, as third  appendix, a quotation of the most important part of `Détermination de l'éclat des taches de la Voie Lactée’  (Determination of the brightness of  spots of the Milky Way), the first attempt to determine photometrically the relative brightness of parts of the Milky Way by J.C.Houzeau. Belgian astronomer, meteorologist and journalist Jean-Charles Hippolyte Joseph Houzeau de Lehai (1820--1888), known for his reviews of astronomical literature and leader of solar eclipse and Venus transition expeditions, was Director of the Koninklijke Sterrenwacht van België (Brussels Observatory).

There is only one place where Easton could possibly have conducted this historical research -- the library of Leiden Observatory. It is probable that this discussion on historical studies and descriptions of the Milky Way was written after  van Oven had sent the `completed manuscript’ to Leiden, and the historical chapter only was researched and added while the book was in preparation for publishing.

\begin{figure}[t]
\begin{center}
\includegraphics[width=0.90\textwidth]{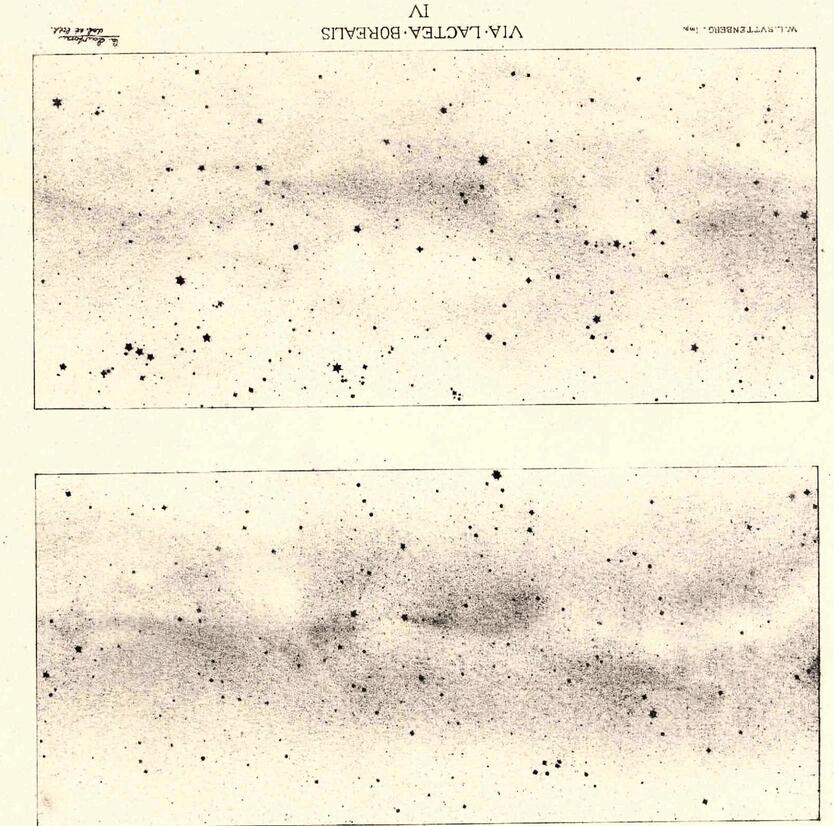}
\end{center}
\label{fig:}
\end{figure}

\begin{figure}[t]
\begin{center}
\includegraphics[width=0.28\textwidth]{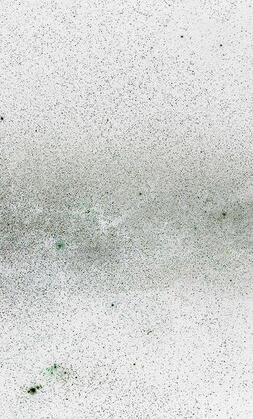}\includegraphics[width=0.56\textwidth]{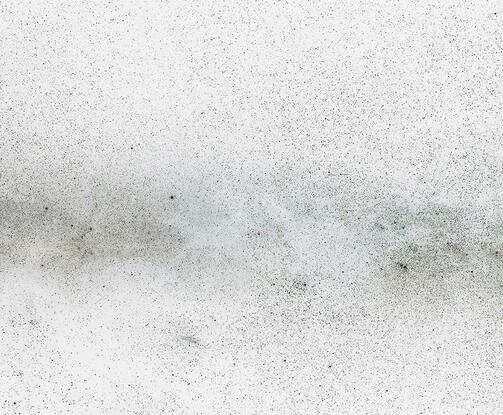}
  
\includegraphics[width=0.84\textwidth]{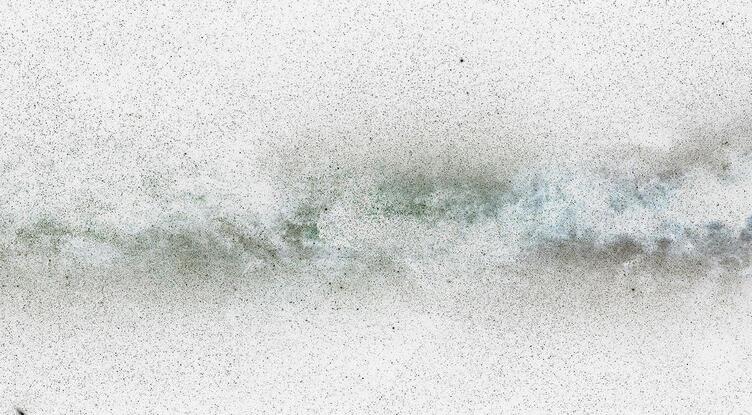}
\end{center}
\caption{\normalsize Easton’s `general map’ published in his \textit{La Voie Lactée dans l’hémisphère boréal} (Easton, 1893). The maps in the publication have west on the left and south up, so I have rotated them by 180\degs\  and turned them into reverse video.  The positions of the foreground stars are plotted by Easton using Marth (1872, 1873a). For comparison I have cut the same areas from the ESO Panorama (ESO, 2009). The agreement is excellent. Credit: ESO/S. Brunier. }
\label{fig:}
\end{figure}

Then follows  a very detailed description of the northern Milky Way including a {catalog} of luminous and dark objects and features, respectively 164 luminous and 47 dark. The main result is a set of maps. First three detailed maps with shadings indicating surface brightness,  together covering the northern Milky Way, followed by a general (two-part) map showing the northern Milky Way in a single overview. The latter is reproduced in Fig.~15. This map is followed by more or less transparent overlays with the same layout-out as the ‘general map’, which identifies all features in his {catalog}. Finally there is what Easton called an analytic chart, on the same scales as the general map with star positions and constellations. The positions of stars were plotted using the list of stars on or near the Milky Way given by Marth (1872, 1873a,b). Albert Marth (1828--1897) was a German astronomer, who worked mostly in London, Malta and Ireland. He published a list of more than 750 positions of bright stars to draw a map of stars in rectangular coordinates along the entire Milky Way.  In a sense these are forerunners of Galactic coordinates, which were not available in {catalogs} and the like. Marth’s stars are plotted on the analytical chart, but also on the brightness maps of Easton for orientation (see Fig.~15). I have also reproduced for easy comparison the same areas from the ESO Panorama (ESO, 2009). The agreement is remarkable and excellent. This is a major accomplishment and testimony to Easton’s carefulness, dedication, perseverance, draftsmanship and patience.

Altogether the book contains 63 pages of text, {catalog} and table of contents (and another two with a Post-scriptum with errata and additional information added in print), plus the six figures. There is a dedication of the work to Dr. D.S. van Oven and a short Preface by H.G. van de Sande Bakhuyzen.

Easton, in his {\it Notebook}, noted (p.163):
\begmarg
[...] that I completed the publication of my La Voie Lactée on the night of 3/4 November 1893. That evening I put the finishing touches to the retouches of the first 20 copies that I had to send to the publisher Gauthier-Villars in Paris. `Completing’ is the right word here, because observing, drawing, putting on stone, it was all my work; I also had to update the prints and I had taken care of the entire edition myself, after all, Gauthiers-Villars only had them ‘in deposit’. It was the work of my hands.
\endmarg
\bigskip

The preparation had taken a year and a half (Easton remarked that it was in addition to his editorial employment and marriage, together taking up a major portion of his time). From  the {\it Notebook} (p.161): 
\begmarg
 That firm  [Blussé \&\ Cie.] also had the great goodwill to set the costs of the publication -- which I  had to bear myself for the time being -- as low as possible: I believe that the printing costs amounted to only fl. 175.--. The maps, my travels in connection with the publication and the other expenses were of course also at my expense. Thus the 200 portfolios alone cost me fl. 50.--.
\endmarg
The notation `fl.’ for `florin’ indicates Dutch guilders. In terms of current purchasing power fl. 175,-- in 1895 corresponds to of the order of  \textgreek{\euro} 1800 (estimated with the {\it Historical Currency Converter}, see www.historicalstatistics.org/Currencyconverter.\linebreak[4]
html) and fl. 50,-- to  around  \textgreek{\euro} 500. The travel probably concerns trips to Leiden Observatory and van de Sande Bakhuyzen, undoubtedly in large part for the research for the historical section. During one of these visits he met Anton Pannekoek, who also was heavily involved in drawing maps of the Milky Way and who would later compare his and Easton’s maps quantitatively.  

From the {\it Notebook}(p.163): 
\begmarg
It had been one of my small triumphs that the world-famous firm of Gauthier-Villars accepted my work. But they also took their share of the proceeds: a mere 43 per cent! so that the price of 18 francs  was actually not high enough to bring much into my coffers. About 200 copies were printed; in retrospect that circulation was too small, but I did not think at the time of the possibility of a ‘bookseller’s success’ with a work like this. Strangely enough, to a certain extent it has been just that: in ten years the first print run has been virtually sold out, [...] Prof. van de Sande Bakhuyzen supported me indirectly by purchasing 10 copies for the Sterrewacht, it still took years before I had more or less covered my main expenses. And my circumstances certainly did not permit me to play Lick or Carnegie!
\endmarg 
In current purchasing power the price of the book would be about 90  \textgreek{\euro}.
\bigskip

\begin{figure}[t]
\begin{center}
\includegraphics[width=0.31\textwidth]{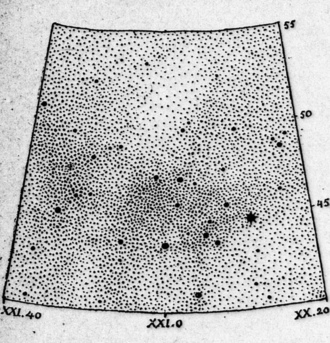}
\includegraphics[width=0.32\textwidth]{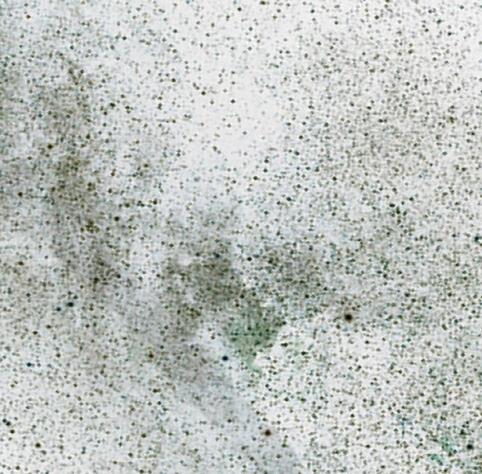}
\includegraphics[width=0.30\textwidth]{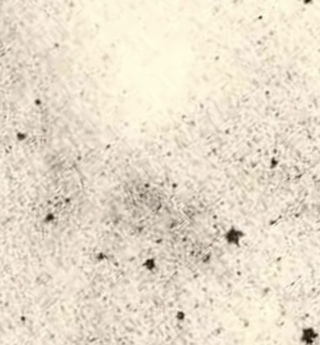}
\end{center}
\caption{\normalsize  A part of the Milky Way near the star Deneb or $\alpha$ Cygni (somewhat to the right of and below the middle). On the left, part of the map using pointillizing from Easton (1895a), in the middle roughly the same part from the ESO Milky Way Panorama and the lithographic reproduction on the right from the \textit{La Voie Lactée dans l’hémisphère boréal} (Easton, 1893).}
\label{fig:}
\end{figure}

In the maps the surface brightness distribution was represented by shading. Later, Easton regretted he had chosen for the lithographic procedure, and in hindsight felt he should {have} chosen the method of `pointillizing’ the maps with Chinese ink. Then the surface brightness is represented with little circles of a size corresponding to the brightness. These could after all be reproduced simply black-and-white, so this would have avoided many difficulties. The other advantage according to Easton was that reproduction errors are almost excluded. He used the pointillizing in a paper `La distance de la Voie Lactée’, published in the {\it Bulletin de {la} Société Astronomique de France} (Easton, 1895b). This journal, in full the {\it Bulletin de la Société Astronomique de France et revue mensuelle d'astronomie, de méteorologie et de physique du globe}, has a somewhat complicated history. It in fact is the forerunner of the well-known {\it l’Astronomie}, and was published between 1887 and 1910 after the founding of the {\it Société Astronomique de France} by Camille Flammarion. From 1912 on the journal was renamed {\it l’Astronomie}, after a journal founded also by Flammarion in 1882. The two were published in parallel until 1894, when the original {\it l’Astronomie} had been terminated, but it reappeared in a different form in 1912, preserving the volume numbering of the {\it Bulletin}.

A comparison between pointillizing and shading  is shown in Fig.~16, which compares the region shown in Easton (1895b) with the corresponding area in the published {\it Voie Lactée}. Also the corresponding area from the {\it ESO Milky Way  Panorama} has been included. The correspondence with the sky is not bad in broad terms, but although much easier to produce, pointillizing from the viewpoint of presentation seems to me inferior. 
\bigskip

The questions remain what other work there was in comparison and what impact the book made. In fact there was very little comparable work, certainly in 1888 when Easton finished his manuscript. There were atlases with usually no more then an outline of the Milky Way. Two very notable examples are the {\it Atlas Coelestis Novus} by Heis (1872), and the {\it Uranométrie Générale} by Houzeau (1878). Eduard Heis (1806–1877) was a German astronomer (and mathematician), who worked at the Königliche Theologische und Philosophische Akademie in Münster, predecessor of the Westfälische Wilhelms-Universität. He was interested in observing the Milky Way, but also in zodiacal light. The Heis Atlas has no  link in ADS to a digitized version,  but it has been digitized by the University of Latvia (see the Reference Section). It has beautiful maps, but no detailed drawings of the Milky Way. The Houzeau {catalog} and atlas are available through ADS; it has detailed star maps, but only some outlines of the Milky Way.

A beautiful set of fifteen astronomical drawings was published by Trouvelot (1882), of which one was a large part of the Milky Way. Étienne Léopold Trouvelot (1827--1895) was an artist with an astronomical interest. After fleeing France after the Napoleon coup to the USA, he worked as artist and astronomer at Harvard Observatory producing beautiful illustrations. In 1882 he returned to France and worked at the Observatoire de Paris (Meudon). The accompanying {\it Manual} to fifteen of his pictures (Trouvelot, 1882) has a chapter on the one of the Milky Way, detailing aspects of the picture and history of research on the  sidereal universe. According to Easton (1893), in my translation from the French  (p.23):
\begmarg
The size of the plates (in chromo-lithography) of this series is 50 x 70 centimeters. The part represented on this plate gives only half of the Milky Way visible from Cambridge (Massachusetts) which extends from the constellation of Cassiopeia to that of Scorpius, the part visible in summer.
\endmarg

By the time the {\it Voie Lactée} appeared in print (but not when Easton finished his manuscript, as he stressed himself in his {\it Notebook}) the comparable publication of Boeddicker  had become available. The publication (Boeddicker, 1892) consisted mainly of four lithographic reproductions of his  drawings with only a short introduction. From Easton (1893; p.23, my translation):
\begmarg 
The most important publication that has appeared to date is undoubtedly the large atlas of Mr. Dr. Otto Boeddicker, at Birr-Castle, Parsonstown (Ireland). These drawings, the result of observations made between 1884 and 1889, were exhibited at Burlington House in November 1889, before members of the Royal Society, and published in 1892 [...] The drawings of Mr. Boeddicker were lithographed by Mr. W.-H. Wesley, adjunct secretary of the Royal Society.
\endmarg

Of what had been published in terms of similar material before Easton’s {\it Voie Lactée} really only the maps by Trouvelot and Boeddicker are worth considering. The quality of Easton’s work is definitely comparable to or better than these earlier publications, which lack the detailed description, inventory of features, both bright and dark, and very thorough description of what has been published before on the Milky Way as a phenomenon. That Easton’s work is of high quality and very important for studying the Milky Way, was noted in the {\it Preface} written by H.G. van de Sande Bakhuyzen (Easton, 1893; p.7; my translation):
\begmarg
Even after Mr. Boeddicker’s superb drawings, Mr. Easton’s work is of great value, because the author, possessing the various qualities necessary for the success of what he had set out to do, has been able to guard against the influence so harmful of preconceived ideas, either those arising from some theory on the constitution of the Milky Way, or those which arise quite naturally from the study of the drawings of other astronomers. It was only after he had completed all his own drawings that he studied similar ones published by other astronomers.

	I am convinced that Mr. Easton's publication will be a great help to all those involved in the interesting but too long neglected study of the structure of the Milky Way.
\endmarg

The study of the surface brightness distribution on the Milky Way requires calibrated maps, containing isophotes, such as presented quite extensively by Pannekoek (van der Kruit, 2024). To this Easton would contribute as well.
\bigskip

The reception of the book was very positive. Van de Sande Bakhuyzen made a favorable announcement at the Royal Academy of Arts and Sciences (KNAW), which became quoted by a number of Dutch newspapers. Easton must have sent around a fair number of complementary copies which gave rise to positive letters. In the {\it Notebook} he gave a long list of quotations. There were very positive remarks by Jacobus Kapteyn, with whom Easton would develop a closer relationship. But he had also very positive letters from his competitors, Boeddicker and Trouvelot, and from his great inspirator Flammarion. Trouvelot mentioned he had contacted A. Couper Ranyard, the editor of the magazine {\it  Knowledge}, which led to an excellent relationship of Easton with Ranyard, and after his death with Walter Maunder, and to a number of articles in that magazine by Easton.  {\it Knowledge: An illustrated magazine of science} was a British popular magazine about science, published between 1881 and 1918, founded by astronomer Richard Proctor. {\it Knowledge} began as a weekly but became a monthly magazine. In a sense, it was the European counterpart of {\it Scientific American}. Arthur Cooper Ranyard (1845--1894), an English astronomer, was editor of {\it Knowledge}   for many years after Proctor’s death.
\bigskip

From the {\it Notebook} (p.176):
\begmarg
Gauthier-Villard, who advertised the work annually in the {\it  Anniversaire du Bureau de Longitude}, now and then received a new shipment from me in deposit. And around 1905 the work appeared to be as good as sold out, both with me and with them. -- Now and then I still see it “antiquarian” in an auction or fund {catalog}.
\endmarg

The in total  200 printed copies were  essentially sold out in ten years. Not bad for a book of \textgreek{\euro} 90 in current prices on a subject attracting a limited audience.

\section{Foreground stars and background Milky Way}

One would have thought that as soon as the {\it Voie Lactée} had appeared, Easton would have turned to his other manuscript on the spiral structure in the Galaxy. But his attention was captivated much more by another matter. While working on the historical chapter of his book, he had the Atlas that had been published as part of the {\it Bonner Durchmusterung} effort, on loan from Leiden Observatory. Let me briefly recall, that the {\it B.D.} had been constructed from transit observations at the Bonner  Sternwarte under the initiative and leadership of its director Friedrich Argelander. The telescope used was not a meridian circle, but a refractor on a parallactic mount, with a 16 cm aperture and 1.9 meters focal length. The {catalog} had been published in three parts between 1859 to 1863 (an extension was published in 1886). The {\it Durchmusterung}, being a systematic survey, resulted in a {catalog} of the positions and magnitudes of 342,198 stars brighter than approximately apparent magnitude 9.5, covering the sky from the North Pole to -2\degs\ declination. It had been accompanied by an atlas: {\it Atlas des nördlichen Himmels f\"ur den Anfang des Jahes 1855,  nach der in den Jahren 1852 bis 1862 auf der Königlichen Universitäts-Sternwarte zu Bonn durchgeführten Durchmusterung des Nordlichen Himmels} (Argelander, 1863). This consisted of 40 maps on a fairly large scale (the size of the atlas is 52 by 73 cm).

Easton had noted that there seemed a correspondence between the surface density of stars and the surface brightness of the Milky Way. If true, such a correlation would have consequences for the distance of the Milky Way, because it would have to be similar to that of the stars the correlation was holding up for. He decided to study this first.

\begin{figure}[t]
\begin{center}
\includegraphics[width=0.60\textwidth]{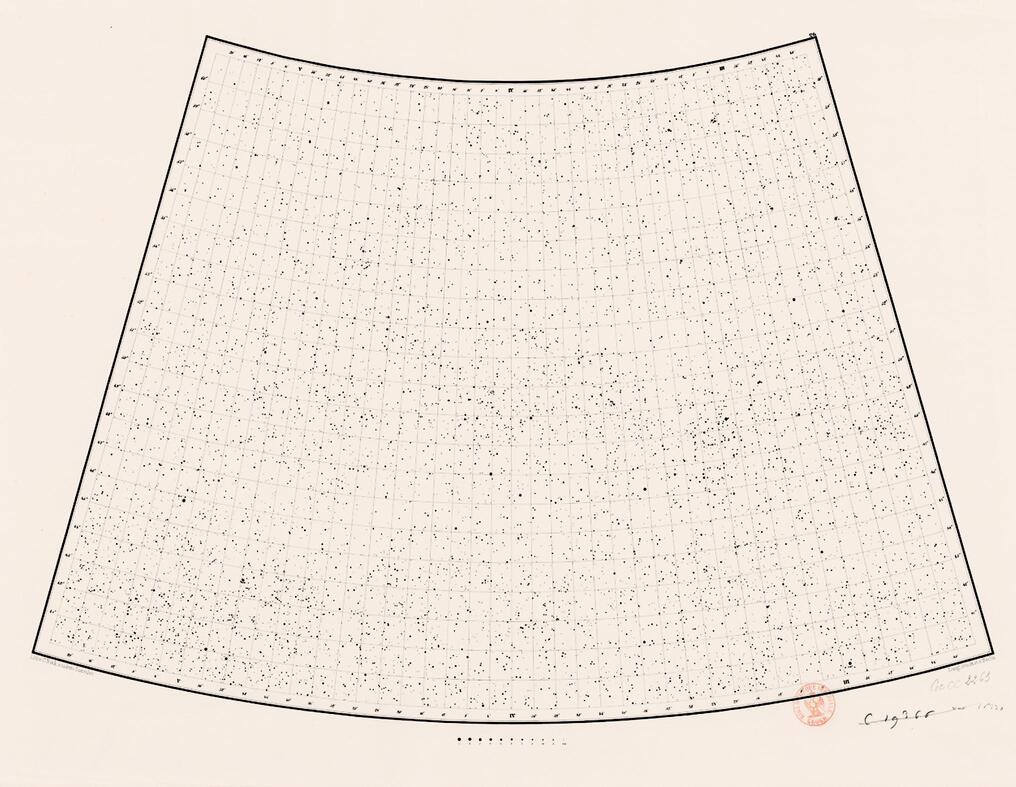}
\includegraphics[width=0.60\textwidth]{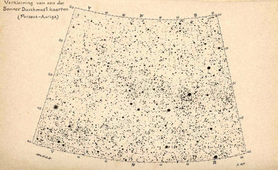}
\end{center}
\caption{\normalsize Easton’s first attempt to study the possible spatial relation of the surface density of stars on the sky and the surface brightness of the Milky Way. To this end Easton reproduced some maps or parts of it from the \textit{Atlas des nördlichen Himmels} reduced in size by a factor  4, while leaving the size of the symbols intact. Top: the original map of the Perseus-Auriga region; below: the reduced map from the  \textit{Notebook}.}

\label{fig:}
\end{figure}

He started out in a way that he himself called `clumsy’, but which was certainly very time consuming. He  reproduced some maps of the Argelander atlas on a much reduced size (1/16; in area, so the side to 1/4),  but left the size of the star symbols {intact}. These were black dots with decreasing size with fainter magnitudes. This made it possible for Easton to more clearly see patterns in the distribution of stars that could be compared to his maps. Fig.~17 shows an area he had pasted in his {\it Notebook} with the corresponding map in the original {\it Atlas des nördlichen Himmels}. The reduction to 1/16 reduces a page in the {\it Atlas} to about 13 by a little over 18 cm. Easton wrote in his {\it Notebook} that he had `completed the time-consuming reduction of fairly large chunks of the Bonner maps to 1/16 and -- with Luttenberg’s help again -- put them on stone' (p.179). As far as I am aware he used only a very small part in an astronomical publication, and this is an article in {\it Knowledge} (Easton, 1895b) that will be discussed below. The version of Fig.~17 did get published in a popular Dutch article by Easton in {\it Elseviers Geïllustreerd Maandschrift} `What are the stars?’ (Easton 1906) merely to illustrate the crowding of stars fainter than the naked eye can see (the crossed-out letters `Str.K.C.E.', star chart C. Easton, betrays the fact the figure in the {\it Notebook} is a copy from the {\it Elsevier}-issue). The `comparison with the Milky Way maps would follow how much the distribution of Milky Way light and the distribution of bright stars correspond with one another’ yielded the result that the `degree of coincidence thus obtained was of course small, and the result could not be generalized’ (from the {\it Notebook}, p.180). Easton referred to the publication by Proctor (1888), for the entire Northern Hemisphere, which did not present any conclusive result.

Easton attended the meeting of the Astronomische Gesellschaft in Utrecht in August of 1894, and did show them there. He referred to the fact that they were also shown at the meeting of the `Association of Friends of Astronomy and Cosmic Physics’ in Münster in 1893
 by Joseph Plassmann (1859--1940), a Gymnasium teacher there, who together with Wilhelm Julius Förster (1832--1921) of the Berlin Observatory had founded this association. Pannekoek had prepared special  new star charts in Galactic coordinates for persons interested in mapping the Milky Way. These had been lithographed by Easton and made available to anyone interested, and these were shown at this Münster meeting by Plassmann. They were in Galactic coordinates and as such not otherwise available. He urged others to use these to map the Milky Way, but Pannekoek reported Easton had not very many requests  (see van der Kruit, 2024).
\bigskip

Having not been able to confirm definitely that there is a correlation between the position of the Milky Way bright spots and the distribution of the stars on the Bonner atlas, Easton looked into more detail into two broad strips, one in Aquila and the other in Cygnus. Dividing these up into smaller areas he classified in these the Milky Way as bright, faint, very bright, etc. and compared this to various star counts from the {\it Bonner Durchmusterung} to progressively fainter counts by Celoria, Epstein, and William Herschel.

The counts by Giovanni Celoria (1842--1920), an Italian astronomer who worked in the field of star counts and statistics and was director of the Brera Observatory near Milan, go to about magnitude 11. Theobald Epstein (1836--1928) was a Frankfurt teacher in mathematics and physics with an interest in astronomy. He counted stars  between 1877 and 1888 in about 2700 fields distributed over the sky with a 6-inch (15-cm) telescope, but these never were published. The limit of the counts was about magnitude 12. For more information on Epstein, see van der Kruit (2024a). Herschel’s limit was assumed 15 (which is up to a magnitude fainter than found in van der Kruit, 1986). In addition Easton had received ‘excellent reproductions of three fine photographs by Max Wolf, one in which the plate was exposed for three hours, still showed stars of magnitude 11.5; the other two (exposure times 11 and 13 hours) those of magnitude 13.5 or 14’ (Easton, 1895a, p.87; my translation). Easton counted stars on these photographs. And he used his ‘reduced’ charts of the Bonner Atlas.

In the end Easton concluded ({\it Notebook}, p.182)   that the group of faintest
\begmarg
stars of Argelander (approximately the stars 9.0-9.5 or 9.8) corresponded in their distribution in a remarkable way with the position of the bright and dark spots in the Milky Way, and that indications of such a correlation already occurred here and there with much brighter stars.

\noindent
[...]

I drew the bold conclusion (in 1894) that the Milky Way stars were therefore in general not considerably further away than the stars of average: 9th to 10th magnitude. Far-reaching `feelers’ full of stars, penetrating deeply into space and radiating away from the Sun, would nevertheless be a very remarkable structure of the galaxy. However, I left open the possibility that the Milky Way could be closer in one part of its run on the sky than in the other, and I even ventured an allusion to a spiral structure, without however elaborating this last idea.
\endmarg

This resulted in his second publication, `Sur la distribution apparente des étoiles dans une partie de la Voie Lactée’ in {\it  Astronomische Nachrichten} (Easton, 1895a). Van de Sande Bakhuyzen communicated an abstract to the Royal Academy {of Arts and Sciences}.
\bigskip

 However, it did not take very long before Kapteyn put a sort of a damper on it (from the {\it Notebook}, p.183):
\begmarg
From prof. Kapteyn I soon received a – salubrious – shower on my enthusiasm. He explained to me that I had neglected to take into account the strong inequality of luminosity of the stars, as a result of which in every distant accumulation, if it is numerous enough, a few stars of extraordinary brightness will show themselves like `nearby’ and in that direction this will increase the normal number of apparent brightnesses of a certain magnitude. Moreover, one may no longer assume that the majority of the {\it Bonner Durchmusterung} stars are so close (the stars visible to the naked eye must for instance be exceptionally bright for more than 90 per cent (Eddington), so are largely quite far away), that the nebular clouds consist mainly of faint stars. The most important thing is: my `average distance of stars of the n-th magnitude’ is actually an expression without meaning; Kapteyn himself in particular has shown that the apparent brightness can only be a rough distance measure for large groups when used in combination with the proper motion.

My conclusions from {\it Astronomische Nachrichten} 3270 have therefore largely been invalidated -- `jumping to conclusions’ is mainly a mistake of beginners.
\endmarg

\begin{figure}[t]
\begin{center}
\includegraphics[width=0.88\textwidth]{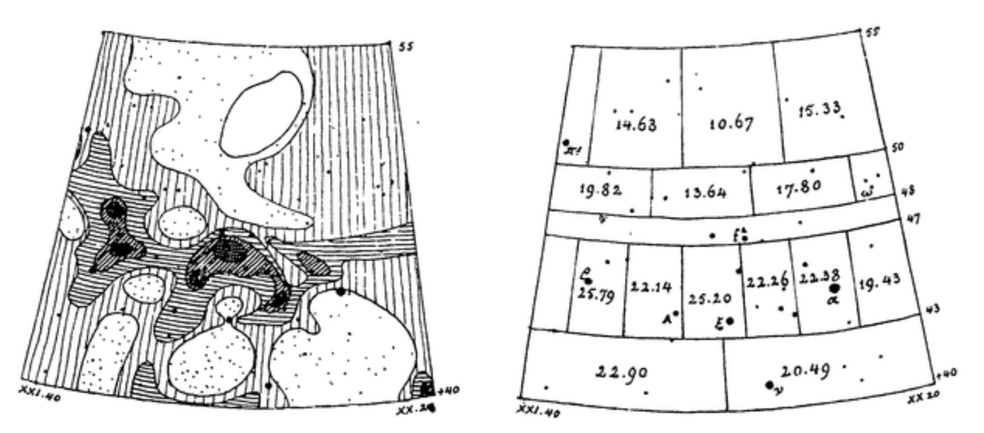}
\end{center}
\caption{\normalsize Comparison of Galactic light to star counts. Left: Isophotes of the Milky Way `in the zone $\alpha$-$\pi$ Cygni'.  Right: Stellar surface density of stars of magnitude 9.1-9.5 in the \textit{Bonner Durchmusterung}. From Easton (1895c).}
\label{fig:}
\end{figure}

The data in themselves and correlations gave rise to further publications.  First he published a paper (Easton, 1895b)  `La distance de la Voie Lactée’ in the {\it Bulletin de la Société Astronomique de France}, in which appeared the pointillists map of Fig.~16, together with a chart (not reproduced here) of the same area with expressed in actual numbers the density of stars of magnitude 9.5 in the {\it Bonner Durchmusterung}. Next he published what really was an English version (although not a literary translation) of the {\it Astronomische Nachrichten} paper, `On the distribution of stars and the distance of the Milky Way  in Aquila and Cygnus’ (Easton 1895c), somewhat rewritten and with two figures comparing Milky Way brightness and star counts in a 15\degs\ area in Cygnus. In these figures (see Fig.~18) he presented the Milky Way in the form of an isophotal map. In the text there is no reference to these figures, nor any explanation of how they were obtained or what the isophote levels are or the unit of the surface densities. Easton repeated his conclusion -- but provided no convincing arguments -- that the stars contributing to the diffuse {Galactic} light should be at a distance `which does not greatly exceed that of 9th or 10th magnitude stars’ (Easton, 1895c, p.220). Here the matter rested for a while.
\bigskip

Easton's  paper {\it Astronomische Nachrichten} 3270 was in his view his second publication, and it is that indeed also in his publication list {toward} the end of his the {\it Notebook}. In ADS, however, there is a publication by Easton in {\it Nature} (Easton, 1894), that precedes this. It is a short note `On the nebula in Andromeda'. It does not concern M31 itself, but companion dwarf elliptical NGC205, designated h44 (cataloged by John Herschel). Easton commented on photographs of the Andromeda Nebula and of NGC205. There are two by Roberts (1888, 1889), who in those years published many articles o`n astrophotography in the {\it Monthly Notices}. Isaac Roberts (1829--1904) was an engineer and businessman, Welsh by birth,  mostly living just across the {English} border in Birkenhead,  but also amateur astronomer working on  astrophotography with his private 20-inch reflector. The photographs are not reproduced in the journal, but `deposited in the Library’, undoubtedly of {Royal Astronomical Society}  R.A.S. But in 1894 Roberts did publish a book, {\it A selection of photographs of stars, star clusters and nebulae together with information concerning the instruments and the methods employed in the pursuit of celestial photography} (Roberts, 1894). Easton must have seen this in Leiden. Fig.~19, left-hand panel, shows part of an 1888 photograph of the Andromeda Nebula.

The other picture was from the set published by Trouvelot (1882) in his collection of drawings referred to above.  It contains a drawing of the Andromeda Nebula from 1874 (Fig.~19, second panel). Now in Easton’s {\it Nature} paper a drawing (Fig.~19, third panel) is presented, which he suggested might show that the nebula had rotated by 15\degs. Easton felt this was significant, even though the Trouvelot image was a drawing and not an actual photograph.  The only justification he gave for trusting Trouvelot’s image is Roberts’ statement that is was excellent. The difference in orientation is noticeable in the first two panels of Fig.19. The question whether or not Easton was right cannot be answered by dismissing  the drawing out of hand. The reason is that such a rotation -- when isophotes at different surface brightness -- is often observed in elliptical galaxies. Surface photometry of NGC205 actually shows that it displays a pronounced example of these so-called `isophote twists’, in that case probably due to gravitational interaction with M31. 

\begin{figure}[t]
\begin{center}
\includegraphics[width=0.22\textwidth]{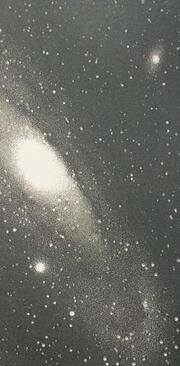}
\includegraphics[width=0.22\textwidth]{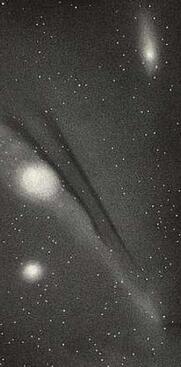}
\includegraphics[width=0.22\textwidth]{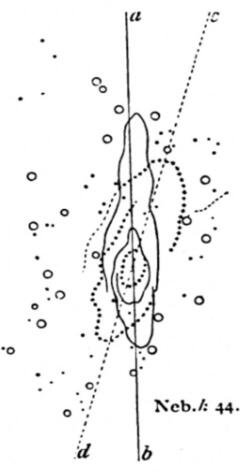}
\includegraphics[width=0.22\textwidth]{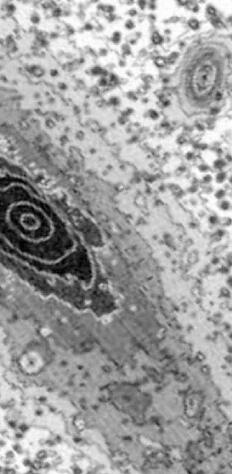}
\end{center}
\caption{\normalsize  Left: Photograph of the Andromeda Nebula (M31) and two of its companions taken in 1888 by I. Roberts with a 20-inch telescope. NGC205 (sometimes referred to as M110, but assigned to the Messier catalog posthumously) is near the top, and M32 projected on M31 (Roberts, 1888, 1889).  Second panel: Drawing after an 1874 photographic image obtained by Trouvelot (1881). Third panel: Drawing of Easton of NGC205 after Roberts (dotted lines) and  Trouvelot (drawn lines). Right: Isophote map of Walterbos \&\ Kennicutt (1987). Contour levels are at 19, 20, 21 B-mag arcsec$^{-2}$ and then in steps of 0.5 mag.}
\label{fig:}
\end{figure}

The Roberts’ photograph on the left goes significantly deeper than the Trouvelot drawing. This is most easily seen by noting that in the former the spiral arm in M31 is seen well curving around in the bottom-right corner, while this feature lacks in the second image, except for the bright part before the strong curving (due to the inclination) sets in. Now comparing this to the contour diagram with optical isophotes in  Fig. 5a of Walterbos \&\ Kennicutt, reproduced for the same part are in the right-hand panel in Fig.~19, shows that the limiting surface brightness in the Trouvelot drawing is about 22.0 to 22.5 J-magnitudes per square arcsecond (mag arcsec$^{-2}$), while for the Roberts photograph it is more like 24.0 to 24.5 in the same units. These are magnitudes defined by the Kodak \mbox{IIIa-J} emulsion and this is rather similar to the Johnson-B band (Kormendy \&\ Bahcall, 1974: van der Kruit, 1979). At a dark site the sky background surface brightness is of order 22.5 B-mag arcsec$^{-2}$, and this means that the galaxy limiting surface brightness as photographed by Roberts is of order a reasonable 20\%\ or so of sky. In this Walterbos \&\ Kennicutt figure the isophote twists in NGC205 can readily be seen.

The surface photometry study of NGC205 by Choi, Guhathakurta \&\ Johnston (2002) show in their Fig.~11 isophotes of NGC205 and the twists are well determined. In the inner part the orientation is close to north-south (fits show a position angle of 30\degs\ with respect the line joining NGC205 to M32, Fig.~12 of Choi {\it et al.}), which decreases to 10\degs\ at about 5 arcmin radius and a surface brightness of about 23.5 B-mag arcsec$^{-2}$. This is indeed in between the limiting surface brightnesses of Trouvelot and Roberts. The outline of NGC205 in the Roberts photograph lines indeed almost up with the line to M32. The angular distance between M32 and NGC205 is about 55 arcmin. The diameter  of NGC205 on the Roberts photograph can be seen to be of order 12 arcmin, consistent with the onset of the twists at a radius of 5 arcmin or so. The Trouvelot outline is some 15 arcmin, much larger than would be expected on the probable limiting surface brightness, and the position angle is more like 40\degs\ or slightly more with respect to the line {toward} M32. The drawing by Trouvelot should not be interpreted quantitatively.

The difference of position angle noted by Easton can on this basis be seen to be real and due to isophote twists in NGC205 as a result of gravitational interaction with M31. That he attributed it to a rotation of the nebula as a whole is not surprising; his deduction based on two pictures is entirely correct.

The {\it Nature} note  states that the drawing in Fig.~19  was `presented May 2, 1894  to the {\it Société Astronomique de France}’, but how is not specified. It is not obvious why Easton did not list this {\it Nature} paper in his  publication list, not even refers to it in the autobiographical text. Maybe he felt it was too short to count, or maybe he felt his speculation was too wild, or maybe he was not too sure about his inference of a rotated image. 

By way of concluding I note a surprising twist to this story (of isophote twists). Roberts (1888) had actually noted when presenting one of his photographs of Andromeda that the orientation of NGC205 had changed by over 25\degs\ since 1847. This was in comparison to a `Bond’s drawing’. This is likely to refer to William Cranch Bond (1789--1859) and/or his son George Phillips Bond (1825--1865). They were first and second director of Harvard Observatory. In 1847, the year quoted by Roberts, the first (15-inch reflecting) telescope was installed at Harvard. The drawing must be the result of visual observing with that telescope, since only in 1850 they were the first in the USA to photograph a star (Vega). Consequently it must be even less deep than the Trouvelot drawing. Roberts reported that in the drawing the major axis of NGC205 made an angle of 45\degs\ with the line {toward} the Andromeda Nebula. It is not clear where the drawing is published, but the fact that they observed the Andromeda Nebula was reported in a letter to Edward Everett (1794--1865), President of Harvard University, and published in the {\it Astronomische Nachrichten} (Bond \&\ Bond, 1850). It appears that Easton only read {(in the library of Leiden Observatory)} the Roberts (1894) publication and never {there} consulted the Roberts (1888, 1889) papers .

\section{The Milky Way seen from its North Pole}

Easton  returned to the spiral hypothesis in 1897; the first presentation related to it was when {Hendricus}  van de Sande Bakhuyzen presented during the ‘ordinary meeting’ of the Division Mathematics and Physics of the Royal Academy of Arts and Sciences on Friday December 24, 1897 on behalf of Easton a report ‘Over de groepeering van de sterren in den Melkweg’ (On the ordering of  stars in the Milky Way). In this he presented a long review of what had been published on this issue by the Herschels, Proctor, Celoria and others. Without further argumentation, so that it seemed to come completely out of the air, he concluded with the following (Easton, 1898a; p.393; my translation):
\begmarg
On the basis of these considerations one might consider a spiral structure of the galaxy probable — somewhat like that in which M. 74 Piscium and 51 Can. Venat. appear in Roberts' photographs — with a central condensation in the direction of $\gamma$ Cygni. The sun would then either form part of a secondary group of stars, or — which is perhaps more likely — be situated in a place where the condensation, {toward} Cygnus, of stars, as bright as our sun, or much brighter [...] is already strongly felt, while then the streams of much smaller stars or star-like bodies would wind themselves, through that cluster, in a spiral form around larger stars.
\endmarg

Tantalizing, but no real discussion. It was not very detailed and far from convincing. Easton next prepared two articles for {\it Knowledge}. The first appeared in January 1898 and this discussed mostly the background  (Easton, 1898b). Although entitled `Richard Proctor’s theory of the Universe', it started by describing William Herschel’s cloven disk and John Herschel’s galactic ring, before going into detail into Proctor’s theory (see section 6 and Fig.~13 above) and ending with the work of Celoria (1877). He wrote (Easton, 1898b; p.14):
\begmarg
Celoria conclude[d] that the Milky Way is composed of two branches, two distinct rings, uninterrupted circumference. One of these rings is represented by the continuous feature of the Milky Way, crossing the sky in Monoceros, Auriga, Sagitta, and Aquila; the other begins in the brilliant stars of Orion, passes through the Hyades, the Pleiades, Perseus, Cygnus, and ends in Ophiuchus. The two rings cross each other,  and are perhaps confounded in one system in the constellation of Cassiopeia; and separating, one part passing through Cygnus and the other through Perseus, they make an angle of about nineteen degrees.
\endmarg

As Easton noted, the outer ring is comparable to John Herschel’s model and the inner one (tilted by 19\degs\ or so to the Milky Way) is in fact what we now know as Gould’s Belt of bright, blue O- and B-stars.

\begin{figure}[t]
\begin{center}
\includegraphics[width=0.40\textwidth]{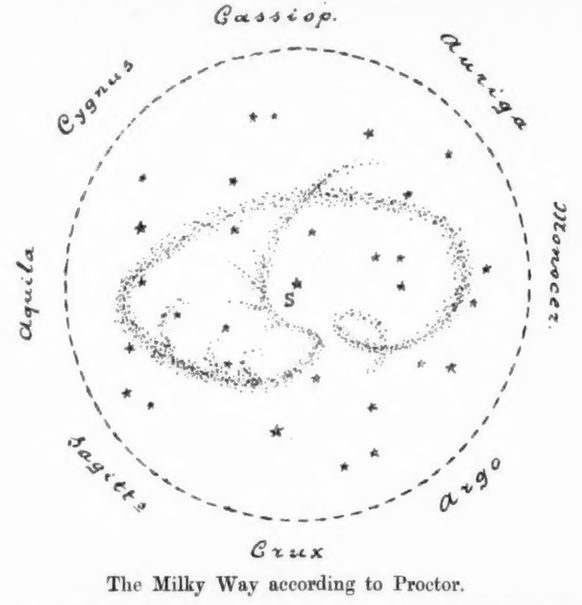}
\includegraphics[width=0.40\textwidth]{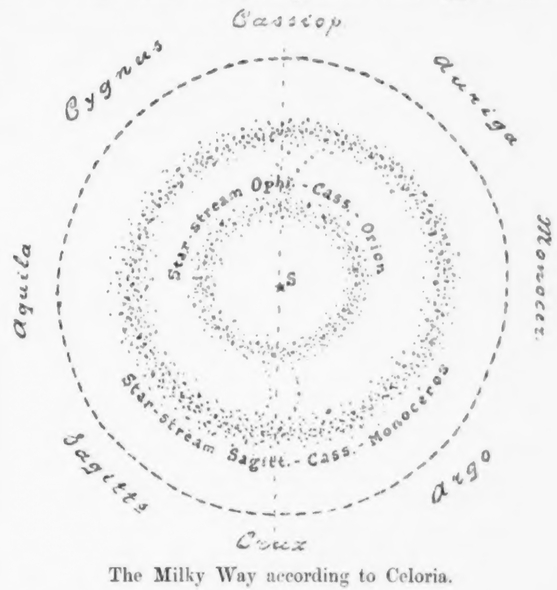}
\end{center}
\caption{\normalsize Sketches of the models of the Milky Way from Proctor and Celoria, according to Easton. The left-hand panel should be the same as Fig.~12, but has been mirrored and rotated for unspecified reasons. From Easton (1898a).}
\label{fig:}
\end{figure}

To illustrate this Easton added two drawings (Fig.~20). The one on the left is Proctor’s proposal (Fig.~13), but adapted. Firstly, he deleted the outer ring with the view of the Milky Way on the sky, and replaced it with written constellation names. But he had also mirrored the image! If we follow the constellations and compare to Fig.~8, we see that taking the clockwise order in the latter figure corresponds to going {\it anti-}clockwise in Fig.~20. This is opposite from Proctor’s original figure in Fig.~12; this has no identification of constellations, but even an astronomer with very modest acquaintance with constellations will easily recognize Cassiopeia at 9 o’clock, Cygnus at 11:30, Orion at 6:30, the Pointers (Centaurus) at 3:00 and Crux at 3:30, and maybe a few more. Easton does not refer to this adapted figure in the text and gives no explanation for presenting a mirror image. The second drawing on the right represents Celoria’s view, drawn by Easton since Celoria himself never produced such an representation. It follows the same convention.

In the Mach 1898 issue, Easton (1898c) expanded on his spiral hypothesis. In the {\it Notebook} (p.188) he explained (see Fig.~20 on the right):
\begmarg
But now this `modified Celoria system’ must be amended immediately on two points.

First the inner ring cannot be a continuous ring, second it must have points of contact with the outer one. The Milky Way drawings, which give an image of the whole, clearly indicate this. Furthermore, it appears from this that the worst gap in the outer ring, the weakest Milky Way spot, must lie in the direction of Perseus. If one then draws the streams and `bridges’ according to the Milky Way maps, one automatically gets the spiral, [...].
\endmarg
\bigskip

There are two things that Easton (1898b) added to this. The first is on the appearance of the Milky Way from its poles (p.59):
\begmarg
As an analogy from what we see in the heavens, I will take, not the nebula of Lyra, but rather the nebula Mess. 101 Urs Maj. [two references to photographs in the literature], or else the celebrated spiral nebula in Canes Venatici, Mess. 51 Can. Venat. [one reference].
\endmarg

And on the location of the center as seen from the Sun (p.60):
\begmarg
Besides, the part of the Milky Way in Cygnus is remarkable from more than one point of view. The luminous spot $\beta$-$\gamma$  Cygni [see Fig.~9] is the only luminous patch situated in the ‘secondary branch,’ but near the dark space. It is an exception to the manner of distribution of brightness over the breadth of the Milky Way, between the Eagle [Aquila]  and Cassiopeia. It is evidently connected with several other very brilliant regions (the spots $\alpha$-A, $\rho$-$\pi$ Cygni, etc., perhaps to the series of spots west of Altair [$\alpha$ Aquilae, off to he right of Fig.~9] ). There are in the Milky Way other more luminous spots, but they are much smaller. Sir William Herschel here found his maximum gauge (588 stars in a telescopic field of 15\minspt4). Not far from here, Kapteyn placed the {center} of the agglomeration of bright stars in the {neighborhood} of the sun. Without wishing to dogmatize, it is here that I would place the central condensation of a galactic spiral; the sun is thus found between this central nucleus and the spirals directed {toward} Monoceros, in a region relatively sparse. 
\endmarg
\bigskip

The definite presentation `A new theory of the Milky Way’ took place in the year 1900, just before the nineteenth century ended, in the {\it Astrophysical Journal}. His manuscript `La Voie Lactée dans l’espace, vue de son pôle boréal’ ended up in a probably much adapted form, in any case with a different title and in a different language, and in one of the most prestigious professional journals.

\begin{figure}[t]
\begin{center}
\includegraphics[width=0.40\textwidth]{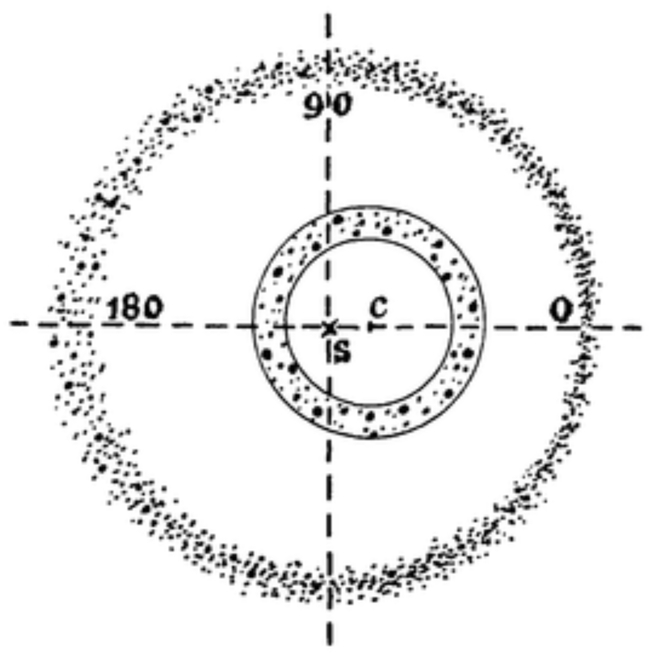}
\includegraphics[width=0.40\textwidth]{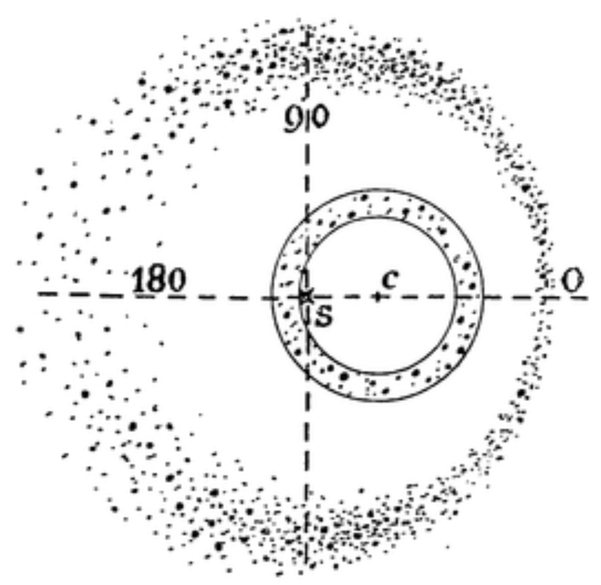}
\includegraphics[width=0.40\textwidth]{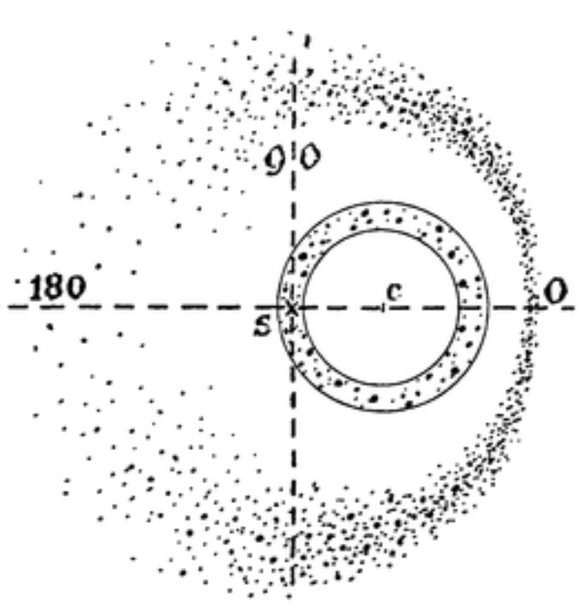}
\includegraphics[width=0.40\textwidth]{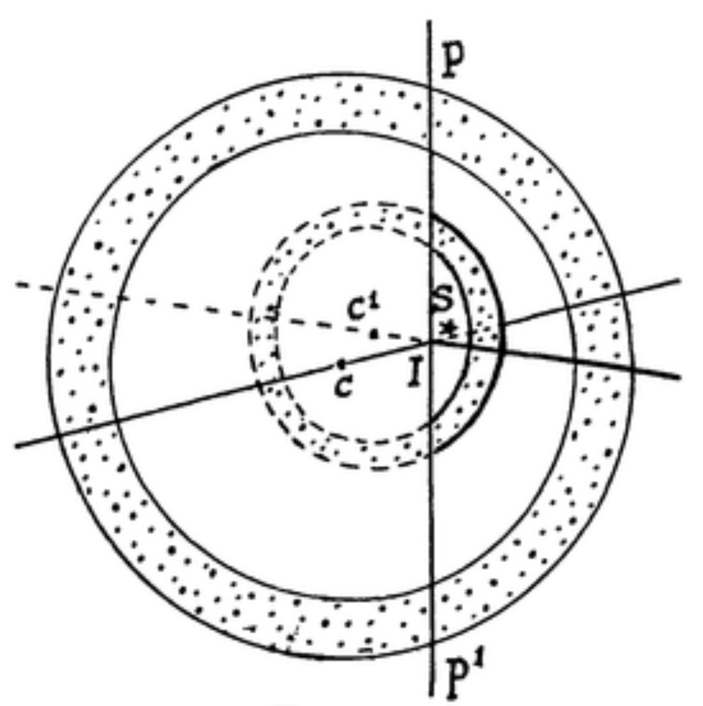}
\end{center}
\caption{\normalsize  Four possible configurations and positions of the Sun in the case of a system with two rings. Easton ruled these out as inconsistent with the general brightness and width of the Milky Way. From Easton (1900).}
\label{fig:}
\end{figure}

The paper starts referring to his earlier {\it Astrophysical Journal} paper (Easton, 1895c) which `seemed to indicate that it has a roughly annular structure’ and Easton concluded that this is no longer feasible in view of the large inequality of surface brightness in parts as Aquila and Monoceros. The view he then adopted is an inner ring surrounded by a distant stratum.  He then went into an involved discussion of what the position of the Sun would be. He distinguished five cases, illustrating three (see Fig.~21). The Sun is either in the center (position c in the top-left panel), inside but eccentric (position s in that panel), at the inner edge (top-right panel), inside (bottom-left panel) or at the outer edge of an inner ring.  He concluded that none of these was tenable, because inconsistent with the brightnesses and widths of the Milky Way. He then looked into the suggestion of {\it two} rings, as proposed in different forms by John Herschel and Celoria, and  (e.g, the bottom-right panel in Fig.~21) concluded that this may be more reasonable, and refers to it as a `provisional theory’. Proctor is dismissed, saying that  `Proctor’s `spiral', moreover, explains none of the principal features of the galactic phenomenon, although it led its author to make interesting remarks’ (p.142, in the footnote). The different planes introduced by Herschel and Celoria are dismissed with the statement (p.147): `Such an arrangement of the greater part of the stars in two planes, slightly inclined to each other, would appear hardly compatible with the idea of a purely fortuitous distribution of the stars in the galactic layer.’

\begin{figure}[t]
 \begin{center}
\includegraphics[width=0.24\textwidth]{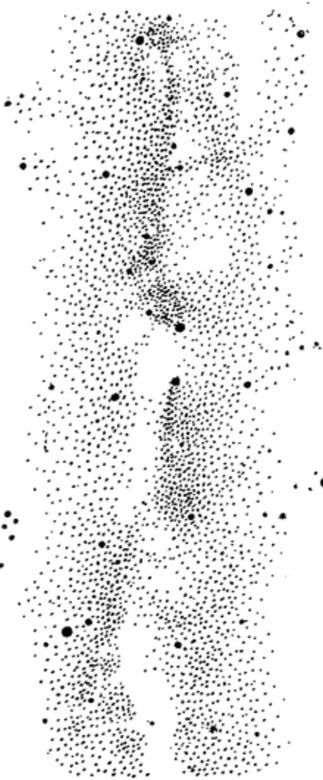}
\includegraphics[width=0.58\textwidth]{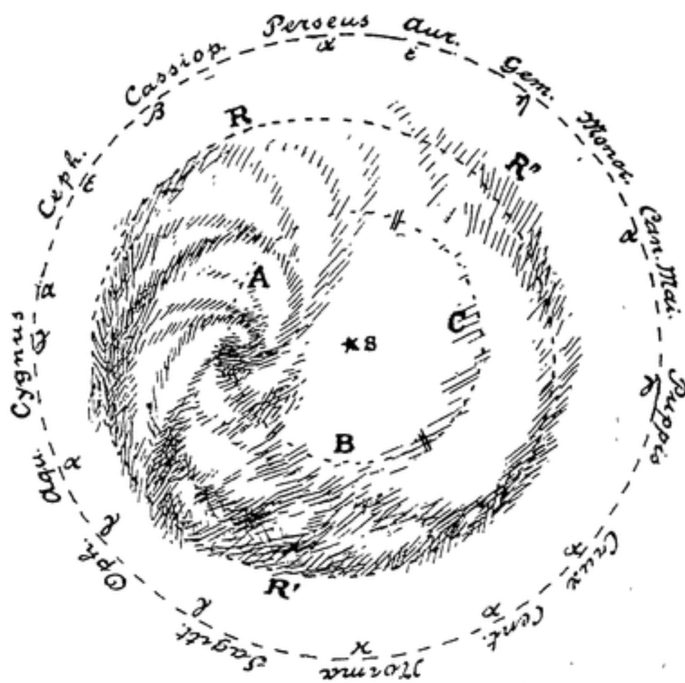}
 \end{center}
\caption{\normalsize Left: The Milky Way between $\gamma$ Ophiuchi (lower edge) and $\beta$ Cassiopeiae based on Easton’s \textit{Voie Lactée}. Right: Sketch of what the stellar system might look like, based on a variation of the two-ring model. The letters refer to considerations that have been quoted in the text. For more details see text. From Easton (1900).}
\label{fig:}
\end{figure}

Easton concluded from the appearance of the Milky Way and various features in it, that there must be much structure in the system, in the form of thin strata or streams of stars. Next he argued that the Cygnus area is exceptional, in view of its brightness. And then presents a sketch of the Milky Way System (see Fig.~22) with the following justification (Easton, 1900; p.155-156):
\begmarg 
In order to simplify the drawing I have left unbroken the exterior ring {\bf RR’R’’} (principal branch of the Milky Way) except the very faint part between {\bf R} and {\bf R’’} (Perseus). As for the interior ring, it must divide into at least three principal parts:

\noindent
{\bf A}, the bright part between $\gamma$  Ophiuchi and Cassiopeia, considered as an appendage of the principal ring [...]

\noindent
{\bf B}, the secondary branch in Serpens, Scorpius, Lupus; closely related rather to the principal branch in this region than to the secondary branch in Ophiuchus (north of $\gamma$) and Cygnus.

\noindent
{\bf C}, the belt of bright stars, projected upon a very faint nebulosity. Certain details between Aquila and Cassiopeia, the ‘luminous bridges’  which are projected upon the ‘rift’ between the two branches, etc., have been inserted from the galactic chart of this region ([left-hand panel]) --

The representation of the Milky Way thus obtained curiously resembles the spiral nebulae, of which Dr. Isaac Roberts has given such beautiful photographs.  To facilitate the comparison I have sketched in [the right-hand panel of Fig.~22]  the principal features  of the nebula M. 74  Piscium. It is unnecessary to remark that the distortion of the spiral in  is due to the preconceived idea of the two rings (in reality the cluster in Cygnus, and not the Sun, is at the center of the system).

From what precedes it follows, furthermore, that the convolutions of this `galactic spiral’ would not be situated in a single plane, but principally in two planes forming an angle of about 20\degs.
\endmarg

\begin{figure}[t]
\begin{center}
\includegraphics[width=0.385\textwidth]{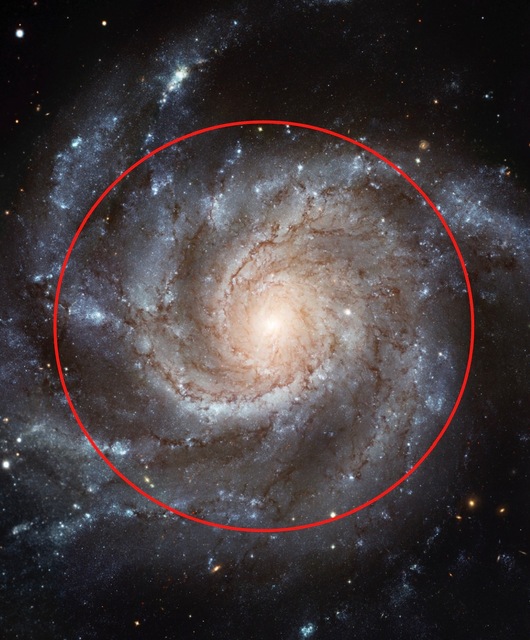}\includegraphics[width=0.46\textwidth]{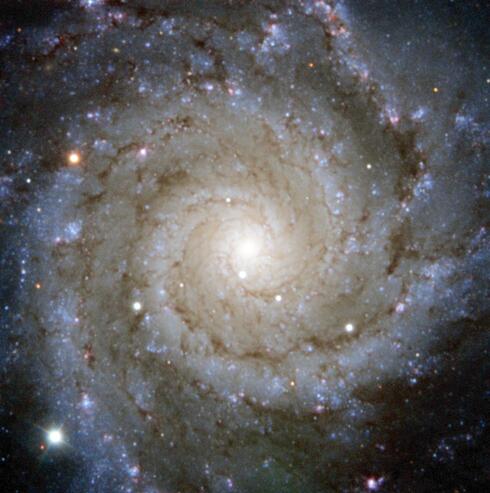}
\end{center}
\caption{\normalsize Modern photographs of two bright spiral galaxies Easton referred to in his papers on the subject  of the spiral structure of our Galaxy (the third one, M51, has been reproduced in Fig.~13). These are M101 (left, see text for meaning of the red circle),  taken with NASA/ESA Hubble Space Telescope, and NGC628 (also sometimes referred to as M74), taken with the New Technology Telescope at the ESO La Silla  Observatory. Credits: M101: commons.wiki\linebreak[4]
  media.org/wiki/File:M101\_hires\_STScI-PRC2006-10a.jpg; NGC628: commons.wikimedia.org/wi\linebreak[4]
  ki/File:PESSTO\_Snaps\_Supernova\_in\_Messier\_74.jpg lower-right SN2013ej. PESSTO stands for Public ESO Spectroscopic Survey for Transient Objects.}
\label{fig:}
\end{figure}

The spiral nebula M74 (among astronomers in the field of nearby galaxies known as NGC628) was sketched by Easton and added as a separate figure. I have included a picture of that galaxy in Fig.~23 together with M101, which Easton (1898b) referred to (see above). The third spiral he liked our Galaxy to, M51, has been illustrated in Fig.~13 above. NGC628 is well-chosen for a view straight from above; it is remarkable for being exceptionally close to face- on, namely seen at only about 5-7\degs \ (Shostak \&\ van der Kruit, 1984). It is clear now also that our Galaxy is of somewhat earlier Hubble type (Sb rather than the Sc’s chosen by Easton), and will in particular very probably have, seen from outside, somewhat more tightly-wound, less-developed spiral arms (van der Kruit, 1990). It is remarkable that Easton used a few bright spiral galaxies to illustrate what our Milky Way system would look like from the outside This implies that he believed that  these were outside the sidereal system our Sun belongs to and that  the nebulae had to be similar to our local system and of roughly equal size and at corresponding distances. He might have felt too unsure about this (he stressed repeatedly that his `new theory’ was no more than a working hypothesis and Kapteyn for example had not been convinced about it) to make too sweeping statements and weaken his case. It is remarkable that Easton chose NGC628 (M74) to illustrate in a drawing in his 1900-paper. It is a beautiful, very regular and symmetric two-armed spiral that is certainly not the usual spiral structure encountered in nature. The spiral displayed in Fig.~22 is far from this. His other example, M51, likewise is grand-design two-armed, a result of the interaction with the companion NGC5195. His choice in 1898b of M101 is interesting, since M101, even on the photographs of Roberts referred to, is rather disturbed, being in the outer parts clearly three-armed and much more resembling Fig.~22 , while in the inner parts (within the red circle in Fig.~23) M101 is two-armed and much more regular.

\section{Isophotes and nebulae}

\begmarg
In the summer of 1900, during vacation in the dunes of Scheveningen, I conceived a great plan: to investigate the distribution of all celestial objects in connection with my Milky Way maps: stars, star clusters, novae, peculiar spectra, etc. as a function of galactic longitude. The global distribution according to galactic latitude had been and is now being worked on sufficiently by others. […] but soon I understood that I had to reduce it considerably. First of all: compare the light distribution in the Milky Way with the distribution of the stars (i.e. relatively bright stars; there was still too little data for faint ones) as completely as possible, over the entire northern Milky Way -- an extension of my correlation investigations, partially, in Aquila, in Cygnus from 1894 and later.
\endmarg

In this passage from the {\it Notebook} (p.197) Easton described his research plans after publication of his Milky Way charts, correlation studies and spiral galaxy model. The distribution of various objects with longitude, he hoped,  might provide more information on the structure of the Galactic system. The first thing to do was turn his chart into an isophotal one. For this he took the area of the Milky Way from Orion to Aquila. 
\bigskip

Easton produced isophotes in two steps. First he determined from a new set of direct observations the large-scale isophote  shapes and then filled in the details using his maps in the {\it Voie Lactée}. He used six levels of brightness, from $a$ to $f$, but of course the problem was how to calibrate them, in terms of differences in contour level. An absolute calibration was not necessary for his purpose of comparisons, but these was also no way for Easton how to attempt this in the first place. For the relative calibration he reviewed what before him had been done and decided it was not reliable. So he invented his own method (Easton, 1903a, p.11):
\begmarg
The following assumptions were necessary: {\it a}, that the Weber-Fechner law applies to this case, where it is a question of comparing luminous surfaces just at the limit of visibility; {\it b}, that the ratio between the sums of actinic and visual light is the same; {\it c}, that all the stars visible in the photograph participate in the formation of the galactic image, and that stars too faint to be separately visible on the plate have no appreciable influence, nor does the nebulae that may be found in these regions; {\it d} , that no appreciable influence is exerted by photographic images of bright stars large enough to mask small stars.

The objections that can be raised against assumption {\it b} are the most serious. Indeed, the rich regions of the Milky Way probably contain an excessive proportion of stars whose rays are very actinic. And as for {\it c}, we certainly do not have the right to admit that the sensitivity limit of our retina coincides exactly with that of any photographic plate. […] during these investigations, it was found that [...]  bright stars do not play an important role in the formation of total light, and -- as regards assumptions {\it c} and {\it d}  -- that the share of the faintest stars ($\lt$13.5) is relatively small [...]
\endmarg
`Actinic’ here means photographic, referring to the actinic effect light has on a silver-based photographic emulsion.

The steps between the isophote levels $a$ to $f$ are supposed to be equal, which according to the Weber-Fechner `law’ means logarithmic. The first part named after Ernst Heinrich Weber (1795-1878) states that the perception of an increase of a stimulus is proportional to that stimulus, and the second part by Gustav Theodor Fechner (1801–1887) that -- as an extension -- the sensation of a perception is proportional to the logarithm of the stimulus. Thus Easton’s six levels each subsequently differed by a factor $d$ and he needed to determine the value of $d$. His condition {\it b}  is indeed the most questionable, since it assumes in fact that the stellar content of the part of the line-of-sight where individual stars can by distinguished is proportional to that part that presents itself to us as diffuse background.

Now Max Wolf had put at Easton’s  disposal a photographic plate of the vicinity of $\gamma$ Cygni, and Easton chose two rectangular area’s, of equal size and at approximately the same distance from the center of the plate; one very bright (level $f$) and one relatively faint (level $b$) and determined of all stars their brightness to half a magnitude and added up to their total collective brightness. This way he determined $\log d = 0.1371$, so the spacing of the isophotes would be 0.33 magnitudes.

\begin{figure}[t]
\begin{center}
  \includegraphics[width=0.93\textwidth]{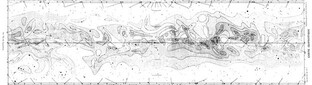}
  
\includegraphics[width=0.147\textwidth]{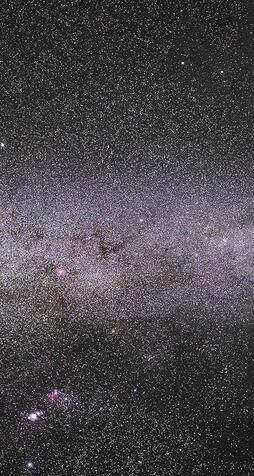}\includegraphics[width=0.73\textwidth]{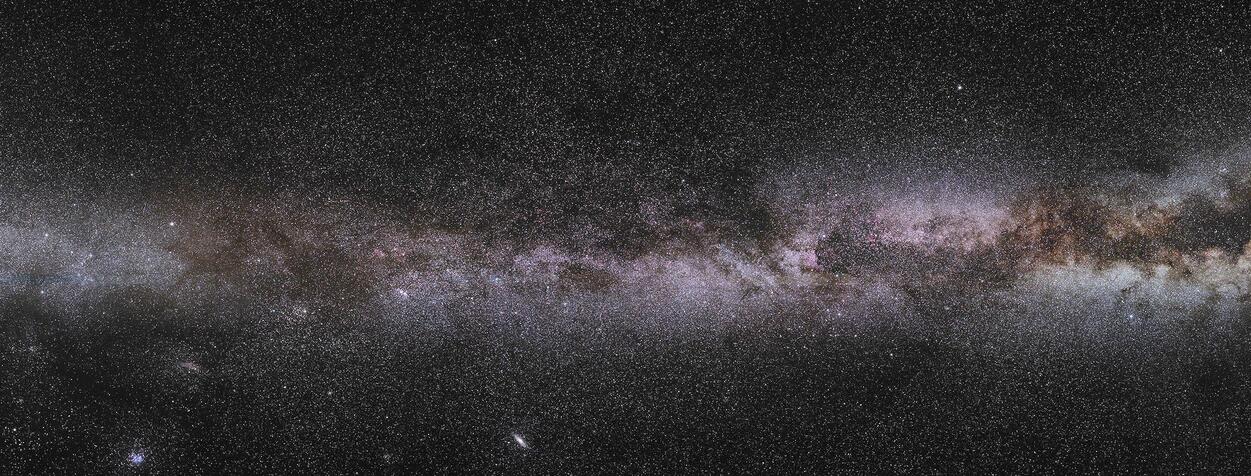}
\end{center}
\caption{\normalsize Top: Isophote map of the Milky Way between Orion and Aquila, as published in Easton (1903a). Bottom: The corresponding area of the ESO Panorama, where now the parts at either side of longitude 180\degs\ (lower-right and upper-left in Fig~8 have been joined (the joint can be discerned on the left). Credit: ESO/S. Brunier..}
\label{fig:}
\end{figure}

Easton’s surface brightness levels were five steps, together 1.65 magnitudes, which corresponds to a factor of about 5 when expressed in linear units. The scale itself is not unreasonable when compared for example to Pannekoek’s photographic isophotic chart of the northern Milky Way (Pannekoek, 1933; see Fig.~28 in van der Kruit, 2024a). The difference between the surface brightness in the area between $\beta$ and $\gamma$ Cygni  and the dark area near $\alpha$ Cygni is about 5 contour steps in Easton’ map (1.3 magnitudes) and a factor about four or so (1.5 magnitudes) in Pannekoek’s chart.  So the comparison is rather good. Of course, this conclusion has already been reached by Pannekoek, when he made the comparison (Pannekoek, 1920, see also van der Kruit, 2024a)

The resulting chart is in Fig.~24 (Easton, 1903a), compared with the same area from the ESO Panorama of the Milky Way. A careful inspection reveals that many details drawn by Easton can be identified also on the photographic panorama.
\bigskip

The first thing to compare the chart to would be stars in the {\it Bonner Durchmusterung}. Hendricus van de Sande Bakhuyzen, chairman of the Section Physics of the Royal Academy of Arts and a Sciences,  insisted on a ‘communication’ to the Academy, but when it grew to too large a manuscript he  suggested that the Academy should publish it as a `Verhandeling' (treatise). That was a special publication and to be asked or allowed to publish such a contribution was an honor. But then it had to be written in French, van de Sande Bakhuyzen felt.

Easton divided stars in the {\it Bonner Durchmusterung} into four groups: group I embracing stars of magnitude 0-6.5; group II, 6.6-8.0; group III, 8.1-9.0; group IV, 9.1-9.5. Then he divided the area in the chart, which runs from (old) Galactic longitude 0\degs\ to 180\degs, latitude -18\degs\ to +18\degs\ in 15\degs$\times$15\degs\ rectangles and classified these as follows: {\it A}: the density decreases gradually from group I to group IV ({\it A?}: the decrease is approximately gradual); {\it B}: the density increases gradually from I to IV ({\it B?}: almost gradual). Of the 108 rectangles only 61 could be classified: 17 {\it A}, 19 {\it A?},   8 {\it B} and 17 {\it B?}. 

{Having} come this far, Easton felt he should have advice from Kapteyn. Easton’s acquaintance with Kapteyn after they first met in 1893 in Dordrecht had developed well. Easton had published an article in the {\it Nieuwe Rotterdamsche Courant} in October 1900 on Kapteyn’s laboratory, and mentioned that institution in {\it Knowledge} articles. Kapteyn had reciprocated with the gift of the {\it Cape Photographic Durchmusterung} and had asked him to visit Groningen. Easton made use of this invitation during a three-day visit to Groningen in the summer of 1901, followed by a visit by Kapteyn to the Eastons  one evening in Rotterdam, where they now lived, in November of that year. Kapteyn's advice has not been recorded.

Easton next supposed  that in  a region which has only a few relatively faint stars, but is very rich in bright stars, we are seeing a part of the Milky Way that is closer to us, and when the number of bright and relatively bright stars is gradually diminishing and at the same time that of fairly faint and very faint stars increasing, it would be a region of the Milky Way which is further away. Easton concluded that the main condensed region (Cassiopeia-Aquila) of the northern Milky Way consists mainly of two stellar layers, situated at different distances. The {Galactic} region near $\gamma$ Cygni forms the core of an enormous stellar agglomeration. However, he did not discuss in detail how this fits into his spiral theory and no coherent, convincing picture emerged. 

Kapteyn and   Ernst Frederik van de Sande Bakhuyzen (1848--1918; the younger brother of the director of Leiden Observatory)  reported on the article at the February 1902 meeting of the Academy {of Arts and Sciences}. After much correction, etc. – which took the full year 1902,  it finally appeared in January 1903 in printed form (Easton, 1903a) . Easton `received the treatise, bundled in 50 copies (of which 25 at my own expense at 40 cts.) at home for distribution’ ({\it Notebook}, p.200).

Easton  presented this study also in {\it Astronomische Nachrichten} (Easton, 1902c). And he published three articles in {\it Knowledge}. In the first two -- a two part description of past and ongoing research on the Milky Way -- with the title {\it Distant Worlds} (Easton, 1902a,b) --, he presented a version of his isophotal contour map, the part from Cassiopeia to Aquila in the second paper. And in Easton (1903b), {\it What is the Milky Way?}, he published the map of the Cygnus region covered by the photographic plate Max Wolf put at his disposal, with 1760 stars  indicated at the measured position with different  symbols indicating apparent magnitude. In the {\it Notebook} he stressed that publishing in {\it Knowledge}  was an important means to publish important figures to complement his papers in the {\it Astronomische Nachrichten} that  could not include illustrations.
\bigskip

The matter of the distribution of various celestial objects relative to the Milky Way was still on Easton’s mind and he next turned to nebulae. Nebulae had been cataloged visually, but had increasingly been subject of study by photography. However, Easton was interested in their distribution and {catalogs} were his primary sources of information. After the first {catalog} by Charles Messier (1730--1817), the most extensive inventory had been started by William Herschel, and continued by his sister Caroline and son John, leading up the the {\it General Catalogue of  nebulae and clusters of stars} published by John Herschel, and the final form, the {\it New General Catalogue of nebulae and clusters of stars}, the {\it NGC}, by Dreyer (1888). John Louis Emil Dreyer (1852--1926), was a Danish astronomer who spent most of his career working in Ireland.

Now the distribution with respect to the Milky Way had been studied first by Cleveland Abbe (1838--1916), an American astronomer and prominent meteorologist, and Julius Bauschinger (1860--1934), for most of his career  director of the Astronomisches Rechen-Institut in Berlin. Abbe (1867) used the Herschel precursor of the {\it NGC}, and Bauschinger (1889) the {\it NGC} itself (the paper is written as a book review of the {catalog}), to come to some general conclusions that I quote here in my translation from the Bauschinger paper (1889, p.51.):
\begmarg
 1. The faint nebulae avoid the Milky Way; the largest concentrations are found near the poles of the Milky Way; from these poles the number of nebulae decreases the nearer one comes to the Milky Way. In addition, independent concentrations are found in the southern sky in the Cape clouds, and in the northern sky in the constellation of Andromeda. 2. The bright nebulae show exactly the same behavior as the faint ones, which proves that the general brightness of the Milky Way is not the sole reason for the characterized distribution. 3. The planetary nebulae, with very few exceptions, lie in or near the Milky Way. 4. The star clusters, with the exception of isolated objects and the region of the two Cape clouds, are all in the Milky Way or near it.
\endmarg
It is this that Easton followed up, especially the first two points.

 Now W. Stratonoff of {Tashkent} Observatory in Uzbekistan  published an extensive study of the distribution of stars, nebulae and star clusters (Stratonoff, 1900). It contained twenty-three maps with numerical and contour representations of the distributions on (mostly) both hemisphere of stars of various apparent magnitudes, spectral type, and of star clusters, but particularly of nebulae in general, bright and faint or small and extended.  For the maps Stratonoff had used the Dreyer (1888) {\it NGC}  and its (first)  {\it IC}  extension (Dreyer, 1895), combined with a large number of visual and photographic studies of nebulae,  ending up with  9264 nebulae (divided into 1345 bright ones and 7919 faint ones, or 1723 extended ones and 7541 small ones).

 It should be remarked here that of the 7840 objects in the {\it NGC} and 5386 objects in the two {\it IC} {catalogs} (the second one was published later as Dreyer, 1910) are now known to be galaxies for respectively 77\%\ and 74\%\ (see SEDS, 2025); this in contrast to the 37\%\ for the Messier {catalog}. And away from the the Milky Way and what was later dubbed the Zone of Avoidance (for galaxies) almost all objects are galaxies (a very nice graphical illustration of this has been presented by Lowe \&\ North, 2025). Close to the Galactic north-pole there is at least for brighter galaxies in the {\it NGC} a definite concentration in the form of the Virgo Cluster.

It is this material of Stratonoff that Easton used to study the distribution of nebulae with respect to the Milky Way. He concluded that there is no clear accumulation of nebulae to the Galactic south-pole. He offers no reason for this asymmetry. Easton regarded the faint nebulae part of the Sidereal System and attributed the relative lack at low latitudes to them being more difficult to isolate in the bright background of the Milky Way. The most remarkable result was obtained when he considered all nebulae with a latitude (positive or negative) less than 50\degs. He divided this up in 24 ranges of longitude. Then Easton claims significant maxima around longitudes 100\degs\ and 280\degs. This he considers a result of the spiral structure theory he had proposed and support for his idea. 

This was published in the {\it Astronomische Nachrichten} (Easton, 1904c). He had treated all of this in much more detail in two contributions to the Royal Academy {of Arts and Sciences}, delivered by Hendricus van de Sande Bakhuyzen before the {\it A.N.} paper (Easton, 1904a,b). He presented his longitude distributions in graphical form. The tabular data do not show anything very significant, except for a peak in the area the Large Magellanic Cloud. Easton’s supposed maxima are at the constellations Perseus and Norma. The data are presented graphically in Fig.~9. of Easton (1904b); allowing for the peak in the LMC area, however there is not much in terms of a systematic and significant nature. Easton had over-interpreted his data.

\section{Doctor ‘honoris causa’}

In Fig.~25 I show Easton on an undated photograph. This picture has been provided by Kristian Vlaardingerbroek from the material still available to his mother Titia, great-granddaughter of Easton. There is a different, but somewhat  similar to the photograph of Easton that I used in my article about Kapteyn’s honorary doctors (van der Kruit, 2022), but that seems a slightly younger Easton. That other picture came from the  {\it Album Amicorum}, presented to H.G. van de Sande Bakhuyzen at his retirement in 1908 (Leiden Observatory, 1908).

\begin{figure}[t]
\sidecaption[t]
\includegraphics[width=0.62\textwidth]{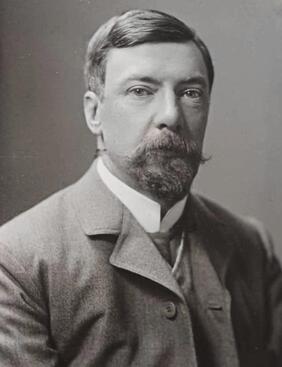}
\caption{\normalsize Easton some time after 1900. This picture has been made available by his great-granddaughter Titia Vlaardingerbroek-de Knegt through her son Kristian. It has been produced by Atelier C.E. Mögle at Rotterdam. This is the studio of German-born photographer Carl Emile Mögle (1857--1934), who was considered one of the best artistic photographers of his days.}
\label{fig:}
\end{figure}

In 1903 Easton was awarded an honorary doctorate by the University of Groningen with Kapteyn as honorary `promotor’. This was a very special honor, not often offered by the university, and also an expression of the extraordinary position Kapteyn held already at that time in Groningen. 

Jacobus Kapteyn had studied mathematics, physics and astronomy in Utrecht, where he had in 1975 obtained a PhD with a thesis in the area of applied mathematics. Since that he had been employed at the Sterrewacht Leiden, where he was an `observator’. He next had been appointed in 1878 in Groningen in the position of  professor of astronomy, statistics and theoretical mechanics. This appointment was a direct consequence of the Higher Education Act, which had been passed the year before.  As a result, each of the three government-funded universities in Leiden, Groningen and Utrecht would have its own professor of astronomy. The  number of professors, as well as the total budget for the universities roughly doubled. While Leiden and Utrecht had observatories to support the astronomy professor,  in Groningen there were no facilities whatsoever.

Against these enormous odds Kapteyn had succeeded to build up a world reputation, by establishing an `Astronomical Laboratory’, where plates taken elsewhere where measured and analyzed. He prepared there the first comprehensive {catalog} of stars in the southern hemisphere with the same depth and quality as the {\it Bonner Durchmusterung} in the north. He had accomplished this by measuring plates of the southern skies by David Gill (1843--1914), director of the Cape Royal Observatory in South-Africa. This resulted in the {\it Cape Photographic Durchmusterung} (Gill \&\ Kapteyn, 1897, 1897, 1900) of some 450,000 stars. This work had given Kapteyn a very high international reputation, and had made him the most prominent astronomer of the Netherlands by far, and probably the most prominent professor  at the University of Groningen. In 1902 Kapteyn was awarded the Gold Medal of the Royal Astronomical Society of the UK, the highest honor in astronomy.

This brief outline of Kapteyn’s career up to the end of the nineteenth century serves as explanation why when the 25-th anniversary of his appointment as professor in Groningen in 1903 was approaching, the university agreed to honor his exemplary status with an honorary doctorate for someone of Kapteyn’s choice. And this would be Cornelis Easton.  The University of Groningen had been very modest in bestowing honorary doctorates. During Kapteyn’s first 25 years in Groningen, only five had been awarded. The first, in 1884, was an internationally renowned physician and microbiologist Heinrich Hermann Robert Koch (1843--1910), one of the founders of bacteriology and the concept of infectious disease. He would receive the Nobel Prize in 1905 for identifying the bacterium that caused tuberculosis. But the other four where of the ‘corrective’ type,  to Dutch persons that had important scientific  contributions on their record, but who for some reason had never been in a position to present a PhD thesis. An example is Jacobus Laurens Sirks (1837--1905), who had studied language and physics and had become teacher at gymnasia. He had obtained a doctorate in arts, but while a teacher did important work in physics. He received an honorable doctorate in 1892 in mathematics and physics for work on interference microscopy.

Easton must be seen in this category as well, a person that would not qualify for submitting a thesis, but whose contributions nevertheless were significant.  For more on this see van der Kruit (2021c), including the Kapteyn honorary doctorates of Karl Schwarzschild (1873--1916) and Andrew Carnegie (1836--1929) in 1914, and of Annie Jump Cannon (1863--1941) in 1922. From this it follows that the honorary doctorate awarded to Easton fitted in an existing tradition of rare occasions and in recognition of important and significant research. 
\bigskip

Kapteyn had shown very positive appreciation of Easton’s work on the structure of the Milky Way on the sky and the publication of his {\it Voie Lactée}, but had been more reserved about the suggestion of spiral structure. Easton had been very impressed by Kapteyn and admired and worshiped him. It is in the context the honorary promotion worthwhile to consider the relationship between the two men. From Easton’s side we have the obituary he wrote in {\it Hemel \&\ Dampkring} shortly after Kapteyn’s death. At that time Easton was chairman of the Board of the amateur  society that published the periodical and chair of its editorial board. It is logical that Easton wrote this obituary, whose title in English was `Personal memories of J.C. Kapteyn’ (Easton, 1922a). An English translation of parts is available as Appendix C. in my Kapteyn biography (van der Kruit, 2015), and a full translation with annotations on the accompanying Website.

The `dedication’ to this article is remarkable, where Easton compared himself to the presenter of recollections of two other famous men: 
\begmarg
 The privilege to be the Eckermann of this Goethe! The Forster of this Dickens! ...
\endmarg

Johann Peter Eckermann (1792--1854) served for a number of years as Goethe’s personal secretary and published after Goethe’s death his conversations with him in the form of a book {\it Conversations with Goethe.}  He was also involved in preparing his posthumous works for publication. John Forster (1812--1876) is well-known for his biography {\it Life of Charles Dickens},  with whom he had been well acquainted. He is reported to have been in the possession of the originals of Dickens’ novels, which he later bequeathed to the South Kensington Museum.

In this obituary Easton described their first meeting as having taken place in 1894, when Kapteyn served as external examiner of final exams in Dordrecht secondary schools. Such exams were organized nation-wide by appointing ‘gecommiteerden’ (to be understood as delegated person) as independent guardians of the quality, correct procedure and fairness of examinations by the school teachers. Easton at that time was editor of the {\it Dordrechtsche Courant}. 
\begmarg
... and that same evening he paid me a return visit at St. Jorisweg 41. It was a warm evening and we sat with a glass of wine and cigars talking about the stars until about eleven o’clock, rambling on for a long time in the depths of the universe, doing my utmost to keep up with the astronomer. I found it enjoyable, wonderful and honoring.
\endmarg

Easton and Kapteyn from then on kept up a close relationship, leading up to a friendship, that resulted in them keeping in touch on various matters, particularly Easton's reseach activities and they visited each other every now and then. Kapteyn must have had much admiration for Easton's work (particularly his mapping of the Milky Way), so the {choice} for Easton when his jubilee of 25-years professorship approached seemed natural. However, it appears that a role in this was also played by an unexpected person, one of Easton's colleagues at the newspaper he worked for at the time, \textit{Nieuwe Rotterdamsche Courant}. This was Gerardus Pekelharing, (1857--1924), who would later become director of the \textit{Bijbank} (secondary branch) of the \textit{Nederlandsche Bank} (the state-owned national bank) in Rotterdam. He had read some tributes in the astronomical literature to Easton and was impressed. From the {\it Notebook} (p.202):
\begmarg
When Kapteyn [came] to stay with us in the Kortenaerstraat in the first days of 1903, the ‘Pekels’ asked him to dine with us, and in those days it happened that, entering our miniature drawing room, I surprised Kapteyn and Pekelharing in mysterious conversation, and remarked that they had a special interest in the tablecloth. Apparently {\it I} had been the subject of their conversation. But the connection between these conversations, and what Kapteyn intended to do in Groningen, and his request for my `iron universe', escaped me at that time.
\endmarg
The `iron universe' concerns the following ({\it Notebook}, p.201): 
\begmarg Just before my treatise came off the press [this is Easton {(1903)],} Kapteyn stayed with us for a few days. [...] When he left, Kapteyn asked casually whether I could not make a model of that spiral of mine out of iron wire? I made attempts for such an `iron universe', but I did not succeed, my dexterity for such things is too small. However, the question had a special purpose -- which I did not suspect at the time.
\endmarg
Pekelharing had also played a role in organizing the manner in which Easton was notified about the degree. In the summer of 1903, Easton's wife and children had already gone to Scheveningen for the summer vacation, with Easton set to come later, temporarily staying with the Pekelharings. On Friday before Easton would join them in Scheveningen (June 7, 1903) Easton worked at the newspaper and opened some telegrams. One of these read ({\it Notebook}, p.203)
\begmarg
The Senate has decided to grant you a doctorate in mathematics and astronomy. Promotion Saturday, the 13th of this month. Letter to follow -- Kapteyn.

\noindent
[...]
With the `Pekels' -- who knew about it and had the timing set for the eve of my vacation -- great joy and cordiality. After a much-desired cigar I was escorted in a procession to my bedroom with candles, as a rehearsal for the upcoming promotion procession.
\endmarg
\bigskip

\begin{figure}[t]
\sidecaption[t]
\includegraphics[width=0.62\textwidth]{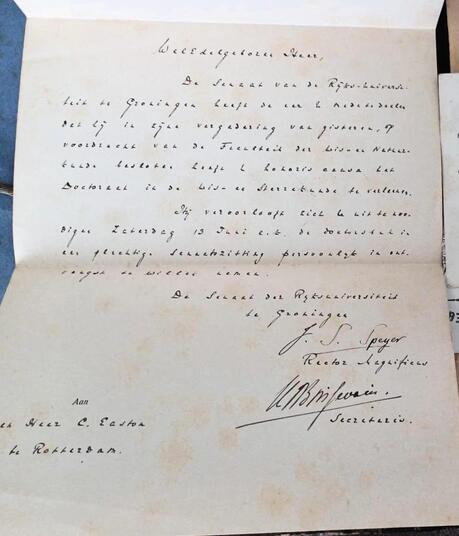}
\caption{\normalsize Letter by the Rector Magnificus of the University of Groningen to notify Easton of the decision to bestow an honorary doctorate upon him. From the private records of Titia Vlaardingerbroek-de Knegt, great-granddaughter of Easton through her son Kristian.}
\label{fig:}
\end{figure}

The ceremony took place as Kapteyn had written on Saturday June 13, 1903. However, in the {\it Archives} a further telegram is present, dated June 12, sent from Groningen at 8:21 received 8:50. It reads in my translation:
\begmarg
\noindent
Easton \ \  Scheveningen \ \  Gevers Deynoutweg 87

\noindent
Letter apparently lost --   Come tonight with your wife to stay with us  -- bring tails -- no paranymphs -- You respond at ceremony with a few lines --  Reply by telegram -- Kapteyn
\endmarg

At a promotion ceremony the candidate (the 'doctorandus') is dressed in tails (at least males; females sometimes too, but usually they wear a long, dark  dress) and in regular cases is accompanied by two paranymphs (assistants).

The lost letter would {be} notification, sent to Easton, by the Rector Magnificus that of the University of Groningen. It apparently turned up again later and is present in the papers of great-granddaughter Titia Vlaardingerbroek-de Knegt.(see Fig.~26). The text of this hand-written document reads:
\begmarg
{Honorable} Esquire,

The Senate of the University of Groningen has the honor of notifying you that the Faculty of Mathematics and Physics will bestow upon you {\it honoris causa} the title of Doctor of Mathematics and Astronomy.

He takes the liberty to invite you to accept in person the Doctor’s diploma this coming Saturday June 13.
\endmarg
It was signed by the Rector Magnificus, Jacob Samuel Speyer (1849--1913), who was professor of Latin in Groningen -- but was that same year appointed professor of Sanskrit in Leiden -- and the secretary Ursul Philip Boissevain (1855--1930), professor of ancient history.

\begin{figure}[t]
\sidecaption[t]
\includegraphics[width=0.62\textwidth]{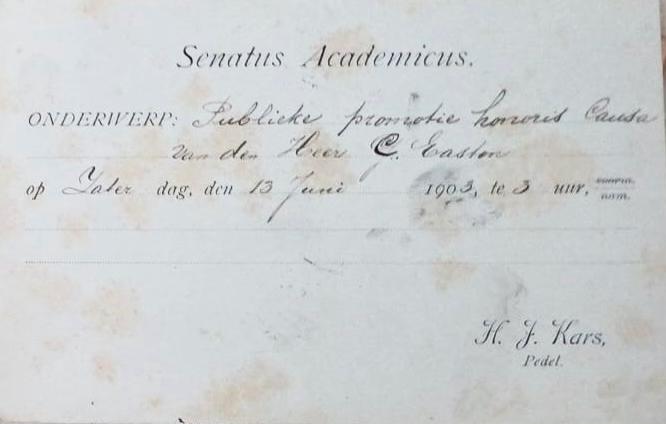}
\caption{\normalsize   Invitation to the public ceremony of Easton’s promotion to honorary doctor in Groningen. From the private records of Titia Vlaardingerbroek-de Knegt, great-granddaughter of Easton and her son Kristian.}
\label{fig:}
\end{figure}

There also is a card with invitation to the ceremony (see Fig.~27), signed by the beadle, who organizes and oversees the proceedings.

The {\it laudatio} Kapteyn read was only published much later in a special issue of {\it Hemel \&\ Dampkring} that appeared on June 13, 1928, celebrating the 25-th anniversary of the bestowing of the honorary degree. Easton pasted a copy of that issue in his {\it Notebook} and {my} English version has an English translation. It expresses much praise for Easton's work, especially the {\it Voie Lact\'ee} ({\it Notebook}, p.235):
\begmarg
And imbued with the modern scientific sense, you began to collect facts. For five years we see you busy, in the hours that your daily work left you alone, observing and drawing, estimating and comparing, until you had arrived at as complete a picture of the Milky Way as you were able to make.  -- And with that the work was not yet finished. The possibilities of reproduction appeared to be greater than one could expect. Characteristic of the way in which you see it your duty is that when the various attempts at the reproduction of the vague, lesser brightness levels did not satisfy you, you yourself did not shy away from the trouble of acquiring the necessary skills to transform the images onto stone yourself and later to retouch each of the separate prints with care. --  Thus was born the {\it La Voie Lactée dans l'Hémisphère Boréal}, a work that, next to the almost identically published work of Boeddicker, now forms the main source of our knowledge of the appearance of the Milky Way in the Northern Hemisphere, ...
\endmarg
Not surprisingly Kapteyn did express reservations concerning Easton's spiral theory ({\it Notebook}, p.237):
\begmarg
And based on this and many other similar data, we see you finally constructing the map of the real Milky Way on the basis of your charts, an image that can be modeled. Does that model correspond to reality? Who can say for the time being? Your spirals may be tempting... You will not expect me to declare that we accept your conclusion unconditionally. You yourself do not do that either. But I think I may say that anyone who will attempt another solution will have to take into account the data you have collected. -- That no solution can be considered conclusive if it does not account for the relationships you have found.
\endmarg

\section{Wrapping up astronomical research}

\begin{figure}[t]
\sidecaption[t]
\includegraphics[width=0.48\textwidth]{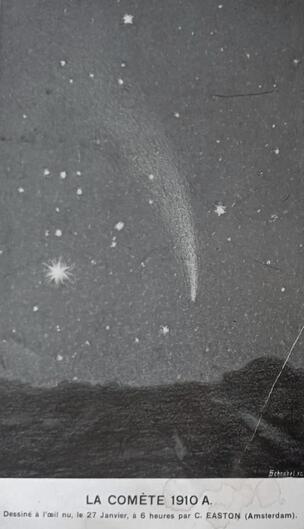}
\caption{\normalsize  Drawing of comet 1910A by Easton. This picture has been supplied by Kristian Vlaardingerbroek and his mother Titia, great-granddaughter of Easton. It is a cliché reproduction (see text underneath). The logo at the lower right is that of the \textit{Dirk Schnabel Photo Chemigrafische Kunstinrichting}, a cliché factory in Amsterdam. Director Dirk Gerardus Andries Schnabel (1870–1933) was a draftsman and photographer. He was born in Dordrecht, and set up the factory in 1900 in Amsterdam.}
\label{fig:}
\end{figure}

After being awarded the honorary doctorate, Easton's astronomical work came more or less to a standstill. As a result of the doctorate, which made him better known among a wider audience, he started giving lectures on astronomical subjects and writing  popularizing articles. One thing remained and that is that he tells us that he has observed and made drawings of every major comet that appeared during his lifetime. Fig.~28 concerns comet 1910A. This very bright comet is also known as the {\it Daylight Comet}. It has an orbital eccentricity of 0.99978 (Kronk, 2007), and will return only after about 9,000 years.

Easton’s research in climatology had started with two presentations in 1904 for the Royal Academy {of Arts and Sciences} on oscillations of the solar activity and the climate. It could have arisen in conversations with Kapteyn, but Easton does not record that. Kapteyn  had in the first years after his appointment in Groningen (which was in 1878) also looked into periodicities in the weather. His motivation had been to search for patterns in the weather such as were known for solar and lunar eclipses (as the Saron), and he had used disks (slices) cut from old trees for this to study the patterns of growth rings. Correlation with records that classify summers as hot or cool, and winters as cold and mild, did not produce useful insight, but he did find a large degree of `parallelism’ between the measurements of ring thickness and amount of rainfall, and he noted a large degree of regularity with cycles of a well-defined period of 12.4 years (see for more on this van der Kruit, 2015, 2021a).  Easton refers to these studies in his extensive obituary of Kapteyn (Easton, 1922a), but does not add any reference to his own work in the area.

\begin{figure}[t]
\sidecaption[t]
\includegraphics[width=0.62\textwidth]{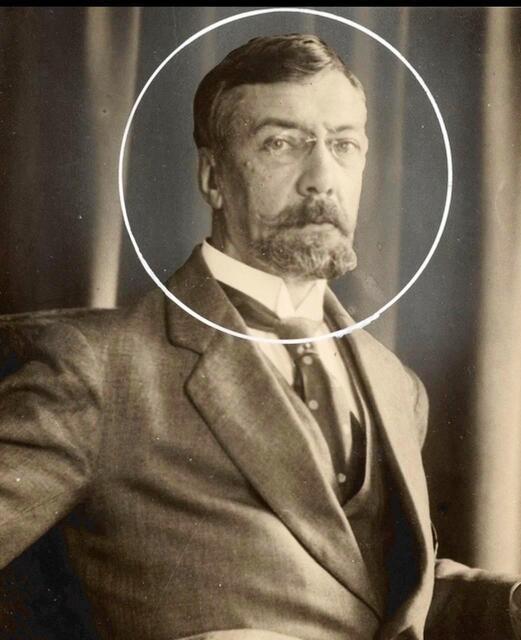}
\caption{\normalsize    Cornelis Easton in 1920. The reason for drawing the annoying white circle around his head is unknown. Credit: commons.wikimedia.org/wiki/ File: Journalisten\_\linebreak[4]
persfotografen,\_SFA\linebreak[4]
022812956.jpg.}
\label{fig:}
\end{figure}

Easton’s approach had been to look for patterns in climatological data associated with the roughly 11 year cycle of sunspots. Collecting data on classifications of severity of winters, eventually back to AD.760(!), he found evidence for a period of 89 years or eight solar cycles. He published two more contributions to the Royal Academy {of Arts and Sciences} in 1912 and 1917, further establishing evidence for this period,  and finally a year before his death a book with a title `Les hivers dans l'Europe occidentale. Étude statistique et historique sur leur temp\'erature; discussion des observations {th\`ermom\'etriques} 1852-1916 et 1757-1851; tableaux comparatifs; classification des {hivers} 1205-1916; notices historiques sur les hivres remarquables’.  The periodicity may be spurious, but the data collecting would be valuable.

For completeness I mention that Easton’s  interests in map-making and geography has led to a number of publications in that area too (see Fig.~29 for an impression of what Easton looked like at that time). An obituary by Julius Christiaan van Oven (1881--1963), jurist and son of Easton’s mentor van Oven, lists an inventory  100 publications by Easton, of which 60 are on astronomy, 16 on climatology, 18 on geography and 5 on other subjects.
\bigskip

After 1905, Easton published only four more research papers on astronomy. The first was only in 1913 in the {\it Astrophysical Journal} and concerned his mapping of the Milky Way and the spiral structure of the Galaxy. Ever since publishing his Milky Way maps based on his visual observations he had been interested in comparing, and possibly improving these on the basis of photographs. In the paper (Easton, 1913), he pointed out that there are a number of advantages in the use of photographs, namely trustworthiness, accuracy and wealth of details. The disadvantage obviously is the (very) limited field of view. For quite some time Easton had examined photographic material and had trained himself drawing maps in a manner similar to what was done usually by observing the Milky Way. Much of this consisted of looking for publications (on loan from the observatories in Leiden and Utrecht), that contained photographs of parts of the Milky Way. Astronomers adding such material to the literature included Edward Emerson Barnard (1857--1923)  from Lick Observatory, Max Wolf from Heidelberg, Henry Chamberlain Russell (1836--1907) from Sydney, Edward Charles  Pickering (1846--1919) from Harvard, and others. For some he had had advance access through Kapteyn, who sometimes received pre-publications. Wolf in fact provided Easton with glass plates. And there were many photographs scattered throughout the astronomical literature.

\begin{figure}[t]
\begin{center}
\includegraphics[width=0.80\textwidth]{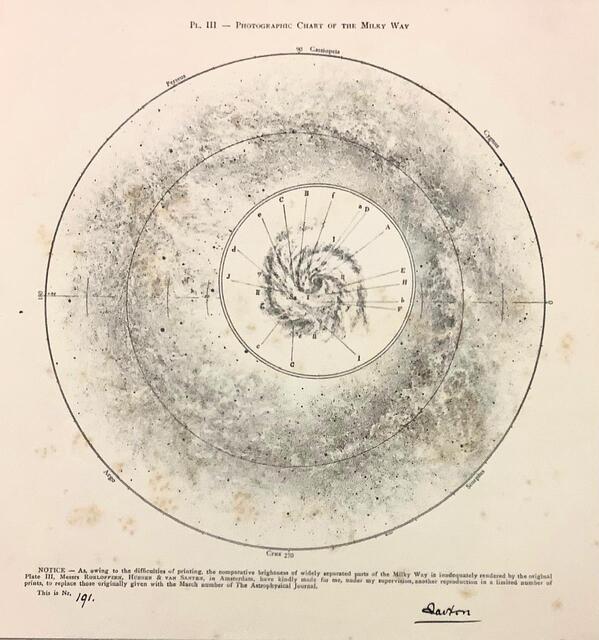}
\end{center}
\caption{\normalsize Easton's spiral structure map of the Milky Way Galaxy, reproduced from a loose version in the \textit{Easton Archives} and based on his `photographic chart of the Milky Way' (Easton, 1913). This version was produced by Easton to include in reprints, since he was not satisfied with the quality of the reproduction in the journal. They were numbered and signed by Easton. The {\it Archives} contains four remaining copies of these inserts, numbered 191, 195, 196 and 197.} 
\label{fig:}
\end{figure}

From this material Easton produced a `photographic’ map of the full Milky Way, consisting of four panels of $10 \times 17$ inch in the projection of Galactic coordinates (Marth, 1873a,b). For this paper he next reworked his four-part map into a circular map and presented that as a fold-out figure. It  is reproduced here as the outer ring in Fig.~30. It depicts the Milky Way between latitudes +20\degs\ and -20\degs. He further stressed the complexity and went into describing specific regions (that he identified with Greek letters)  without much by way of conclusions.

The `photographic map’ was then used to further extend Easton’s hypothesis of the spiral structure of the Galaxy. To this end a possible image of the Galaxy as seen from the pole has been inserted with the supposed position of the Sun in the center of the map. Easton wrote (1913, p.112):
\begmarg
 I have now taken as typical a figure intermediate between two of the best known spiral nebulae: M 51 Canis Venaticum and M 101 Ursae Majoris. The adopted spiral must answer to the condition that the principal features of the Milky Way result freely and plausibly from its projection on the sphere, viewed from the sun ({\bf S}).
\endmarg

This meant that when arms run along our line-of-sight a bright area should result in the form of `brilliant patches’. The lines drawn in Fig.~30 concern lines-of-sight for which Easton made remarks and the Greek letters correspond to the parts of the map he had noted. This is too much detail to consider here. The Milky Way on the sky is over a large part split into two branches divided by a dark lane; this Easton felt is a result of the two major `branches of the spiral’ (we would say the two arms) not being  in the same plane.

It is important to realize that the hypothetical image of the Milky Way Galaxy in Fig.~30 was {\it not inferred} from the map of the Milky Way surrounding it, but followed from an assumed likeness to actual spiral nebulae in the sky.  

Easton discussed two objections to his theory. The first is that the center, which would be in the constellation Cygnus, should be much brighter. The prominence of the central areas in spiral galaxies predicts a much more pronounced difference for the Cygnus region compare to the rest of the Milky Way.  According to Easton the `preponderance’ of the nucleus decreases on better photographs of spiral nebulae. This seems a rather  dubious argument, since better resolution photographs showed generally brighter and smaller nuclei and there is no counterpart of this in the Cygnus area. Easton’s argument that the relative `proportion of mass’ (this is nothing other than brightness) between center and arms is not a serious criterion when the theory is otherwise acceptable, is far from convincing.

Secondly, he rejected the opinion that if his spiral theory were correct that this would mean that the many spiral nebulae discovered in photographic plates, as  by Keeler (1899) with the Crossley Reflector at Lick Observatory, would have to be `external galaxies’. He refers to the discussion of his 1900-paper in the widely-read standard work {\it Handwörterbuch der Astronomie} (Valentiner, 1898-1902). This extensive and comprehensive work, edited  by Karl Wilhelm Friedrich Johannes Valentiner (1845--1931) had been published with four volumes in five tomes  and was the end product of contributions by fourteen leading German astronomers. It contained just over three thousand pages. Valentiner was specialized in astrometry and observations of asteroids. He was director of the observatory of Mannheim and later the astrometry department at Heidelberg (next to Max Wolf’s astrophysics department). Wikipedia pages on Valentiner fail to mention that from 1868 to 1875 he was `observator’ at Leiden Observatory. His departure for Mannheim opened the position that was offered to Kapteyn in the same year. In Valentiner’s {\it Handwörterbuch} (p.123 of Vol.4) the view is promoted that Keeler’s (1899) discovery of large numbers of nebulae at Lick Observatory -- he estimated that the full sky would contain 120,000 new nebulae discoverable with the Crossley reflector, and the large majority of these were spiral nebulae -- combined with Easton’s spiral theory of our sidereal system, would suggest that the universe is full of spiral galaxies, of which the Milky Way system would be one. Easton disagreed with  this; he felt these small spiral systems were part of our system, small eddies in the convolutions of the great one, that are embedded in the larger system. 

Easton added surprisingly little support of his theory. His study of the relation between nebulae and the Milky Way did show two maxima at longitudes about 90\degs\ either way from the supposed center in Cygnus. This result seems to me incorrect, see above. He did note a stratum of nebulous light, except in the `dark singular holes and cracks’. He noted he did see `no evidence whatever of continuous masses of opaque matter lying between us and the stars, although such non-luminous matter must be scattered freely throughout the system’ (p.116). He did speculate that the {\it  nubeculae} (Magellanic Clouds, a term that Easton in fact sometimes used) are small systems as NGC5194 connected to the M51 galaxy. But all of this was not presented as {support} for his hypothesis.

The presentation of  the possible appearance of the Milky Way Galaxy was only one of the aims of the 1913-paper. As the title shows the presentation of the `photographic map’ was a main reason to publish the paper. Easton started out by producing four maps of $10 \times 17$ inches, but did not publish these in the 1913 paper. He would eventually -- in December 1928, the year before his death --   publish two of them in his last paper (in {\it Monthly Notices of the Royal Astronomical Society}). These concerned the northern half of the Milky Way from old Galactic longitudes 0\degs\ to 180\degs, from Orion to Ophiuchus (Easton, 1928b).

In relation to Fig.~30 he remarked in his {\it Notes}: 
\begmarg
I was also unlucky with the publication of the photographic map. Although I had taken all precautions; even sent a platinum print to America, the shading on the prints in the Astroph. Journal were so much incorrect, that I had decided to have a large number of reprints made in Amsterdam, and to send them around. But with these `improvements’ one does not recover the entire circulation of the Astroph. Journal.
\endmarg
Fig.~27 shows one of the remaining inserts in his {\it Archives}.
\bigskip

\begin{figure}[t]
\begin{center}
\includegraphics[width=0.90\textwidth]{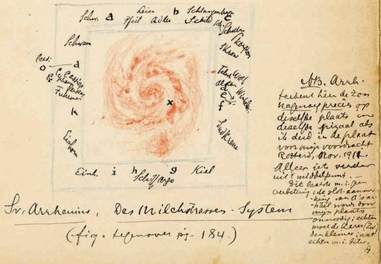}
\end{center}
\caption{\normalsize   The view of our Galaxy according to S.A Arrhenius. This is a sketch by Easton of a well-known figure by Arrhenius. The image is M51 (with south at the top, so NGC5195 appears at the bottom).  However, in the original  the constellation names are in Swedish, and the lower case letters around the border are missing. This sketch is made after a version in an article ‘Das Milchstrassen-System’ in the German magazine \textit{Nord und Süd} (Arrhenius, 1912). From the \textit{Notebook}.}
\label{fig:}
\end{figure}

Ever since his first proposal of spiral structure in 1900, but also after his improved map in 1913, Easton had support for his idea. As we have seen, Kapteyn was  skeptical at the time of the awarding of the honorary degree and, in fact, remained skeptical. Others were more sympathetic to the proposal. An example that very much flattered Easton was Svante August Arrhenius (1859–1927), who received the Nobel Prize in Chemistry in 1903. Arrhenius wrote a number of widely read  books on cosmic physics, astronomy, life in the universe, etc., in which he discussed Easton’s spiral theory. His book ‘The destinies of the stars’ (Arrhenius, 1918), first published in Swedish in 1915, has been translated into a large number of languages. In that book he included a photograph of M51 with the position of the Sun  indicated at what he felt was the most likely situation. In his {\it Notebook} Easton included a drawing of the figure (Fig.~31) and wrote (p.192):
\begmarg
Arrhenius puts the Sun here in almost exactly the same place in the same spiral as I did in my lecture in Rotterdam, November 1912. Only somewhat further from the center. This last thing is hardly an improvement in my opinion; [...] 
\endmarg

In fact, the figure had been published earlier {by Arrhenius} and that is where Easton refers to in the {written} text. It was in a short article {in German} on  `Das Milchstrassen-System’ in a German semi-monthly journal {\it Nord und Süd}  (Arrhenius, 1912), and this is where the figure has been taken from by Easton. 

The spiral structure view of Easton was already during his lifetime seriously challenged. Investigations by Harlow Shapley (1885--1972) showed a much larger dimension of the Galactic System than Easton envisioned, possessing a system of globular clusters with a center of order 15 kpc away in the direction of the constellation Sagittarius, almost perpendicular to that of Cygnus. The notion of interstellar extinction and its role in shaping the features visible in the Milky Way was gaining ground.  But the largest blow came with the discovery of Galactic rotation (Oort, 1927) about a  center in the same direction of Shapley’s system at the border of Sagittarius and Scorpius. Oort had come up with the idea of  differential rotation and his proof of Galactic rotation was based on observations of radial velocities and proper motions of nearby stars and had direct consequences for both local and global structure. The difference of about 90\degs\ with Easton’s center {toward} Cygnus was the death blow. In the {\it Notebook} he reproduced Fig.~30 above with a large star for the position of the Sun opposite Sagittarius ({\it Notebook}, p.211):
\begmarg
The position of the Sun can also be shifted in such a way that the center of a (conceivably large) system then lies in the direction of Sagittarius. 
\endmarg
\bigskip

The other two astronomical papers published by Easton after 1905 concerned his conviction that the distribution of bright stars in the sky correlated with the brightness of the Milky Way. This correlation was challenged by a publication by Pannekoek on the distance to the Milky Way (Pannekoek, 1919), which I discussed in much detail in van der Kruit (2024a). Pannekoek’s assumined that a bright patch in the Milky Way in reality was a cloud of stars with a dimension along the line-of-sight comparable to the dimension projected onto the sky, made up of stars with luminosities distributed like Kapteyn’s local luminosity function. Then the brightest stars at some level would start contributing to star counts and this would be made visible by comparing star counts on and off this patch. Pannekoek used star counts (total numbers of stars to some limiting magnitude) from various sources,  namely the {\it Bonner Durchmusterung}, star counts by  Theobald Epstein (1836--1928) from Frankfurt am Mainz, counts on photographic plates of the {\it  Carte du Ciel} and William Herschel’s star gauges as the deepest. He then calculated gradients, the slope of the counts between the limiting magnitudes, for which he used overall values. This must have been highly inaccurate, certainly not accurate to a tenth of a magnitude, and using gradients where such an accuracy is required even more uncertain. Pannekoek employed this method on two clouds, the Cygnus Cloud between $\beta$ and $\gamma$  Cygni (see Fig.~9) and a bright patch in Aquila. His analysis indicated that stars in these clouds started contributing  at magnitude 11.5 to 12. From this Pannekoek deduced enormous distances of 40 and 60 kpc. This supported Shapley’s model for the Sidereal System.

But it contradicted Easton’s earlier correlation results that indicated that the distribution for {\it Bonner Durchmusterung} stars of magnitude 9 were part of these clouds. With Pannekoek's distances these stars would intrinsically have to be enormously bright, up to a million times the luminosity of the Sun. Easton addressed this by analyzing these stars using other properties. He argued that if there is an excess of bright stars that can be attributed to  Milky Way cloud, this would be accompanied by an excess  (1) of very small parallaxes, (2) of small proper motions, (3) of early-type stars, since B and A stars were known to be brightest. He then compared some bright spots and faint areas in the immediate vicinity in the Milky Way, assuming that in the darkest spots we only see the `field stars’ in our vicinity.

Easton published the results of his investigations in a paper `On the distance of the Galactic star-clouds’ in the {\it Monthly Notices of the Royal Astronomical Society} (Easton, 1921). It is telling for the impression he had made in the professional astronomical world that the article was communicated by the Astronomer Royal, who was then Frank Watson Dyson (1868--1939). Easton selected five bright areas in the Milky Way with dark spots nearby at the same Galactic latitude. The work was rather extensive: it concerned a total of 668 square degrees (331 in the dark, 337 in the bright parts of the Milky Way) and contained 1311 stars (491 in the dark, 820 in the bright parts). Remarkable is Easton’s treatment of outliers (proper motion of 10 arcsec per century or more). He either ignored them or set them equal to 10; and thirdly he did what we would do today: taking medians. I quote from the {\it Notebook} (p.205):
\begmarg
The calculation of all this, after collecting data in the {\it Bonner Durchmusterung} and in the {\it Draper {Catalog}}, was very cumbersome, but through the mediation of W.J. Luyten I had much of that work done at my cost at the Leiden Observatory. [...]  -- The result was not contrary to my expectations, but still contained much less conclusive value than I had hoped to obtain, partly because the dark parts in the MW. are so difficult to delineate well and because one can never prove  that there is no obscuring matter in front of them, which Pannekoek sees a little of everywhere, much too much in my opinion. Nevertheless I thought I could draw the conclusion, for various parts of the galactic strip, that the galactic clouds exert an influence on both the proper motions and on the spectra of the whole of the stars observed there; and that, unless we want to ascribe to many `cloud stars’ a luminosity of millions of times that of the Sun, the galactic clouds there cannot lie at distances of 50,000 or even 10,000 parsecs.
\endmarg

The results were in my opinion not conclusively in  support of either Pannekoek's or Easton’s propositions.
\begmarg 
My paper appeared in the Monthly Notices of the R.A.S. Jan. 1921: On the distance of the galactic star clouds; it was submitted by the Astronomer Royal, Dyson. This is worth mentioning, because in that post-war period the English magazines hardly included work by foreigners any more -- the reason why the B.A.N, was founded; Bulletin of the Astronomical Institutes of the Netherlands -- so that also Luyten did not expect that my article would be published.
\endmarg

There were actually other reasons for founding the {\it Bulletin of the Astronomical Institutes of the Netherlands.} {(B.A.N.)}. Easton concluded that the `cumulation of evidence seems to point rather convincingly to the conclusion’, that the clouds could not be at the large distances of 50 kpc or so advocated by Pannekoek, not even 10 kpc  (Easton, 1921). Actually `convincingly' is by all means an overstatement.

The relation between Pannekoek and Easton had been very warm. Pannekoek had been of great help to Easton in the research leading up to the latter’s first paper on the correlation of brighter stars with Milky Way bright spots (Easton, 1895a). Pannekoek had even chosen Easton’s side in a proposition accompanying his PhD thesis when leading German astronomer Ritter Hugo Hans von Seeliger (1849--1924) had attacked the spiral structure proposal.  Pannekoek however, did disagree with the Easton (1921) result and published an article on the distance to the Cygnus Cloud (Pannekoek, 1922). He analyzed the star counts {toward} this cloud between magnitudes 6.5 and 9.5 following the methods developed by Kapteyn and derived a stellar density along the line-of-sight between 250 and 1600 parsec. Thus showed an agglomeration of stars between 200 and 600 parsecs, with a density 1.5 times that in the vicinity of the Sun. These stars were thus independent of the Cygnus Cloud and that contradicted Easton. Pannekoek mapped this `agglomeration’ by tracing the density of bright stars on the sky relative to the Milky Way surface brightness. This new analysis reduced  the distance of the Cygnus Cloud significantly from his earlier 40 kpc to 18 kpc.

Easton wrote a rebuttal that he wanted to publish in the {\it B.A.N.} also. The problem was that the {\it B.A.N}, being a merger of four institute publication series was open only to members of these institutions through their directors. Pannekoek opposed accepting Easton’s manuscript as an Amsterdam publication, but in the end he agreed to allow Easton’s paper to appear as a contribution from the Amsterdam Astronomical Institute.  In this he presented the superposition of a foreground agglomeration of bright stars and the background Cygnus Cloud as an improbable coincidence (Easton, 1922a). He showed two maps of the Milky Way region between $\alpha$ and $\pi$ Cygni, one for the stars of magnitude 8.1—9.5 and one from Pannekoek’s (1920) surface brightness maps of the northern Milky Way. To this he added as the concluding sentences of the paper:
\begmarg 
We cannot, I concluded, of course expect a perfect agreement, but who could believe that these two diagrams represent two distinct and independent agglomerations of stars, situated respectively at distances of 400 and 18,000 parsecs? 
\endmarg
To our modern eys Easton's argument is far from convincing. The publication of his rebuttel has been described in the {\it Notebook} as follows (p.206):
\begmarg
The publication of this piece in the B.A.N. -- from which the work of those (such as Father Stein and myself) who are not connected to one of the official Dutch institutes is excluded – would have been difficult, but I pressed Pannekoek very much; because he had made remarks on my work, I demanded my defense before the same readership. He finally gave in, but refused to have his diagram from B.A.N. 11 reprinted, which for me contained evidence against himself. I then had this diagram reproduced myself, reduced in size with a caption, and placed it in my reprints!
\endmarg
A stack of these inserts are still present in the Easton {\it Archives}. Pannekoek also did not perform a detailed statistical analysis (using instead statements like `we see at once’) and there is no obvious support for Easton’s case in it. This whole matter seems more a case of wishful thinking than presenting  convincing cases supported by rational arguments.
\bigskip

That same year (1922) an occasion arose to debate the issue when Harlow Shapley visited the Netherlands.  Anton Pannekoek and Willem de Sitter (1872--1934), director of Leiden Observatory,  had organized a special session of the {\it Nederlandse Astronomen Club} (the professional society of astronomers in the Netherlands) on the dimensions of the Sidereal System on 28 May, 1922. This had been preceded by the first General Assembly of the {\it International Astronomical Union} in Rome and would be followed by the centennial of the {\it Royal Astronomical Society}, and Shapley and his wife, astronomer Martha Betz (1890--1981),   had been invited to visit the Netherlands (they had visited Utrecht the day before).  This was weeks before Kapteyn’s death and he of course could not attend. Shapley had justified his acceptance with the statement (see for a very short report: Easton, 1922d):
\begmarg
[...] there is no country where the percentage of students of the structure of the Universe is so great. I am sorry that the greatest of all, Kapteyn, is so ill that he cannot be present.
\endmarg

Easton had made notes on the meeting that have been conserved in his {\it Archives}; they are accompanied by a drawing of Shapley (see Fig.~32). From these notes we know who were present: {Ejnar Hertzsprung, Pieter van Rhijn, Albertus Nij\-land, Jacob de Vos van Steenwijck, Jan van der Bilt, Johannes Stein, Isidore Nort, Egbert Kreiken, Jan Weeder, Willem van den Bos, Jan Schildt, Coert Hins, Alexander Leckie, Peter van de Kamp and Cornelis Easton.} Pretty much all permanent staff of the Netherlands (de Sitter had, according to these notes, not yet returned from Rome) and current students, and some former PhDs, who had written theses on the subject.

\begin{figure}[t]
\sidecaption[t]
\includegraphics[width=0.58\textwidth]{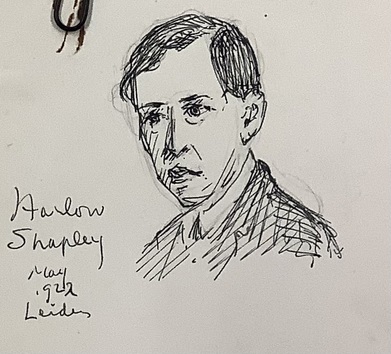}
\caption{\normalsize  Drawing of Shapley by Easton, produced during the meeting in Leiden on 28 May, 1922. From Easton’s \textit{Archives}.}
\label{fig:}
\end{figure}

Willem Johannes Adriaan Schouten (1893--1971), who had written a PhD thesis in 1918 under Kapteyn, and who for five years had been appointed privaat-docent in Amsterdam (he was secondary school teacher), opposed Shapley’s view, but was not present. Schouten had assumed that globular clusters had the same luminosity function as the solar neighborhood and found distances up to eight times smaller than Shapley’s based on pulsating stars (Cepheids, in fact  W Virginis,  and RR Lyrae stars).  Pieter Johannes van Rhijn (1886--1960), also student of Kapteyn and soon to become his successor, had pretty much accepted Shapley’s views. He and Kapteyn had criticized Shapley’s calibration of the period-luminosity relation for pulsating stars,  but van Rhijn had started doubting this analysis (indeed it was flawed; see van der Kruit, 2015,  p.592). Hertzsprung seemed to support Shapley.

Shapley started with an introduction, in which he spent much time on his recent spectroscopic observations and only mentioned distances briefly, saying (as recorded, and presumably paraphrased, in English by Easton):
\begmarg
These distances maybe divided  by 2, perhaps 1.5, but not by a larger factor, unless there be fundamental [sources of error]. We have not investigated f.i. a general -- not selective -- obstruction of light.
\endmarg
This concerned a recalibration of globular cluster distances. Easton then opened the discussion:
\begmarg
Mr. Chairman! I won't take much of your time but I would like to have some additional information on a few points, only slightly insisted upon by  Prof. Shapley, but of great interest: the distribution of the globular clusters and of the galactic stars.
\endmarg
Easton then showed his isophote map and a few other diagrams, describing his correlation work and arguing that so distant a center would result in a much larger brightness difference between the Sagittarius area and the opposite Auriga region of the Milky Way. Only a few months before Easton had published a long article in {\it Hemel \&\ Dampkring} summing up the controversy, including a summary of the famous  Shapley-Curtis debate and his own analysis that ended up being quite definite about the Milky Way being much closer than Shapley had claimed (Easton, 1922c). He would have concluded:
\begmarg
Mr. Chairman, I have submitted those observations to our distinguished guest of this afternoon in order to have his opinion and to see a way out of the apparent conflict between Shapley's ideas on the structure of the universe and the constitution of the stellar zones, but in any way to detract from the high value of Prof. Shapley's splendid work, for which I have the greatest admiration.
\endmarg
Shapley  had responded without presenting any new points of view, calling the Milky Way `a dangerous place', everything a mixture and full of pitfalls. It is full of `obstructing matter'. Easton suggested there are voids. Pannekoek said he could solve the discrepancy and presented his work on the distances of the Milky Way clouds. He defended Shapley’s dimension, arguing that star counts were not yet sufficiently deep to reveal what had to be a shallow gradient. Shapley had justified the direction of the Galactic center being in Sagittarius by pointed at bright areas in there and further  had replied that {\it Kapteyn’s Plan of Selected Areas}, which Easton had used, were unsuited because these had been chosen without regard of the Milky way behind them and had often been situated in or near dark areas in the Milky Way.  Easton reacted according to his notes:
\begmarg Dr. P. has mentioned a \underline{possible} explanation, but there are no indications that force us to believe that it is the true explanation, no indication of a systematic change of the mean magnitude of the Galaxy stars in Galactic longitude .
\endmarg
Here the notes end; as expected the outcome was inconclusive.
\bigskip

\begin{figure}[t]
\sidecaption[t]
\includegraphics[width=0.58\textwidth]{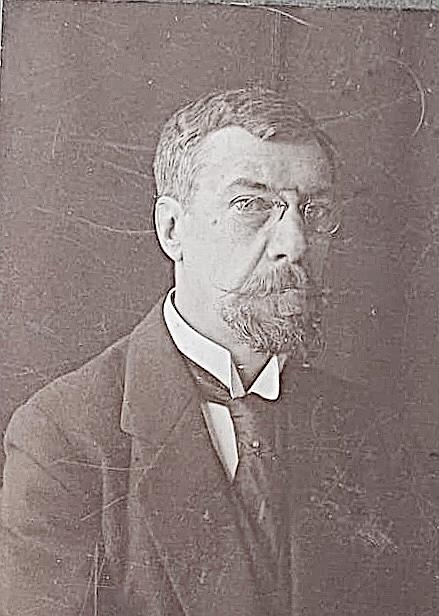}
\caption{\normalsize  Illustration of Easton accompanying the special Easton-issue of {\it Hemel \&\ Dampkring} (June 1928). This version has been supplied by great-granddaughter Titia Vlaardingerbroek-de Knegt through her son Kristian.}
\label{fig:}
\end{figure}

The meeting proceeded further as follows  (from Easton, 1922d):
\begmarg
After the presentation there was a discussion. This was mainly attended by Dr. Easton and Dr. Pannekoek. Then a reception and tea at Prins on the Rapenburg.
\endmarg
Prins refers to a restaurant near the Academy Building in Leiden, which is situated on the canal called Rapenburg. 

In spite of being an active participant to the meeting, Easton complained not being regarded equal to the others of the professional community of astronomers, his honorary doctorate being a second rate degree (from the {\it Notebook}, p.209):
\begmarg 
[...] and when Shapley came to our country in ’22 as a young director of Harvard, he asked for me – my official colleagues hastened to tell him that I was ‘in business’ (as he had been himself, Shapley told me) and left me out of the more intimate company of the famous guest .... 
\endmarg
This probably refers to informal talks outside the official program.

Pannekoek and Easton did not come to an agreement. Formore details see Chaokang Tai's PhD thesis (Tai, 2021; pp.85-90).
\bigskip

The paper in {\it B.A.N.} (Easton, 1922b) was Easton’s final astronomical publication in the professional literature. In the {\it Notebook} he added one more idea. It built on the deflection of light by the Sun’s gravitational field as predicted in Einstein’s General Relativity and confirmed during a solar eclipse in 1919. This gave Easton the idea that if bright spots in the Milky Way are at or near the edge of the Galaxy and therefore have no stars in the background, and are massive conglomerations of stars, then this bending would produce empty areas around it. This could be a partial  explanation for `the curious very dark spots next to many of the most luminous spots in the Milky Way’ ({\it Notebook}, p.207). This idea, published in {\it Hemel \&\ Dampkring} (Easton, 1920), was not meant to be a general explanation of dark clouds, in which ‘we undoubtedly have to do with empty spaces’ (Easton, 1920, p.104). There is no acknowledgment to Willem de Sitter, director of Leiden Observatory and expert in General Relativity, so apparently he was not consulted. Had he been approached, de Sitter very likely would have told Easton that such a deflection over many degrees was out of the question.

\section{Discussion}

\subsection{Easton’s legacy}

For a picture of Easton {toward} the end of his life, see Fig.~33. 
Easton’s three areas of contribution to astronomical research were (1) the mapping of the surface brightness distribution of the Milky Way; (2) the proposal that our Galaxy is a spiral galaxy and the way this reflects in the structure of the Milky Way on the sky; and (3) the correlation of the distribution on the sky of bright (up to apparent magnitude 8 or 9) with the surface brightness  of the Milky Way and the consequences for the distance to the Milky Way.

The first of these was highly successful. The publication of the book {\it   La Voie Lactée dans l’hémisphère boréal} was a milestone. The only other publication of similar quality and importance was Boeddicker’s publication of 1892, although that lacks the extensive discussion of features and structure, and Easton’s book is certainly superior in the complete and detailed discussion of the history of writings on the Milky Way. Anton Pannekoek used Easton’s isophote chart, which was calibrated in a relative sense, to improve his own maps by averaging these. Comparison with modern data showed that Pannekoek’s surface brightnesses are quite accurate and this reflects of course on the Easton work (see van der Kruit, 2024a). This is a major accomplishment and a the main reason for the bestowing of the honorary doctorate onto Easton. However, as concluded in van der Kruit (2024a), little was done with these maps and they did not provide any significant progress in understanding our stellar system.  

The second issue of spiral structure proved correct as a fundamental insight into the nature of the stellar system we live in. His analysis of the structure of the Milky Way on the sky in terms of this idea was frustrated by interstellar extinction, the establishment of which increased the size of the system substantially and rendered the assumption that the spiral structure is reflected in the appearance on the sky incorrect in the first place. He opposed putting the center in the direction of Sagittarius, but conceded that it would be possible. I feel Easton should be credited for being the first to bring up spirals, where others, like leading astronomers, as the Herschels and Celoria, had struck to rings, or Proctor to filaments. When spiral galaxies were shown to be extragalactic many astronomers would bet our Galaxy had spiral structure like so many other galaxies; Easton was decades ahead of them, even while the extragalactic nature of spiral nebulae was not established. 

The third area of research was less successful. Easton in the end was simply wrong. Easton provided detailed maps of stars and the surface brightness of the Milky Way and judged by eye their similarity and used his common sense, nit mathematics, to decide whether there was a correlation or not. This is prone to biased judgment. His contributions to his area of research have been disappointing. His knowledge of physics very likely was rather limited too

Easton distinguished himself from professional astronomers not only by having another primary occupation, but also by being primarily a literary man, not a scientist educated with a strong background in mathematics and physics. Certainly the first, but to some extent also the second aspect mentioned above, as an observer impressed by the starry sky and the Milky Way, were particularly suited to his approach as a non-scientist in the sense described. The third aspect of was more along the lines of a natural scientist and here he was much less successful. Easton should be credited for being a extraordinary contributor to the mapping of our stellar system and thinking about its appearance in space from other aspects. Easton had no involvement in teaching or organization of professional astronomy. He did of course have a major involvement in amateur astronomy in the country.
\bigskip

In his {\it Notes} he gave a comprehensive summary of his professional work. For the newspapers he worked for, he produced in particular articles on science and international affairs and travel. He wrote serials on astronomy, geography and medical issues. Later he often wrote commentaries on international developments. With his wife {and} daughter Titia he made trips to Italy and wrote travel stories (some by Titia); these were collected and re-published in  the form of a book. For {\it Het Nieuws van de Dag} he and his family moved in 1906 to Amsterdam (he had been selected out of 103 applicants), but he was less happy in that position as editor-in-chief that gave him less freedom than simply editor. During War I he wrote very regularly about the military situation and developments, accompanied by maps drawn by himself. Part of these stories and maps have been collected and published after the war in two books. In 1923 the newspaper, although successful with a commendable number of subscribers, was sold to the main competitor  and terminated. Easton was held in high regard as reporter and editor by his colleagues. In a comprehensive study of the history of reporting in the Netherlands between 1850 and 2000 (Wijfjes, 2004), Easton appears regularly as one of a select few leaders. He felt, contrary to many others in the field,  that women were definitely suited for the job as well. He was more than a processor of information, adding mature commentary and was not opposed to travel to find out things.

\begin{figure}[t]
\sidecaption[t]
\includegraphics[width=0.48\textwidth]{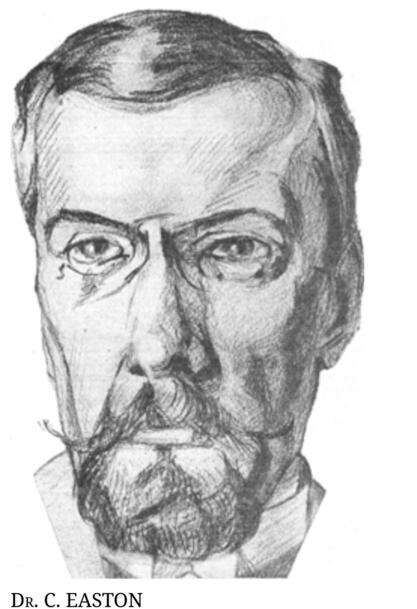}
\caption{\normalsize Drawing of Easton that appeared as part of an obituary in \textit{Den Gulden Winckel}, volume 28 (1929). This was a literary magazine that appeared between 1902 and 1942. See www.dbnl.org/tekst/\_\linebreak[4]
gul00119290\_01/\_\linebreak[4]
gul001192901\_01\_0102.php.}
\label{fig:}
\end{figure}

In addition to a travel guide for Italy (again with contributions by his daughter), for which he and his wife and daughter made various trips to Italy and for which he drew maps. He translated a number of books on societal issues by Theodore Lothrop Stoddard (1883--1950), American historian and political scientist,  Paul de Kruif (1890--1971), American, but Dutch-born, bacteriologist, and Paul Wilberforce Harrison (1883--1962), American `medical missionary to Arabia'. The first may raise some eyebrows, because Stoddard, of whom Easton translated two books, was an outspoken supporter of white supremacy.

His societal  work was based on (from the {\it Notes}): `everything that was good and just and that is not done by taking sides or looking for cliques'. He was board member, chairman or secretary of a fair number of societies and funds, concerning tuberculosis, literature, war and peace, nature preservation, geography, etc.  

Since 1921 Easton (see Fig.~34) was chairman of the {\it Nederlands Vereeniging voor Weer- en  Sterrenkunde} (Dutch Meteorological and Astronomical Association). With his interest and research as an amateur in both disciplines he was an obvious and quite appropriate choice. With his background as a newspaper editor he rescued the Association's magazine {\it Hemel \&\ Dampkring} from its financially precarious situation in which he had  found it. On his initiative, a reorganization took place , which led to the contract in 1922 with a new publisher. In a circular he addressed a number of members, in order to remedy the dubious finances by means of voluntary contributions. This resulted in a more solid basis for  the periodical. In 1922 Easton also became  chairman of the Editorial Committee. To express their gratitude for his efforts and celebrate the 25-th anniversary of his honorary doctorate, the editorial staff paid tribute to him with a special  Easton issue of {\it Hemel \&\ Dampkring} of 13 June 1928, 25 years after obtainoing his honorary degree,

\subsection{Stellar dynamics; arriving fashionably late}

Dynamics, or even kinematics, of stars -- let alone mathematics and physics -- played no significant role in Easton’s work. The dynamical aspects of their work were also not prominent at all  for many Galactic astronomers up until early in the twentieth century and I wish to present some thoughts on this here.

This article is my final  contribution in what became in hindsight a series of biographical studies on Dutch Galactic astronomers: Jacobus Cornelius Kapteyn (van der Kruit, 2015, 2021a), Cornelis Easton (this paper), Anton Pannekoek (van der Kruit, 2024a), Pieter Johannes van Rhijn (van der Kruit, 2022), and Jan Hendrik Oort (van der Kruit, 2019, 2021b). Together with biographies of Frederik Kaiser (van den Berg, 2022), Willem de Sitter (Guichelaar, 2012) and Hendrik Christoffel van de Hulst (1918-2000; van Delft, 2021), and articles that are more than perfunctory biographical notes on Lucas Plaut (1910--1984; Henkes, 2020) and Adriaan Blaauw (1914--2010; Oort, 1985, van Woerden, 1985),  these form a reasonably complete, although scattered account of the growth of Galactic astronomy in the Netherlands, up to roughly halfway the twentieth century. Unfortunately the biographies of Kaiser and van de Hulst have not appeared in English translation. I do not intend to summarize all this material here, but I do want to comment on the lack of attention to gravitational dynamics and instability in Galactic astronomy before the start of the twentieth century. For those readers lacking the necessary background, I have added an appendix with a brief history of stellar dynamics, nowadays commonly referred to as Galactic dynamics when concerning the Milky Way Galaxy; or galactic dynamics when applied to external galaxies.

Newtonian gravity theory was widely applied during the nineteenth century to the motions and stability in our planetary system. In spite of this no attempt was made to incorporate it in the understanding of the structure of the Sidereal System. Of course velocities were hardly known, but that would not have been a reason to not even consider dynamics. How it could possibly  conserve its shape, even approximately so, in view of gravitation, was not really addressed. The next question of how not only the structure as a whole could be stable, but also the system was prevented from breaking up and collapsing on smaller scales, was not even considered until well into the twentieth century. One may argue that at first stability was not an issue in view of known timescales -- the date of Creation was determined to be October 23, 4004 BCE by Bishop James Ussher (1581--1656) so the Universe was only some 6000 years old. On the basis of the Kelvin-Helmholtz time of radiating away all gravitational, potential energy, defined by Lord Kelvin (William Thomson; 1824--1907) and Hermann Ludwig Ferdinand von Helmholtz (1821--1894), Kelvin in 1862 estimated the age of the Sun for linear contraction to be 20 million years. When in 1859 Charles Robert Darwin (1809--1882) published his theory of biological evolution, he pointed out that this  required much longer timescales, possibly billions of years, and this was supported by geological evidence. Stability of the Sidereal System could well have become a serious issue.

The starting point of William Herschel’s  famous crosscut was the assumption of a uniform distribution of stars in space; however, the Herschels did not address the dynamical consequences, as far as I am aware, of abandoning the uniform distribution, while the Newton-Bentley correspondence (see Appendix) had shown that equilibrium required it to be perfectly uniform (and even Newton had to rely on  providential intervention). William Herschel did arrange nebulous objects into a sequence reflecting gravitational contraction into stars, but nothing of the kind on a larger scale, except deducing a ring-shaped distribution of stars. An important step was taken by Richard Proctor, who -- as we have seen in this paper -- used John Herschel’s outline of the Milky Way to deduce that there was organized, large-scale structure (Proctor, 1869). The view of the Galaxy showed in this model long filaments and this could not be long-lived. However, equilibrium and stability was not an issue to Proctor. The only relevant remark was that (Proctor, 1869, p.54)
\begmarg
the spiral is supposed not to lie in one plane, but (as the contorted figure of the Milky Way obviously suggests) to have been swayed out of that plane by varying attractive influences.
\endmarg
\noindent
Closer to dynamics he did not get.

Of the Dutch Galactic astronomers mentioned above, only Kapteyn and Oort were deeply concerned by the questions of dynamics and equilibrium. For Kapteyn this must have been induced by his discovery of the Star Streams, which proved the existence of large-scale organized motions. I quote from a letter Kapteyn wrote to George Ellery Hale (1868–1938) on September 23, 1915. This long letter had for its first part been on the role of deduction versus induction in scientific research (see van der Kruit, 2015; section 4.7). On the ninth page Kapteyn wrote:
\begmarg
 One of the somewhat startling consequences is, that we have to admit that our Solar System seems to be in or near the center of the Universe, or at least to some local center.

Twenty years ago this would have made me very skeptical in regard to the result of the investigation. Now it is not so. -- Seeliger, Schwarzschild, Eddington and myself have found that the number of stars pro unit of volume is greatest near the Sun. I have sometimes felt uneasy in my mind about this result, because in its derivation the consideration of scattering of light in space has been neglected. Still, it appears more and more that the scattering must be too small and also somewhat diﬀerent in character from what would explain the change in apparent density. This change is therefore pretty surely real.

Even more important than the central position of the Sun seems to me to be that our result for the first time shows the evidence of \underline{force} in the great Sidereal System. A rough computation leads to the conclusion that at a distance corresponding with a parallax of 0\secspt02 [or 50 pc] the stars are under the action of a force equal to the attraction of a central mass having 5 million times the Sun’s mass. This number is very considerably higher than the number of stars we assumed up to the present to exist in a sphere of a radius corresponding to [parallax] $\pi$ =0\secspt02. But these of course may be completely dark bodies, of which we know nothing.
\endmarg
This led up to the famous ‘First attempt' paper (Kapteyn, 1922). The first sentence of the abstract says it all: {\it First attempt at a general theory of the distributions of masses, forces and velocities in the stellar system}. This gave rise to Jeans’ equally important paper the same year,  on ‘The motions of stars in a Kapteyn-Universe’, in which he presented his hydrodynamical equations that have been the basis of many studies in stellar dynamics (Jeans, 1922).

The next contribution applying the theory of stellar dynamics to our Galaxy came from Jan Oort, with his discovery of Galactic rotation and theory of stellar dynamics in the Galaxy. In his autobiographical article in the {\it Annual Review of Astronomy \&\ Astrophysics} he wrote (Oort, 1981, p.2):
\begmarg
Practically from the beginning of my study, I had been fascinated by the possibility that dynamics, which had so long been practically a monopoly of the astronomers investigating the solar system, might one day be applied to the so vastly greater systems of stars. An impressive foreshadowing was Kapteyn's ‘star streams’ which Karl Schwarzschild subsequently interpreted as an ellipsoidal distribution of stellar motions.

A strong stimulation had come to me from reading a semi-popular book by Newcomb which Kapteyn had lent me. In later years there were the inspiring books by Eddington, {\it Stellar movements and the structure of the Universe}, and by Jeans, {\it Problems of cosmogony and stellar dynamics}, both of which fundamentally influenced my education as an astronomer. 
\endmarg
Dynamics arrived `fashonably late', as a consideration, a balancing act,  between arriving at the appropriate time and making a grand entrance in style. 
\bigskip

One conclusion from my {biographical} studies is that the relation of models of what the structure of the Sidereal System was, to how it could be in equilibrium and stable was not a subject of inquiry until Kapteyn and continued by Oort. The program to map the surface brightness of the Milky Way and deduce the distribution stars in space from star counts came to a standstill with Shapley’s proof of the large dimension of the stellar system and the growing realization in the 1920s that interstellar extinction was a major factor at low latitudes (e.g. Oort, 1926). This required a completely fresh approach.

In ‘The death of a research programme -- Kapteyn and the Dutch Astronomical Community’, E.R. Paul wrote about this episode (Paul, 1981;  p.89):
\begmarg
The research programme initiated a century-and-a-half earlier by William Herschel, of determining the arrangement of the sidereal universe by counting stars (and, later, measuring stellar magnitudes), had failed in a fundamental way. [...]  As a scientific research programme of global dimensions, statistical astronomy had reached its end with the emergence of Shapley's programme of analyzing globular clusters. 
\endmarg
In broad terms I agree with this. However, in his book  `The Milky Way Galaxy and statistical cosmology, 1890-1924’ Paul wrote (p.157)
\begmarg
Kapteyn's 1920 model represents the culmination of his life's work on the sidereal problem. [...] this model is entirely static, and thus it fails to account for large-scale motion of the stars as represented in his discovery of star-streaming. Because his understanding of star-streaming vis-à-vis his static sidereal model had barely progressed in the years since 1902 and because so many others had attached such importance to this phenomenon, Kapteyn, somewhat anticlimactically, published a dynamic theory of the stellar system in 1922. This theory attempted to explain stellar motions in terms of gravitational forces, in the context of his earlier static, density distribution system.
\endmarg

I do not agree with the statement that Kapteyn’s `culmination of his life's work on the sidereal problem’ was his 1920 model. I believe that this is his last major paper (Kapteyn, 1922) with the inclusion of dynamics in the structure of the Galaxy. Kapteyn wrote only one surviving letter to Oort, dated 29 December, 1921 (van der Kruit, 2018; p.59, 60 and 644):
\begmarg 
I could not help at my departure from Groningen to feel great satisfaction. The many tributes and especially the powerful cordiality made me happy. -- But, believe me in spite of that feeling of satisfaction I still have felt that much has been lacking during the completion of my duties. Not one of the lesser of these shortcomings that I have felt is that in the final year I have not cared sufficiently for my students. I left that for the major part to van Rhijn. -- The reason, you will undoubtedly know, resulted from my irresistible wish to bring my life’s work to some sort of completion. I have taliter qualiter [more or less] reached that completion. You will see in what manner in two articles for the Mount Wilson Contributions that are in the hands of the printers. -- I had to sacrifice much too much for that.
\endmarg

Kapteyn \&\ van Rhijn (1920), on the analysis of starcounts and the model for the sidereal system, had appeared well before  this letter. The two manuscripts referred to by Kapteyn were his `first attempt’ and an article  on some technicalities of the analysis of star counts. Kapteyn himself in this letter to Oort considered his 1922 `first attempt'-paper to be the culmination of his work. And indeed, introducing dynamics was a major step forward in Galactic astronomy. To describe it as an anticlimax after the star count analysis in 1920, is completely off the mark. By including dynamics, Kapteyn ensured that Dutch astronomy blossomed after him, even when his program of statistical astronomy had met its limits.
\bigskip

These studies lead us in my view to the conclusion that the work of Kapteyn, his students and  his contemporaries Easton and Pannekoek led to an enormous boost in Galactic astronomy; it provided detailed maps of the surface brightness of the Milky Way and insight into the distribution and dynamics of stars in space and a growing realization that our system might very well be a spiral galaxy. Kapteyn left two ways to proceed for his successors. One was the {\it Plan of Selected Areas} of collecting more and more data, and the other was the insight to include stellar dynamics as the way to move further. His successor in Groningen, Piet van Rhijn, concentrated on collecting more data through dedication to the {\it Plan of Selected Areas}, which in the end did not bring about too much progress in understanding our Galactic environment and context. The other road through stellar dynamics and integration of structure and kinematics led through Jan Oort to breakthrough progress. It is ironic that not van Rhijn, but Oort did the  definitive analysis of the {\it Selected Areas} data to arrive at a crosscut through the Galaxy (Oort, 1938), in the spirit of William Herschel and Jacobus Kapteyn, that showed evidence for us living in a spiral galaxy, as Easton had envisioned.
\vspace{0.5cm}

\noindent
{\Large {\bf Acknowledgments}}
\vspace{0.5cm}

{\normalsize
  I thank in the first place Kristan Vlaardingerbroek, great-great-grandson of Cornelis Easton, for permission to use the {\it Notebook}, and making available some material from the records of the family, and through him his mother Titia Vlaardingerbroek-de Knegt for making this possible, and his father the late Rob Vlaardingerbroek for putting scans of this {\it Notebook} on the Web, that put me on this trail. I am grateful to professors Jan Willem Pel and Klaas van Berkel, astronomer and historian, for reading a draft of this paper and making remarks that improved it. {\it I wish to record with enormous gratitude that Klaas and Jan Willem have read drafts of \underline{all} my biographical publications, safeguarding me for possible historical and astronomical embarrassment.} I thank Maaike van Rossum of the library of the Rijksmuseum Boerhaave in Leiden for assistance with access to the {\it Easton Archives}. I am happy to acknowledge Careljan Rotteveel Mansveld, President of the Stichting Coen van Oven, for allowing me to reproduce the painting of Dr. van Oven. I thank the staff of the Kapteyn Astronomical Institute for support and help and the previous and current Directors, Professors Léon Koopmans and Inga Kamp, for hospitality extended to an Emeritus Professor as Guest Scientist. This research made extensive use of the SAO/NASA Astrophysics Data System (ADS).

}
\vspace{1cm}

{\Large {\bf References}}
\vspace{0.5cm}

{\normalsize

\ \  Abbe, C., 1867. On the distribution of the nebulae in space. {\it Monthly Notices of the Royal Astronomical Society}, 27, 257-264.

Argelander, F.W.A., 1844. {\it Aufforderung an Freunde der Astronomie, zur Anstellung von eben so interessanten und nützlichen, als leicht auszuführenden Beobachtungen über mehrere wichtige Zweige der Himmelskunde}. Published as part of H.C. Schumacher’s Jahresbuch für 1844. Stuttgart \&\ Tübingen: J.G. Cotta, 122–254 (in German). Translated in 1855 as {\it Handleiding voor vrienden der sterrekunde, tot het volbrengen van belangrijke waarnemingen, die geene werktuigen vorderen, uit het Hoogduitsch vertaald door W.F. Kaiser, met eene voorrede, aanteekeningen en bijvoegsels van F. Kaiser}. Zwolle: Der erven Tijl (in Dutch).

Argelander, F.W.A., 1863. {\it Atlas des nördlichen Himmels fur den Anfang des {Jahres} 1855,  unter Mitwirkung der Herrn Professor Dr. E. Schönfeld und Professor Dr. A. Krueger nach der in den Jahren 1852 bis 1862 auf der Königlichen Universitäts-Sternwarte zu Bonn durchgeführten Durchmusterung des n\"ordlichen Himmels entworfen und im Namen der Sternwarte herausgegeben}. Bonn, Adolph Marcus.

Arrhenius, S.A., 1912.  Das Milchstrassen-System. {\it Nord und Süd}, 36, 180-184.

Arrhenius, S.A., 1918. {\it  The destinies of the stars}, translated from the Swedish by J.E. Fries, New York: Putnam’s Sons.

Bauschinger, J.,  1889. J. L. E. Dreyer, Ph.D., A new general {catalog} of nebulae and clusters of stars. {\it Vierteljahresschrift der Astronomischen Gesellschaft}, 24, 43-51.

Bayer, J., 1603. {\it Uranometria omnium asterismorum continens schemata, nova methodo delineata aereis laminis expressa, by Bayer, Johann; Augsburg: excudit Christophorus Mangus}.

Blaauw, A., 2014. Cornelis Easton, {\it Biographical Encyclopedia of Astronomers}, ed. Hockey, T., Trimble, V., Williams, T.R., Bracher, K., Jarrell, R.A., Marché, J.D., II, Palmeri, J.A. \&\ Green, D.W.E.  New York:  Springer, 2014, 633-634.

Boeddicker, O., 1892. {\it The Milky Way from the north pole to 10\degs\  of south declination, drawn at the Earl of Rosse’s observatory at Birr Castle}, London: Longman Green \&\ Co.

Bond, G.P. \&\ Bond, W.C., 1850. Observations on the belts and satellites of Jupiter, and on certain nebulae by the Messrs. Bond. {\it Astronomische Nachrichten}, 30, 93-96.2

Celoria, G., 1877. Sulla distribuzione generale delle stelle nello spazio. {\it Memorie della Societa Degli Spettroscopisti Italiani},  6, B81-B107.

Choi, P.I., Guhathakurta, P. \&\ Johnston, K.V., 2002. Tidal interaction of M32 and NGC205 with M31: Surface photometry and numerical simulations. {\it Astronomical Journal}, 124, 310-331.

de Wilde, I., 2007. {\it Werk maakt het bestaan draaglijk: Barend Sijmons (1853-1935)}, Groningen: Barkhuis Publishing.

Dreyer, J.L.E., 1888. A new general catalogue of nebulae and clusters of stars, being the catalogue of the late Sir John F. W. Herschel, revised, corrected and enlarged. {\it Memoirs of the Royal Astronomical Society}. 49. 1-237.

Dreyer, J.L.E., 1895. Index catalogue of nebulae, found in the years 1888 to 1894, with notes and  corrections to the New General Catalogue. {\it Memoirs of the Royal Astronomical Society}, 51,  185-228.

Dreyer, J.L.E., 1910. Second Index Catalogue of Nebulae and Clusters of Stars, containing objects found in the years 1895 to 1907; with notes and corrections to the New General Catalogue and to the Index Catalogue for 1888-94. {\it Memoirs of the Royal Astronomical Society}, 59,  105-198.

Easton, C., 1893. {\it La Voie Lactée dans l’hémisphère boréal: cinq planches lithographiées, description detaillée, catalogue et notice historique}. Dordrecht: Blussé {\&} cie; Paris: Gauthier-Villars et fils; abstract in {\it Bulletin Astronomique}, Observatoire de Paris, 11, 364 (1894).

Easton, C., 1894. The great nebula in Andromeda. {\it Nature}, 50, 547-548.

Easton, C., 1895a. Sur la distribution apparente des étoiles dans une partie de la Voie Lactée.  {\it Astronomische Nachrichten}, 137, 81-90.

Easton, C., 1895b.  La distance de la Voie Lactée. {\it Bulletin de Société Astronomique de France}, 9, 49-54.

Easton, C., 1895d. On the distance of the stars in the Milky Way, {\it Knowledge}, 18, 179-182.

Easton, C., 1895c. On the distribution of the stars and the distance of the Milky Way in Aquila and Cygnus. {\it Astrophysical Journal}, 1, 216-221.

Easton, C. 1898a. Over de groepeering van de sterren in den Melkweg, {\it Verslagen van de gewone vergaderingen van de Wis- en Natuurkundge Afdeeling van 29 Mei 1897 tot 23 April 1898}, VI. Koninklijke Academie van Kunsten en Wetenschappen. Amsterdam, Johannes Müller.

Easton, C., 1898b.  Richard Proctor’s theory of Universe. {\it Knowledge}, 21,  12-15.

Easton, C., 1898c. A new theory of the Milky Way. {\it Knowledge}, 21, 57-60.

Easton, C., 1900. A new theory of the Milky Way. {\it Astrophysical Journal}, 12, 136–158.

Easton, C., 1902a. Distant worlds. -- A review of some recent studies in stellar distribution, I. {\it Knowledge}, 25, 154-156. 

Easton, C., 1902b. Distant worlds. -- A review of some recent studies in stellar distribution, II. {\it Knowledge}, 25, 176-178. 

Easton, C., 1903a, La distribution de la lumière galactique, comparée à la distribution des étoiles cataloguées, dans la Voie Lactée boréale. {\it Verhandelingen der Koninklijke Akademie van Wetenschappen te Amsterdam}, (eerste sectie ) VIII. N ; 3.

Easton, C., 1903b. What is the Milky Way?. {\it Knowledge}, 26, 151-154.

Easton, C., 1902c. La distribution de la lumière galactique comparée à celle des étoiles relativement brillantes, dans la Voie Lactée boréal. {\it Astronomische Nachrichten}, 159, 169-176.

Easton, C., 1904a. Over de schijnbare verdeeling der nevelvlekken. {\it Verslagen van de gewone vergaderingen der Wis- en Natuurkundige Afdeeling van de Koninklijke Akademie van Wetenschappen}, 13, 82-89. Also: On the apparent distribution of the nebulae. {\it Proceedings of the Royal Netherlands Academy of Arts and Sciences 1904-1905}, 7, 1904-1905, 117-124.

Easton, C., 1904b.  De nevelvlekken in betrekking tot het Melkwegstelsel. {\it Verslagen van de gewone vergaderingen der Wis- en Natuurkundige Afdeeling van de Koninklijke Akademie van Wetenschappen}, 13, 90-96. Also: The nebulae considered in relation to the galactic system. {\it  Proceedings of the Royal Netherlands Academy of Arts and Sciences (KNAW) 1904-1905}, 7, 125-134.

Easton, C., 1904c. La distribution des nébuleuses et leurs relations avec le Système Galactique. {\it Astronomische Nachrichten}, 166, 129-134.

Easton, C., 1913. A photographic chart of the Milky Way and the spiral theory of the Galactic System. {\it  Astrophysical Journal}, 37, 105–118.

Easton, C., 1920. De relativiteitsleer en de donkere plekken in de Melkweg. {\it Hemel \&\ Dampkring}, 18, 102-104. 

Easton, C., 1921. On the distance of the Galactic star-clouds. {\it Monthly Notices of the Royal Astronomical Society}, 81, 215-226.

Easton, C., 1922a. Persoonlijke herinneringen aan J.C. Kapteyn. {\it Hemel \&\ Dampkring}, 20, 112–117 and 151–164. English translation, Personal memories of J.C. Kapteyn,  available on the Website accompanying van der Kruit (2015) under ‘Obituaries’. 

Easton, C., 1922b.  Correlation of the distribution of bright stars and galactic light in Cygnus. {\it Bulletin of the Astronomical Institutes of the Netherlands}, 1, 157-159.

Easton, C., 1922c. Drieduizend, 30.000 of 300.000 lichtjaar. {\it Hemel \&\ Dampkring}, 20, 45-59.

Easton, C., 1922d. Harlow Shapley in ons land. {\it Hemel \&\ Dampkring}, 20, 87-88.

Easton, C. 1928a. {\it Les hivers dans l'Europe occidentale: étude statistique et historique sur leur temp\'erature, discussion des observations thermom\'etriques 1852–1916 et 1757–1851, tableaux comparatifs, classification des hivers 1205–1916, notices historiques sur les hivers remarquables, bibliographie}. Leiden: Brill.

Easton, C., 1928b. A photographic chart of the northern Milky Way. {\it Monthly Notices of the Royal Astronomical Society}, 89, 207-209.

Eddington, A.S., 1914. {\it Stellar Movements and the Structure of the Universe}. London: Macmillan.

ESO (European Southern Observatory), 2009. {\it The Milky Way panorama}. www.eso.org/public/\linebreak[4]images/eso0932a/.

Flammarion, C., 1880, {\it L'Astronomie populaire -- Description générale du ciel}, Paris: C. Marpon et E. Flammarion.

Flammarion, C., 1882. {\it Les étoiles et les curiosit\'es du ciel — Description complète du ciel visible a l’oeil nu et de tous les objets célestes faciles a observer; Supplément de L’Astronomie populaire}. Paris: C. Marpon et E. Flammarion.

Flammarion. C., 1884. {\it De wonderen des hemels}. Zutphen: Thieme.

Flammarion. C., 1885. {\it  Het rijk der sterren}. Zutphen: Thieme.

Flamsteed, J.1725, {\it Historia Coelestis Britannicae, tribus Voluminibus contenta} (1675-1689), (1689-1720), vol. 1, 2, 3 London, H. Meere.

Gill, D. \&\ Kapteyn, J. C. 1896, 1897, 1900. The Cape Photographic Durchmusterung for the equinox 1875. Part I. Zones -18\degs\ to -37\degs; Part II. Zones -38\degs\ to -52\degs; Part III. Zones. -53\degs\ to -89\degs. {\it Annals of the Cape Observatory, South Africa}, 3, 1–845; 4, 1–702; 5, 1–757.

Guichelaar, J., 2009. {\it De Sitter: een alternatief voor Einsteins heelalmodel}. Utrecht: Veen Magazines. ISBN 978-9085711810. Translated as {\it Willem de Sitter. Einstrein’s friend and opponent}. Cham: Springer. ISBN 978-3-319-98336-3.

Heis, E, 1872. Atlas Coelestis Novus. Cologne, DuMont-Schauberg (in German).
(www.lu.lv/en/\linebreak[4]
muzejs/petnieciba/izstades/zemes-miti-debesis-zvaigznu-atlantu-virtuala-izstade/9-eduards-heiss-1872-atlas coeles tis-novus-neuer-himmels-atlas-jaunais-debess-atlants/).

Henkes, B., 2020. Even after the War we stand alone. In Henkes, B. (ed.), {\it Negotiating racial politics in the family: Transnational histories touched by National Socialism and Apartheid}. Leiden: Brill, ISBN 978-9- 004399-66-2, 86-118.

Herschel, J.F.W., 1833.  Observations of nebulae and clusters of stars, made at Slough, with a twenty-feet reflector, between the years 1825 and 1833. {\it Philosophical Transactions of the Royal Society of London}, 123, 359-505 (downloaded from https://www.jstor.org/stable/108003).

Hodge, P.W., 1973. The structure and content of NGC205. {\it Astrophysical Journal}, 182, 671-695.

Hoskin, M.A. 1977. Newton, providence and the universe of stars, {\it Journal of the History of Astronomy}, 8, 77-101.

Hoskin, M. 1982. The first drawing of a spiral nebula, {\it Journal for the History of Astronomy},  13, 97-101.

Hoskin, M.A., 2008. Nebulae, star clusters and the Milky Way: From Galileo to William Herschel. {\it Journal for the History of Astronomy}, 39, 363-396.

{Houzeau}, J.-C., 1878. Uranométrie Générale. {\it Annales de l’Observatoire Royal de Bruxelles. Astronomie, Nouveau  Series}, I.

IAU (International Astronomical Union), 2022. {\it The constellations}. www.iau.org/public/themes/\linebreak[4]
constellations/.

Jeans, J.H., 1902. The stability of a spherical nebula. {\it Philosophical Transactions of the Royal Society of London. Series A, Containing Papers of a Mathematical or Physical Character}, 199, 1-53.

Jeans, J.H., 1919. {\it Problems of cosmogony and stellar dynamics}. Cambridge University Press.

Jeans, J.H., 1922. The motions of stars in a Kapteyn-Universe. {\it Monthly Notices of the Royal Astronomical Society}, 82, 122–132.

Kapteyn., J.C., \&\ van Rhijn, 1920. On the distribution of the stars in space especially in the high Galactic latitudes. {\it Astrophysical Journal}, 52, 23-38.

Kapteyn, J.C., 1922. First attempt at a theory of the arrangement and motion of the Sidereal System. {\it Astrophysical Journal, 55, 302–328}.

Keeler, J.E., 1899. New nebulæ discovered photographically with the Crossley reflector of the Lick Observatory. {\it Monthly Notices of the Royal Astronomical Society}, 60, p.128-130.

Klein, H.J., 1867. Beobachtungen über die scheinbaree Ausdehnung der Milchstrasse. {\it Wochenschrift für Astronomie, Meteorologie und Geographie}, 10, 285-288. books.google.nl/books/about/\linebreak[4]
Wochenschrift\_für\_Astronomie\_Meteorolog.html?id=vWw5AAAAcAAJ\&redir\_esc=y.

Kormendy, J. \&\ Bahcall, J.N., 1974. Faint envelopes of galaxies. {\it Astronomical Journal},  79, 671-677.

Kronk, G.W., 2003. {\it Cometography: A Catalog of Comets. Vol. 2: 1800–1899}. Cambridge University Press.

Kronk, G.W., 2007. {\it Cometography: A Catalog of Comets. Vol. 3: 1900–1932}. Cambridge University Press.

Leiden Observatory, 1908.{\it Album Amicorum}, presented to H.G. van de Sande Bakhuyzen. See digitalcollections.universiteitleiden.nl/view/item/1987454.

Liouville, J., 1838. Sur la théorie de la variation des constantes arbitraires. {\it Journal de mathématiques pures et appliquées}, 3, 342–349.

Lindblad, B., 1925. Star-streaming and the structure of the stellar system. {\it Arkiv för Matematik, Astronomi och Fysik}, Band 19A, No. 21, 1-8.

Lindblad, B., 1926. Star-streaming and the structure of the stellar system (Second paper). {\it Arkiv för Matematik, Astronomi och Fysik}, Band 19B, No. 7, 1-6.

Lowe, S. \&\ North, C., 2025. {\it  Cosmos: The Infographic Book of Space}. See cosmos-book.github.io/\linebreak[4]
catalogue/NGC/.

Marth, A., 1872. Co-ordinates of stars within and near the Milky Way. {\it Monthly Notices of the Royal Astronomical Society},  33, 1-14.

Marth, A., 1873a. Co-ordinates of stars within and near the Milky Way. {\it Monthly Notices of the Royal Astronomical Society}, 33, 517-527.

Marth, A., 1873b. Supplementary list of co-ordinates of stars within or near the Milky Way. {\it Monthly Notices of the Royal Astronomical Society},  34, 77-82.

Oort, J.H., 1926. {\it Niet-lichtgevende materie in het sterrenstelsel}. University of Leiden. Reprinted in Dutch in {\it Hemel \&\ Dampkring} 25, 13–21 and 60–70. Translated by P.C. van der Kruit into English as {\it Non-light emitting matter in the stellar system} in van der Kruit \&\ van Berkel (2000), Appendix A. 

Oort, J.H., 1927. Observational evidence confirming Lindblad’s hypothesis of a rotation of the Galactic System. {\it Bulletin of the Astronomical Institutes of the Netherlands}, 3, 275–282.

Oort, J.H., 1936. Bouw der sterrenstelsels, {\it Hemel \&\ Dampkring}, 34, 1–16.

Oort, J.H., 1938. Absorption and density distribution in the Galactic System. {\it Bulletin of the Astronomical Institutes of the Netherlands},  8, 233-263.

Oort, J.H., 1981. Some notes on my life as an astronomer. {\it Annual Review of Astronomy and Astrophysics}, 19, 1–5.

Oort, J.H. 1985.  Blaauw's scientific work. In: {\it Birth and evolution of massive stars and stellar groups}. Ed. W. Boland and H. van Woerden. Dordrecht: Reidel.   ISBN 978-90-277-2135-8, p.295-304.

Pannekoek, A., 1898a. II. New charts for inserting the Milky Way. {\it Popular Astronomy}, 5, 485–488 (also, {\it Journal of the British Astronomical Association}, 8, 80–82, 1897).

Pannekoek, A., 1919. The distance of the Milky Way. {\it Monthly Notices of the Royal Astronomical Society}, 79, 500–507.

Pannekoek, A., 1920. Die Nördliche Milchstrasse. {\it Annalen van de Sterrewacht te Leiden}, 11, 1–89 (ADS incorrectly listed the year as 1929) (in German).

Pannekoek, A., 1922. The distance of the Galaxy in Cygnus. {\it Bulletin of the Astronomical Institutes of the Netherlands}, 1, 54–56.

Pannekoek, A.,1933. Photographische photometrie der nördlichen Milchstrasse nach negativen auf der Sternwarte Heidelberg (Königsstuhl) aufgenommen von Max Wolf. {\it Publications of the Astronomical Institute of the University of Amsterdam}, 3, 1-71, Tafel I-III, Karte I-VIII.

Paul, E.R., 1981. The Death of a Research Programme - Kapteyn and the Dutch Astronomical Community. {\it Journal for the History of Astronomy}, 12, 77-94.

Paul, E.R., 1993. {\it The Milky Way Galaxy and statistical cosmology, 1890-1924}, Cambridge University Press, ISBN 0-521-353-637.

Poisson, S.D., 1823). Mémoire sur la théorie du magnétisme en mouvement. {\it Mémoires de l'Académie Royale des Sciences de l'Institut de France}, 6, 441–570.

Proctor, R.A., 1869. A new theory of the Milky Way, {\it Monthly Notices of the Royal Astronomical Society}, 30, 50-56.

Proctor, R.A., 1873. Statement of views respecting the Sidereal Universe. {\it Monthly Notices of the Royal Astronomical Society}, 33, 539-552.

Roberts, I., 1888. Photographs of the nebulae M31, h44, and h51 Andromedae, and M 27 Vulpecula. {\it Monthly Notices of the Royal Astronomical Society}, 49, 65-66.

Roberts, I., 1889. Photographs of the nebulae in the Pleiades and in Andromeda. {\it Monthly Notices of the Royal Astronomical Society}, 49, 120-121.

Roberts, I., 1894. {\it A Selection of photographs of stars, star clusters and nebulae together with information concerning the Instruments and the methods employed in the pursuit of celestial photography}. London: The Universal Press.

Rosse, 1850. Observations on the nebulae, {\it Philosophical Transactions of the Royal Society of London}, 140, 499-514 (downloaded from www.jstor.org/stable/108449).

SEDS (Students for the Exploration and Development of Space), 2025. Website spider.seds.org/\linebreak[4]
ngc/ngc.html.

Shostak, G.S. \&\ van der Kruit, P.C., 1984. Studies of nearly face-on spiral galaxies. II. HI synthesis observations and optical surface photometry of NGC628. {\it Astronomy and Astrophysics}, 132, 20-32.

Stein, J.W.J.A, SJ, 1929. Dr. C. Easton, in memoriam, {\it Hemel \&\ Dampkring},  27, 225-236 and 257-271.

Stratonoff, W., 1900. Études sur la structure de l’Universe, {\it Publications de l’Observatoire Astronomique et Physique de Tachkent}, 2,  1-114, 23 plates.

Tai, C., 2021. {\it Anton Pannekoek, Marxist astronomer: Photography, epistemic virtues, and political philosophy in early twentieth-century astronomy}. PhD thesis, University of Amsterdam, the Netherlands

Tobin, W. \&\ Holberg, J.B., 2008. A newly-discovered accurate early drawing of M51, the Whirlpool Nebula. {\it Journal of Astronomical History and Heritage}, 11, 107-115.

Toomre, A., 1964. On the gravitational stability of a disk of stars. {\it Astrophysical Journal}, 139, 1217-1238.

Trouvelot,  L., 1882. {\it The Trouvelot astronomical drawings}. Drawings available at publicdomain\linebreak[4]
review.org/collection/the-trouvelot-astronomical-drawings-1882/. Manual published by C. Scribner’s sone,  New York, available at https://archive.org/details/trouvelotastrono00trourich/\linebreak[4]
trouvelotastrono00trourich/.

Valentiner, W., 1897 - 1902. {\it   Handwörterbuch der Astronomie} (4 Bände in 5 Büchern), Encyklopaedie der Naturwissenschaften,  Abtheilung III, Theil II. Breslau: E. Trewendt.

van Delft, D., 2021.  {\it Reiziger in het wereldruim. De astronoom Henk van de Hulst 1918-2000}.   Amsterdam: Prometheus. ISBN 97-8904-4647-723.

van den Berg, R., 2022. {\it Een passie voor precisie. Frederik Kaiser (1808-1872). Vader van de Leidse Sterrewacht}. Amsterdam: Prometheus, ISBN 97-8904-4651-478.

van der Kruit, P.C., 1979. Optical surface photometry of eight spiral galaxies studied in Westerbork. {\it Astronomy and Astrophysics, Supplement Series}, 38, 15-38.

van der Kruit, P.C., 1986. Surface photometry of edge-on spiral galaxies. V. The distribution of luminosity in the disk of the Galaxy derived from the Pioneer 10 background experiment. {\it Astronomy and Astrophysics}, 157, 230–244.

van der Kruit, P.C., 1990. The Galaxy in relation to other galaxies. In: {\it The Milky Way as a galaxy}, by I. King, G.Gilmore, \&\ P.C. van der Kruit. Mill Valley, University Science Books. 331-346.

van der Kruit, P.C., and van Berkel, K., 2000. {\it The Legacy of J.C. Kapteyn: Studies on Kapteyn and the development of modern astronomy}. Dordrecht: Kluwer. ISBN-13: 978-1-4020-0374-5.

van der Kruit, P.C,. and Freeman, K.C., 2011. Galaxy disks. {\it Annual Review of Astronomy and Astrophysics}, 49, 301–371.

van der Kruit, P.C., 2015. {\it Jacobus Cornelius Kapteyn: Born investigator of the heavens}. ChaM: Springer (accompanying Webpage www.astro.rug.nl/JCKapteyn).

van der Kruit, P.C., 2019. {\it Jan Hendrik Oort: Master of the Galactic System}. Cham: Springer (accompanying Webpage www.astro.rug.nl/JHOort).

van der Kruit, P.C., 2021a. {\it Pioneer of Galactic Astronomy: A biography of Jacobus Cornelius Kapteyn}. Cham: Springer.

van der Kruit, P.C., 2021b. {\it Master of Galactic Astronomy: A biography of Jan Hendrik Oort}. Cham: Springer.

van der Kruit, P.C., 2021c. Karl Schwarzschild, Annie J. Cannon and Cornelis Easton: the honorary PhDs of Jacobus C. Kapteyn. {\it Journal for Astronomical History and Heritage}, 24, 521–543.

van der Kruit, P.C., 2022. Pieter Johannes van Rhijn, Kapteyn’s Astronomical Laboratory and the Plan of Selected Areas. {\it Journal of Astronomical History and Heritage}, 25, 341–438.

van der Kruit, P.C., 2024a. Pannekoek’s Galaxy. {\it Journal of Astronomical History and Heritage}, 27, 135–199.

van der Kruit, P.C., 2024b. Jan Veth's paintings of Jacobus Kapteyn,
{\it Journal of Astronomical History and Heritage}, 27, 559-578.

van Oven, J.C.,  1930. Levensbericht van Cornelis Easton, 10 September 1864-3 Juni 1929. {\it Jaarboek van de Maatschappij der Nederlandse Letterkunde} 1930,  43-48.

van Woerden, H. 1985.  Adriaan Blaauw and the revival of Groningen astronomy. In: {\it Birth and evolution of massive stars and stellar groups}. Ed. W. Boland and H. van Woerden. Dordrecht: Reidel.   ISBN 978-90-277-2135-8, p.321-345.

von Humboldt, F.W.H.A., 1845-1862. {\it Kosmos -- Entwurf einer physischen Weltbeschreibung}, Stuttgart: Johann Georg Cotta.

Walterbos, R.A.M, Kennicutt, R.C.,  1987. Multi-color photographic surface photometry of the Andromeda galaxy. {\it Astronomy and Astrophysics, Supplement Series}, 69, 311-332.

Wijfjes, H., 2004. {\it Journalistiek in Nederland 1850 - 2000. Beroep, cultuur en organistatie}. Amsterdam: Boom, ISBN 978-90-535-2949-2.

Willink, B., 1991. Origins of the second Golden Age of Dutch science after 1860. {\it Social Studies of Science}, Vol.21(3), doi.org/10.1177/030631291021003004.

Willink, B., 1998. {\it De tweede Gouden Eeuw: Nederland en de Nobelprijzen voor natuurwetenschappen 1870–1940}. Amsterdam: Bakker.

Wikimedia, 2019. See commons.wikimedia.org/wiki/File:Portret van de schilder Herman \mbox{Gunneweg} en zijn echtgenote Anna Maria Caroline Roterman, 1895 - 1905.jpg.

}

\vspace{1cm}

\noindent
{\LARGE {\bf Appendix: Stellar dynamics}}
\bigskip

{\large

  This appendix summarizes for those not familiar with this the history and development of stellar dynamics as a background to the discussion in section 13.2.
  
Our Galaxy is a spiral galaxy. The distribution of matter in the disks of these systems is not completely smooth, but shows {\it structure} on all scales.  Studies of our and other spiral nebulae have shown that the disk component maintains a large-scale equilibrium configuration supported almost entirely by rotation and to a lesser extent by the random motions of stars. The rotation velocity of a class of objects (old stars, young stars, gas clouds, etc.) is somewhat smaller than the circular velocity corresponding to a centrifugal force equal to the gravitational force. This difference (asymmetric drift) is larger for objects with larger random motions because of the dynamical effect of the latter. The old disk stars have larger random motions (of order 50 km/sec) than the interstellar gas or very young stars (of order 10 km/sec), so the neutral hydrogen gas observed with the 21-cm line rotates faster than the collective rotation of the old stars.

Galactic rotation was discovered by Jan Hendrik Oort (1927). He followed up the suggestion by Bertil Lindblad (1895--1965) that the Galaxy consists of ellipsoids with different rotation speeds (Lindblad, 1925, 1926). With this Oort found an explanation for the phenomenon of high-velocity stars. These are part of a more spherical distribution than the disk, supported mainly by high random motions and with much less systematic rotation and these stars  therefore move at high velocity with respect to us and from directions restricted to one hemisphere. Lindblad assumed these ellipsoidal components to have uniform density so that the rotation velocity necessary to support them was proportional to distance from the center and they would consequently rotate as a solid body. Oort felt that this was unjustified and that mass should concentrate {toward} the center and the angular rotation velocity should decrease with galactocentric distance. This differential rotation would result in a systematic pattern of radial velocities and proper motions in the solar neighborhood, for which he indeed found observational evidence.

In addition to this dynamical support there is the matter of the gravitational instability, the tendency for the formation of local condensations that would grow by gravitational contraction. On smaller scales stability against gravitational collapse is provided by the random motions as first described by James Hopwood Jeans (1877--1946) in the form of his well-known Jeans criterion for a gaseous sphere, which says that for a specific density and temperature (random motion) stability against collapse is possible only up to a particular size and corresponding mass, the Jeans length and mass. Alar Toomre (b. 1937) has supplemented this with the notion that on large scales stability against gravitational collapse in a galactic disk is provided by the shear of differential rotation. His famous Toomre criterion then states  that for stability in galactic disks to hold {\it on all  scales} the minimum scale for local stability by rotational shear has to be equal to or smaller than the Jeans length (Toomre, 1964). Galactic disks appear to be marginally stable in the Toomre sense. Marginal stability ensures that structure in disks is possible and can be  maintained, but is  prevented from growing out of control. For more on this and references, see van der Kruit \&\ Freeman (2011).

In  studies of the structure of the sidereal system in the nineteenth century dynamics, equilibrium and gravitational instabilities, that are part of what we now would demand of a ‘theory’,  played no role. Every textbook starts with two fundamental equations of stellar dynamics that were formulated already  early in the nineteenth century. The first is the Liouville equation (Liouville, 1838) -- sometimes collisionless Boltzmann equation -- that relates the time evolution of the distribution of gravitating particles and their velocities to the gravitational potential. Joseph Liouville (1809--1882) did the fundamental work for it in 1838. The second is the Poisson equation (Poisson, 1823), which relates the gravitational potential to the density distribution of gravitating matter. Baron  Siméon Denis Poisson (1781--1840) published it in 1823. As to gravitational stability: the Jeans mass dates from 1902; it was applied extensively in theories of star formation, but application to disk stability entered much later, as did Toomre’s criterion.

That the matter of gravitational stability was not an issue earlier is interesting. The work of Isaac Newton (1642--1727) contained the basics. Induced by correspondence, starting in 1692, with Richard Bentley (1662--1742), theologian and classical scholar, he addressed the question why the stars did not collapse onto each other in view of the gravitational pull each star would exert on every other one (see e.g. Hoskin, 1977). Newton argued that the pull of distant stars was small. But while the pull from each star diminishes with distance, their number increases compensating for this, so that the resulting pull from a distant shell uniformly filled with stars was nil. For nearer stars this required a high degree of regularity in the distribution. In Hoskin’s words: `the Sun can be regarded as surrounded by other stars in such a way that the stars at a given distance taken together are dynamically similar to a uniform shell, and so exercise at most a slight resultant gravitational force on the Sun’. Even then there was a `need for providential intervention to preserve the system in good order, since God has not in fact chosen to give it the perfect symmetry’ required. It took until much later before it was realized, particularly in Jeans’ work, that the distribution being not perfectly symmetrical was compensated for by the random motions of the stars.
\vspace{1cm}

\noindent
{\LARGE {\bf Notes}}
\vspace{0.5cm}

One of the referees of this paper urged me to  supply literary translations into English  of titles of publications, journals, newspapers, books, etc., that were in Dutch, German, French, etc. I give below a list of these to honor this urgent request. He/she also requested a list of acronyms with explanation. This follows the list of titles.
\bigskip

\noindent
-- {\it Album Amicorum}: Album from Friends.

\noindent
-- {\it  Anniversaire du Bureau de Longitude}: Yearbook of the Office of Longitudes.

\noindent
-- {\it Astronomie populaire}: Popular Astronomy.

\noindent
-- {\it Astronomische Nachrichten}; Astronomical Messages.

\noindent
-- {\it Atlas Coelestis Novus}: New Atlas of the Heavens.

\noindent
-- {\it Atlas des nördlichen Himmels f\"ur den Anfang des Jahes 1855,  nach der in den Jahren 1852 bis 1862 auf der Königlichen Universitäts-Sternwarte zu Bonn durchgeführten Durchmusterung des Nordlichen Himmels}: Atlas of the northern sky at the beginning of the year 1855, after the in the years 1852 to 1862  at the Royal University Observatory in Bonn produced Survey of the Northern Sky.

\noindent
-- {\it Aufforderung an Freunde der Astronomie}: Request to friends of astronomy.

\noindent
-- {\it Bonner Durchmusterung}: Bonn Survey.

\noindent
-- {\it Bulletin de la Société Astronomique de France et revue mensuelle d'astronomie, de méteorologie et de physique du globe}: Bulletin of the Astronomical Society of France and monthly review of astronomy, meteorology and physics of the Earth.

\noindent
-- {\it Das Milchstrassen-System}: The Milky Way System.

\noindent
-- {\it Dordtsche Courant}: Dordrecht Newspaper.

\noindent
-- {\it Elseviers Geïllustreerd Maandschrift}: Elsevier's [a publisher] Illustrated Monthly. 

\noindent
-- {\it Haagsche Maandblad}: The Hague Monthly.

\noindent
-- {\it Handwörterbuch der Astronomie}: Concise Dictionary of Astronomy.

\noindent
-- {\it Hemel \&\ Dampkring}: Sky \&\ Atmosphere.

\noindent
-- {\it Historia Coelestis Britannica}: History of the Heavens of Britain.

\noindent
-- {\it Kosmos -- Entwurf einer physischen Weltbeschreibung}: Cosmos -- Draft of a physical description of the world.

\noindent
-- {\it l’Astronomie}: Astronomy.

\noindent
-- {\it La Voie Lactée boréal}; The northern Milky Way.

\noindent
-- {\it La Voie Lact\'ee dans l’espace, vue de son pôle boréal}: The Milky Way in space as seen from its north pole.

\noindent
-- {\it La Voie Lactée dans l’hémisphère boréal: cinq planches lithographiées, des\-cription detaillée, catalogue et notice historique}: The Milky Way in the northern hemisphere; five lithographic plates, detailed description, catalog and historical notes.

\noindent
-- {\it Les étoiles et les curiosités du Ciel}: The stars and the curiosities of the sky.

\noindent
-- {\it Les hivers dans l'Europe occidentale. Étude statistique et historique sur leur temp\'erature; discussion des observations th\`ermom\'etriques 1852-1916 et 1757-1851; tableaux comparatifs; classification des hivers 1205-1916; notices historiques sur les hivres remarquables}: The winters in Western Europe. Statistical and historical study of their temperatures; discussion of the temperature observations 1851-1916 and 1757-1851; tables  for comparison; classification of winters 1205-1916; historical remarks on the remarkable winters.

\noindent
-- {\it Nieuwe Rotterdamsche Courant}: New Rotterdam Newspaper.

\noindent
-- {\it Nieuws van den Dag}: News of the Day.

\noindent
-- {\it Nord und Süd}: North and South.

\noindent
-- {\it  Uranometria Omnium Asterismorum}: Uranometry (chart)  of all constellations.

\noindent
-- {\it Uranométrie Générale}: General uranometry (chart).
\bigskip

\noindent
Acronymns:
\bigskip

\noindent
-- {B.A.N.}: Bulletin of the Astronomical Institutes of the Netherlands

\noindent
-- {HBS}: Hoogere Burgerschool (Higher Citizens School)

\noindent
-- {N.R.C.}: Nieuwe Rotterdamsche Courant

\noindent
-- {R.A.S.}:  Royal Astronomical Society

\noindent
-- {PhD}:  Doctor of Philosophy

}
\end{document}